\documentclass[useAMS,usenatbib]{mn2e}
\usepackage{graphicx,txfonts}

\title{Characterization of Dark-Matter-induced anisotropies in the diffuse gamma-ray background}

\author[Fornasa et al.]{
  \parbox{\textwidth}{
    Mattia Fornasa$^{1,2,3}$,
    Jes\'us Zavala$^{4,5,6}$,
    Miguel A. S\'anchez-Conde$^{7}$, \\
    Jennifer M. Siegal-Gaskins$^{8,9}$,
    Timur Delahaye$^{10}$,
    Francisco Prada$^{1,10,11}$, \\
    Mark Vogelsberger$^{12}$, 
    Fabio Zandanel$^{1}$ and
    Carlos S. Frenk$^{13}$
  }
  \vspace{0.4cm} \\
  \parbox{\textwidth}{
    $^{1}$ Instituto de Astrof\'isica de Andaluc\'ia (IAA - CSIC), Glorieta de la Astronom\'ia, Granada, Spain \\
    $^{2}$ MultiDark fellow, E-mail:{\ttfamily mattia.fornasa@nottingham.ac.uk} \\
    $^{3}$ School of Physics and Astronomy, University of Nottingham, University Park, Nottingham, NG7 2RD, United Kingdom \\
    $^{4}$ Department of Physics and Astrophysics, University of Waterloo, 200 University Avenue West, Waterloo, Canada \\
    $^{5}$ Perimeter Institute for Theoretical Physics, 31 Caroline St. N., Waterloo, ON, N2L 2Y5, Canada \\
    $^{6}$ CITA National Fellow, E-mail:{\ttfamily jzavalaf@uwaterloo.ca} \\
    $^{7}$ SLAC National Accelerator Laboratory \& Kavli Institute for Particle Astrophysics and Cosmology, 2575 Sand Hill Road, Menlo Park, CA, 94025, USA \\
    $^{8}$ Einstein fellow \\
    $^{9}$ California Institute of Technology, Pasadena, CA 91125 USA \\
    $^{10}$ Instituto de F\'isica Te\'orica UAM/CSIC, Universidad Aut\'onoma de Madrid Cantoblanco, 28049, Madrid, Spain \\
    $^{11}$ Campus of International Excellence UAM/CSIC, Cantoblanco, E-28049 Madrid, Spain \\
    $^{12}$ Harvard-Smithsonian Center for Astrophysics, 60 Garden St., Cambridge, MA 02138, USA \\
    $^{13}$ Institute for Computational Cosmology, Dept. of Physics, University of Durham, South Road, Durham, DM1 3LE, United Kingdom \\
  }
}

\begin{document}


\maketitle

\begin{abstract}
The Fermi-LAT collaboration has recently reported the detection of angular 
power above the photon noise level in the diffuse gamma-ray background 
between 1 and 50 GeV. Such signal can be used to constrain a possible 
contribution from Dark-Matter-induced photons. We estimate the intensity and 
features of the angular power spectrum (APS) of this potential Dark Matter 
(DM) signal, for both decaying and annihilating DM candidates, by constructing 
template all-sky gamma-ray maps for the emission produced in the galactic 
halo and its substructures, as well as in extragalactic (sub)halos. The DM 
distribution is given by state-of-the-art $N$-body simulations of cosmic 
structure formation, namely Millennium-II for extragalactic (sub)halos, and 
Aquarius for the galactic halo and its subhalos. 
We use a hybrid method of extrapolation to account for (sub)structures that 
are below the resolution limit of the simulations, allowing us to estimate 
the total emission all the way down to the minimal self-bound halo mass.
We describe in detail the features appearing in the APS of our template
maps and we estimate the effect of various uncertainties such as the value of 
the minimal halo mass, the fraction of substructures hosted in a halo 
and the shape of the DM density profile. Our results indicate that the 
fluctuation APS of the DM-induced emission is of the same order as the 
Fermi-LAT APS, suggesting that one can constrain this hypothetical emission 
from the comparison with the measured anisotropy. We also quantify the 
uncertainties affecting our results, finding ``theoretical error bands'' 
spanning more than two orders of magnitude and dominated (for a given particle 
physics model) by our lack of knowledge of the abundance of low-mass 
(sub)halos.
\end{abstract}

\section{Introduction}
\label{sec:Introduction}
The isotropic gamma-ray background (IGRB) is the radiation that remains after 
the resolved sources (both extended and point-like) and the galactic 
foreground (produced by the interaction of cosmic rays with the interstellar 
medium) are subtracted from the all-sky gamma-ray emission.
A {\it guaranteed} component of the IGRB is the emission of unresolved known 
sources, whose contribution has been estimated from population studies
of their resolved counterparts: {\it blazars} \citep{Stecker:1993ni,
Stecker:1996ma,Muecke:1998cs,Narumoto:2006qg,Dermer:2007fg,Pavlidou:2007dv,
Inoue:2008pk,Collaboration:2010gqa,Abazajian:2010pc,Stecker:2010di,
Singal:2011yi}, {\it star-forming galaxies} \citep{Bhattacharya:2009yv,
Fields:2010bw,Makiya:2010zt,Ackermann:2012by,Lacki:2012si,Chakraborty:2012sh}, 
{\it radio galaxies} \citep{Stawarz:2005tq,Massaro:2011ww,Inoue:2011bm}, 
{\it pulsars and milli-second pulsars} \citep{FaucherGiguere:2009df,
SiegalGaskins:2010mp}, {\it Gamma-Ray Bursts} \citep{Casanova:2008zz} and 
{\it Type Ia Supernovae} \citep{Lien:2012gz}. Additional processes may also 
contribute to the IGRB such as cosmological structure formation shocks 
\citep[e.g.][]{Loeb:2000na,Gabici:2002fg}, and interactions of cosmic rays 
(CRs) with the extragalactic background light (EBL) \citep{Kalashev:2007sn} 
or with small solar system bodies \citep{Moskalenko:2009tv}.

Current estimates, however, suggest that the total unresolved emission from 
the classes listed above is not able to account for the whole IGRB intensity
\citep[e.g.][]{Ajello:2011}, which strengthens the possibility that 
additional, unconfirmed sources are required to match the data. 
Gamma rays from Dark Matter (DM) annihilation or decay could explain the 
missing emission.

DM is the dominant matter component of the Universe, responsible for 
approximately one quarter of the energy density today 
\citep[e.g.][]{Jarosik:2010iu}. 
We know little about its nature, apart from the fact that it has to be
non baryonic. A well-studied class of DM candidates is that of Weakly 
Interacting Massive Particles (WIMPs), whose masses and interactions (set by 
the scale of weak interactions), offer promising non-gravitational signals for 
their detection in the near future. Within the context of annihilating DM, 
WIMPs are favoured by the fact that they naturally have a relic density that 
matches the observed DM abundance \citep[e.g.][]{Kolb,Bertone:2004pz}, while 
for decaying DM, it has been shown that WIMPs can have a decay lifetime larger 
than the age of the Universe, and are therefore viable DM candidates (see e.g. 
\citealt{Bolz:2000fu}, \citealt{Arvanitaki:2008hq}). WIMPs are also appealing 
because their existence is predicted by fundamental theories beyond the 
Standard Model of Particle Physics, such as Supersymmetry (SUSY), Universal 
Extra-Dimensions or models with $T$-parity. In this paper we assume that DM 
is made of WIMPs, without making a specific assumption about the theoretical 
particle physics model from which WIMPs arise. 

This work is concerned with {\it indirect detection} of DM, i.e., the 
possibility of revealing the presence of DM from detection of its annihilation 
or decay products. In particular, we focus here on the case of gamma rays as 
by-products, studying the possible contribution to the IGRB coming from the 
DM annihilations (or decays) in the smooth DM halo of the Milky Way (MW) and 
its galactic subhalos, as well as from extragalactic (sub)halos. These 
contributions have already been estimated in the past using analytical and 
numerical techniques \citep[e.g.][]{Ullio:2002pj,Taylor:2002zd,Ando:2005hr,
Ando:2005xg,Ando:2006cr,SiegalGaskins:2008ge,Ando:2009fp,Fornasa:2009qh,
Zavala:2009zr,Ibarra:2009nw,Hutsi:2010ai,Cirelli:2010xx,Zavala:2011tt}. The 
recent Fermi-LAT measurement of the energy spectrum of the IGRB has been used 
to put constraints on the nature of the DM candidate by requiring that the 
DM-induced emission should not exceed the observed IGRB \citep{Abdo:2010dk,
Hutsi:2010ai,Zavala:2011tt,Calore:2011bt}. The constraints derived are quite 
competitive: for instance, the most optimistic scenario considered by 
\citet{Abdo:2010dk} puts a strong upper limit to the annihilation cross 
section, which is already of the order of the thermal relic value for a DM 
particle lighter than 200-300 GeV.

The energy spectrum is not the only piece of information we can extract from 
the IGRB. Thanks to the good angular resolution of Fermi-LAT, it is also 
possible to measure its angular power spectrum (APS) of anisotropies. 
\citet{Ackermann:2012uf} reported a detection of angular power in the 
multipole range between $\ell=155$ and 504 with a significance that goes from 
7.2$\sigma$ (in the energy bin between 2 and 5 GeV) to 2.7$\sigma$ (between 
10 and 50 GeV), which represents the first detection of intrinsic anisotropies 
in the IGRB.

There are different predictions for the normalization and shape of the APS 
produced by different populations of unresolved sources, both astrophysical 
\citep{Ando:2006mt,Ando:2009nk,SiegalGaskins:2010mp} and associated with DM 
\citep{Ando:2005hr,Ando:2005xg,Ando:2006cr,Cuoco:2006tr,Cuoco:2007sh,
Taoso:2008qz,SiegalGaskins:2008ge,Fornasa:2009qh,SiegalGaskins:2009ux,
Ando:2009fp,Zavala:2009zr,Ibarra:2009nw,Cuoco:2010jb}. The comparison of 
these predictions with the Fermi-LAT APS data can, in principle, constrain 
the contribution of each source class to the IGRB \citep{Cuoco:2012yf}. The 
analysis from \citet{Ackermann:2012uf} seems to suggest an interpretation in 
terms of a single population of unresolved, unclustered objects, due to the 
fact that the APS is roughly scale-independent over the energy range 
analysed. This recent measurement can then be used to complement other 
constraints on a possible DM contribution to the IGRB. In the present paper, 
we take a first step in obtaining such constraints by revisiting and updating 
the prediction of the DM-induced emission (through decay and annihilation) 
and its associated APS, as well as estimating the uncertainties involved. The 
comparison of these predictions with the Fermi-LAT APS data will be done 
in a follow-up study.

In order to compute the DM-induced APS we combine the results of two $N$-body 
simulations of the galactic (Aquarius, hereafter 
AQ, \citealt{Springel:2008cc}) and extragalactic (Millennium-II, hereafter 
MS-II, \citealt{BoylanKolchin:2009nc}) DM structures, to construct all-sky maps 
of the gamma-ray emission coming from the annihilation and decay of DM in the 
Universe around us. Although we only focus here on the study of the 
anisotropy patterns in the gamma-ray emission, these maps represent 
{\it per se} a useful tool for future projects on indirect DM detection and we
plan to make them available shortly after the publication of the follow-up
paper dedicated to the comparison with the Fermi-LAT APS data.

The extragalactic component is expected to be almost isotropic 
\citep[see, e.g.,][]{Zavala:2009zr}, while the smooth galactic one is 
characterized by an intrinsic anisotropy, as a consequence of our position in 
the MW halo. The presence of galactic subhalos, however, reduces the expected 
gradient of the DM-induced gamma-ray flux as one moves away from the Galactic 
Center (GC). In fact, due to the large abundance of substructures and their 
more extended distribution, strong gamma-ray emission is also expected quite 
far away from the GC \citep[as it can be seen, e.g., in][]{Springel:2008by,
Fornasa:2009qh,Cuesta:2010ex,SanchezConde:2011ap}.

Even though numerical simulations represent the most reliable method to model 
the non-linear evolution of DM, they are limited by resolution. Since the 
minimum self-bound mass ($M_{\rm min}$) of DM halos is expected to be many 
orders of magnitude below the capabilities of current simulations\footnote{The 
actual value of $M_{\rm min}$ is related to the nature of the DM particle, with 
typical values covering a quite large range, approximately between 
$10^{-12} M_\odot$ and $10^{-3} M_\odot$ \citep[e.g.][]{Profumo:2006bv,
Bringmann:2009vf}.}, this poses a challenge for an accurate prediction and 
represents one of our largest sources of uncertainty 
\citep[e.g.][]{Taylor:2002zd,Springel:2008by,SiegalGaskins:2008ge,
Ando:2009fp,Fornasa:2009qh,Zavala:2009zr,Kamionkowski:2010mi,
SanchezConde:2011ap,Pinzke:2011ek,Gao:2011rf}. To address this problem, we use 
a hybrid method that models the (sub)halo population below the mass 
resolution of the simulations by extrapolating the behaviour of the resolved 
structures in the MS-II and AQ simulations towards lower masses.
Furthermore, we compute multiple sky maps with different values of $M_{\rm min}$ 
to determine with more precision what is the impact of this parameter on the 
the DM-induced emission. We also consider possible effects due to different DM
density profiles for the smooth halo of the MW.

Such a detailed study of the uncertainties associated with the APS allows us to 
quantify, in addition to the normalization and shape of the APS, a 
``theoretical uncertainty band'', that will prove to be useful in the 
comparison of our predictions with the Fermi-LAT APS data.

The paper is organized as follows. In Sec. \ref{sec:particle_physics} we 
describe the mechanisms responsible for the gamma-ray emission from DM 
annihilation or decay. We then present how the data from the MS-II and 
AQ simulations are used to construct template maps of DM-induced 
gamma-ray emission from extragalactic DM (sub)halos 
(Sec. \ref{sec:Extragalactic}) and from the smooth galactic halo and its 
subhalos (Sec. \ref{sec:Galactic}).
In Sec. \ref{sec:energy_APS_spectra} we present the energy and angular power 
spectra, discussing the different components and estimating their 
uncertainties. We discuss the implications of our results in 
Sec. \ref{sec:discussion}, while Sec. \ref{sec:conclusion} is devoted to a 
summary and our conclusions.

\section{Dark-Matter-induced gamma-ray emission}
\label{sec:particle_physics}
In the case of DM annihilation, the gamma-ray intensity (defined as the number 
of photons collected by a detector per unit of area, time, solid angle and 
energy) produced in a direction $\Psi$ is:
\begin{eqnarray}
\label{eqn:annihilation_flux}
\frac{d\Phi}{dE}(E_\gamma,\Psi) & = &
\frac{(\sigma_{\rm ann}v)}{8\pi m_\chi^2} \int_{\rm l.o.s.} d\lambda \, \sum_i B_i
\frac{dN^i_\gamma(E_\gamma(1+z))}{dE} \times \\
& & \rho^2(\lambda(z),\Psi) \, e^{-\tau_{\rm EBL}(z,E_\gamma)}, \nonumber
\end{eqnarray}
where $E_\gamma$ is the observed photon energy, $m_\chi$ is the mass of the 
DM particle and $(\sigma_{\rm ann}v)$ its annihilation cross section. The sum 
runs over all annihilation channels, each one characterized by a branching 
ratio, $B^i$, and photon spectrum (yield), $dN_\gamma^i/dE$, computed at the 
energy of emission. The integration is over the line of sight (parameterized 
by $\lambda$) to account for the redshift-dependent DM density field 
$d\lambda=c \, dz \, H(z)^{-1}$. The exponential factor accounts for photon 
absorption from pair production due to interactions with the EBL along the 
line of sight, parametrized by an optical depth $\tau_{\rm EBL}(z,E_\gamma)$, 
which we take from the model developed in 
\citet{Dominguez:2010bv}\footnote{We have not checked the effect of other EBL 
attenuation models, since for the energies we consider in this work (from 0.5 
GeV to 50 GeV), the contribution of the damping $e^{-\tau_{\rm EBL}}$ factor is 
marginal.}. 
The first part of the integrand in Eq. \ref{eqn:annihilation_flux} is usually 
referred to as the ``particle physics factor'' and only depends on the 
properties of DM as a particle, whereas the second part is called the 
``astrophysical factor'' and depends on how DM is distributed in 
space\footnote{The particle physics and astrophysical factors are not 
completely independent from each other: the presence of $M_{\rm min}$, which is 
fixed by the particle physics nature of the DM candidate, determines the 
minimum (sub)halo mass scale to be considered. Moreover, the dependence on 
redshift is both for the DM distribution and for the energy in the photon 
yield $dN_\gamma/dE$.}.

In the case of DM decay, Eq. \ref{eqn:annihilation_flux} should be re-written 
as:
\begin{eqnarray}
\label{eqn:decay_flux}
\frac{d\Phi}{dE}(E_\gamma,\Psi) & = &
\frac{1}{4\pi m_\chi \tau} \, \int_{\rm l.o.s.} d\lambda 
\sum_i B_i \frac{dN^i_\gamma(E_\gamma(1+z))}{dE} \times \\
& & \rho(\lambda(z),\Psi) \, e^{-\tau_{\rm EBL}(z,E_\gamma)}, \nonumber
\end{eqnarray}
where the decay lifetime $\tau$ is used instead of the annihilation cross 
section and the dependence on density is linear instead of quadratic.

In the current section we describe the particle physics factor, introducing 
the mechanisms of gamma-ray production considered. 
Secs. \ref{sec:Extragalactic} and \ref{sec:Galactic} are devoted to the 
astrophysical factor.

As mentioned in the Introduction, rather than considering a specific particle
physics model, we focus on a general WIMP candidate, which, for our purposes, 
is completely defined by $m_\chi$, $(\sigma_{\rm ann} v)$ or $\tau$, and its 
gamma-ray photon yield. The latter receives contributions from three 
different mechanisms of emission:
\begin{itemize}
\item {\it Prompt emission}: This radiation comes from the products of DM 
annihilation/decay directly, without any interaction with external particles. 
Within this first category, one can distinguish three different processes: 
$i)$ gamma-ray lines from direct annihilation/decay into photons, $ii)$ 
hadronization of quarks followed by neutral pion decay into photons and $iii)$ 
gamma rays from final state radiation and internal bremsstrahlung whenever 
there are charged final states. For DM annihilation, the branching ratios for 
monochromatic lines are usually subdominant and quite model-dependent (at 
least for SUSY models), while for DM decay, emission lines may be more 
prominent \citep{Choi:2009ng,Vertongen:2011mu,GomezVargas:2011ph}. In this 
work, we do not consider the emission from monochromatic lines, instead, we 
focus on mechanisms $ii)$ and $iii)$, which are characterized by a continuum 
emission \citep[e.g.][]{Fornengo:2004kj,Bertone:2005xz,Bergstrom:2004cy,
Bringmann:2007nk}. The continuum emission induced by hadronization shows some 
dependence on the DM mass and the particular annihilation/decay channel, but 
it is a mild one and the shape is more or less universal. Finally, internal 
bremsstrahlung may also contribute inducing harder spectra and the possibility 
of bumps near the energy cut-off set by $m_\chi$ (see 
e.g. \citealt{Bringmann:2007nk,Bringmann:2008kj,Bringmann:2012vr}).
\bigskip
\item {\it Inverse Compton (IC) up-scattering}: This secondary radiation 
originates when low energy background photons are up-scattered by the leptons 
produced by DM annihilation/decay. Since large $\gamma$ factors are required, 
usually one focuses on the case of electrons and positrons interacting with 
the Cosmic Microwave Background (CMB) photons and with starlight (either 
directly or re-scattered by dust). The amplitude of the IC emission and its 
energy spectrum depends on the injection spectrum of $e^+/e^-$ and on the 
energy density of the background radiation fields 
\citep{Colafrancesco:2005ji,Profumo:2009uf,Zavala:2011tt}. 
For massive DM candidates, those IC photons can fall within the energy range 
detected by Fermi-LAT and, in some cases, represent a significant 
contribution to the DM-induced emission \citep{Profumo:2009uf,Meade:2009iu,
Hutsi:2010ai,Pinzke:2011ek}. See Appendix \ref{sec:IC_emission} for details on 
the computation of the IC emission. We note that for the case of extragalactic 
DM (sub)halos (Sec. \ref{sec:Extragalactic}) we only consider the CMB as a 
background source. This is mainly because the bulk of the emission comes 
from small (sub)halos (see Sec. \ref{sec:energy_spectrum}) that are 
essentially empty of stars and therefore lack any starlight background 
\citep[see e.g.][]{Profumo:2009uf,Zavala:2011tt}. 
On the other hand, for the case of the MW smooth halo, a complete model for 
the MW radiation field is used (see Sec. \ref{sec:smooth_halo}).
\bigskip
\item {\it Hadronic emission}: This radiation comes from the interaction of 
hadrons produced by DM annihilation/decay with the interstellar gas, and 
its contribution depends on the injection spectrum of hadrons and on the 
spatial distribution of ambient gas. To implement this component we follow 
the method described in \citet{Timur:2011vv}
\citep[see also][]{Cholis:2011un,Vladimirov:2010aq} and present the details 
of the calculation in Appendix \ref{sec:hadronic_emission}. We only consider 
this additional component for the case of the MW smooth DM halo.
\end{itemize}
\begin{figure}
\includegraphics[width=0.45\textwidth]{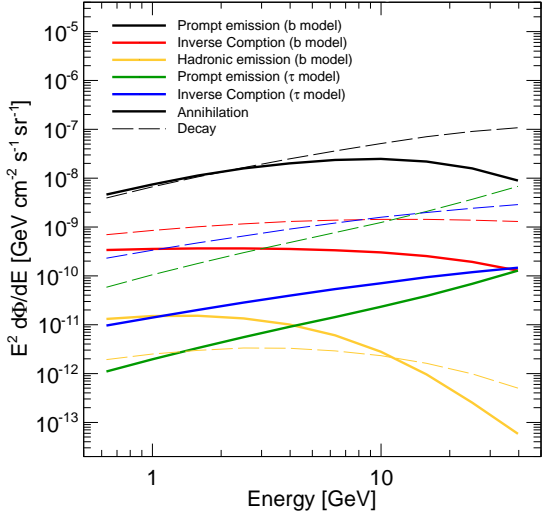}
\caption{\label{fig:energy_spectra} Gamma-ray intensity from DM annihilation (solid lines) and decay (dashed lines) coming from the MW smooth halo (see Sec. \ref{sec:smooth_halo}). For the ``$b$-model'' (black, red and yellow lines) the mass of the DM particle is 200 GeV for the case of annihilation and 2 TeV for decay. We assume $(\sigma_{\rm ann}v)=3 \times 10^{-26}$cm$^{3}$s$^{-1}$ and $\tau=2 \times 10^{27}$s, respectively. For the ``$\tau$-model'' (blue and green lines) the parameters are the same except for the mass which is 2 TeV for both annihilating and decaying DM. Black and blue lines indicate prompt-emission, red and green IC emission, and yellow hadronic emission. The latter is not shown for the $\tau$-model.}
\end{figure}
As benchmarks, in the remainder of this paper, we consider two commonly-used 
annihilation/decay channels, with which we illustrate the role of the 
different mechanisms: a ``$b$-model'' for annihilation/decay entirely into 
$b\bar{b}$ quarks ($B_b=1$) and a ``$\tau$-model'' for annihilation/decay 
into $\tau^+\tau^-$ ($B_\tau=1$). The photon and $e^+/e^-$ yields are computed 
using the tables presented in \citet{Cirelli:2010xx}. For both cases, we fix 
the annihilation cross section and decay life time to 
$3 \times 10^{-26}$cm$^{3}$s$^{-1}$ and $2 \times 10^{27}$s, respectively. The
DM mass is selected to be 200 GeV for the $b$ channel in the case of 
annihilating DM and 2 TeV otherwise. These values are chosen to be slightly 
below the most recent exclusion limits set by the Fermi-LAT data 
\citep{collaboration:2011wa,Dugger:2010ys,Huang:2011xr}.

In Fig. \ref{fig:energy_spectra} we compare the gamma-ray production 
mechanisms listed above. The lines indicate the energy spectrum of the 
emission from annihilation (solid) or decay (dashed) of DM in the MW smooth 
halo (see Sec. \ref{sec:smooth_halo}). For the $b$-model, prompt emission 
(black lines) always dominates over IC (red lines) and hadronic emission 
(yellow lines), both for annihilation and decay. On the other hand, for the 
$\tau$-model, IC (blue lines) overcomes the prompt-emission (green lines) at 
low energies. For the $\tau$-model, hadronic emission is negligible and is 
not plotted.

\section{The gamma-ray emission from extragalactic (sub)halos}
\label{sec:Extragalactic}

\subsection{Resolved (sub)halos in the Millennium-II simulation (EG-MSII)}
\label{sec:Millennium_II}
The MS-II follows the formation and evolution of DM structures in a comoving 
cube of $L=100$\,Mpc$/h$ on a side and a total of $(2160)^3$ simulation 
particles \citep{BoylanKolchin:2009nc}. The simulation is done within the 
context of a WMAP1 cosmology with the following parameters: $\Omega_m=0.25$, 
$\Omega_\Lambda=0.75$, $h=0.73$, $\sigma_8=0.9$ and $n_s=1$; where $\Omega_m$ 
and $\Omega_{\Lambda}$ are the contribution from matter and cosmological 
constant to the mass/energy density of the Universe, respectively, $h$ is the 
dimensionless Hubble constant parameter at redshift zero, $n_s$ is the 
spectral index of the primordial power spectrum, and $\sigma_8$ is the rms 
amplitude of linear mass fluctuations in 8 Mpc$/h$ spheres at redshift zero. 
Its mass resolution is $6.89 \times 10^6 M_\odot/h$ and there are 68 snapshots 
recording the particle distribution at different redshifts between $z=127$ 
and $z=0$.

Instead of working directly with the particles in the simulations, we use
the MS-II (sub)halo catalogs, which are constructed using a friend-of-friends 
(FOF) algorithm \citep{Davis:1985rj} and the SUBFIND code 
\citep{Springel:2000qu} that identifies self-bound substructures within FOF 
halos. Dealing with the (sub)halo catalogs, instead of the particle data,
has two advantages: it is much less expensive computationally and, more 
importantly, it avoids resolution effects near the centre of DM (sub)halos, 
where the simulation particles severely underestimate the DM density (note 
that this is precisely the region with the highest gamma-ray production rate). 
On the other hand, we are neglecting the contribution from the DM mass that 
does not belong to (sub)halos. The emission rate from unclustered regions, 
however, is likely to be negligible especially for DM annihilations (see, 
e.g. \citealt{Angulo:2009hf} who analytically estimated that between 80-95\% 
of the mass is in collapsed objects. In the case of decaying DM this suggests
that, by neglecting unbound particles, we underestimate the luminosity by,
at most, 20\%).

The MS-II (sub)halo catalog contains the global properties needed for each 
object: its virial mass $M_{200}$ (defined as the mass up to $r_{200}$, where 
the enclosed density is 200 times the critical density), its maximum circular 
velocity $V_{\rm max}$ and the radius $r_{\rm max}$ where this velocity is 
attained. The latter two quantities completely determine the 
annihilation/decay luminosity for each halo if we assume that they have a 
spherically symmetric density distribution given by a Navarro-Frenk-White 
(NFW) profile \citep{Navarro:1996gj}. The number of gamma rays (per unit of 
time and energy) coming from a (sub)halo with a boundary at $r_{200}$ is then 
given by $L=f_{\rm PP}L'$, where:
\begin{equation}
f_{\rm PP} = \frac{(\sigma_{\rm ann} v)}{2 m_{\chi}^2} \sum_i B_i
\frac{dN^i_\gamma}{dE},
\end{equation}
\begin{equation}
L' \equiv L_{\rm ann} = 1.23 \frac{V_{\rm max}^4}{G^2 r_{\rm max}} 
\left[ 1 - \frac{1}{(1+c_{200})^3} \right],
\label{eqn:annihilation_luminosity}
\end{equation}
for the case of annihilation, and
\begin{equation}
f_{\rm PP} = \frac{1}{m_{\chi}\tau} \sum_i B_i \frac{dN^i_\gamma}{dE},
\end{equation}
\begin{equation}
L' \equiv L_{\rm decay} = 2.14 \frac{V_{\rm max}^2 r_{\rm max}}{G}
\left[ \ln(1+c_{200}) - \frac{c_{200}}{1+c_{200}} \right],
\label{eqn:decay_luminosity}
\end{equation}
for the case of decay.

The concentration $c_{200}$ is also determined from $V_{\rm max}$ and 
$r_{\rm max}$ inverting the following relation \citep[e.g.][]{Springel:2008cc}:
\begin{equation}
14.426 \left( \frac{V_{max}}{H(z) \, r_{max}} \right)^2 = 
\frac{200}{3} \frac{c_{200}^3}{\ln(1+c_{200})-c_{200}/(1+c_{200})}.
\end{equation}

The choice of the NFW density profile is motivated by its universality and by 
the fact that it gives a good fit to simulated DM (sub)halos over a large 
mass range. However, assuming other DM density profiles could have an impact 
on the total gamma-ray emission, as well as on the shape and normalization of 
the APS. A discussion about this possible source of uncertainty is left for 
Sec. \ref{sec:discussion}.

We define as EG-MSII the signal coming from (sub)halos in the MS-II catalogs 
with at least 100 particles; below this number, the mass and abundance of DM 
objects in the MS-II can become unreliable. This sets an ``effective'' mass 
resolution of $M_{\rm res}=6.89 \times 10^{8}M_\odot/h$ for the extragalactic 
contribution. DM structures with less than a few thousand particles can be 
affected by numerical effects (gravitational softening and two-body 
relaxation, see e.g. \citealt{Diemand:2003uv}) that could influence the values 
of $V_{\rm max}$ and $r_{\rm max}$ and, as a consequence, $L_{\rm ann}$ or 
$L_{\rm decay}$. We implement the prescription described in \citet{Zavala:2009zr} 
to correct for these effects.

In order to simulate the past light cone we need to probe a volume which 
is much larger than the MS-II box. To do this, we follow closely the procedure 
given in \citet{Zavala:2009zr} which can be summarized as follows. The region 
around the observer is divided into concentric shells, each of them centered 
in redshift space on the discrete values $z_i$ corresponding to each 
simulation output. 
The volume defined by each shell has a fixed size in redshift space and a 
corresponding comoving thickness which is filled with identical, 
non-overlapping copies of the MS-II box at the redshift $z_i$ 
\citep[see Fig. 9 of][]{Zavala:2009zr}. 
In order to compute the DM-induced gamma-ray emission from a given direction 
$\Psi$, we follow the line of sight defined by $\Psi$ that crosses the MS-II 
replicas, and sum up the emission produced in all the (sub)halos encountered. 
The projection into a 2-dimensional map is done with the HEALPix 
package\footnote{http://healpix.jpl.nasa.gov} \citep{Gorski:2004by}, assuming 
{\ttfamily N\_side}=512, corresponding to an angular area of approximately 
$4 \times 10^{-6}$ sr for each pixel. If a given halo subtends an area larger 
than this value, then it is considered as an extended source. In this case, 
each of the pixels covered by the particular halo is filled with a fraction 
of the total halo luminosity, assuming the corresponding projected surface 
density profile.

To avoid the repetition of the same structures along the line of sight (which 
would introduce spurious periodicity along this direction), 
\citet{Zavala:2009zr} used an independent random rotation and translation of 
the pattern of boxes that tessellates each shell. This method, however, still 
leaves a spurious angular correlation at a scale $\Delta\theta$ corresponding 
to the comoving size of the simulation box, which mainly manifests itself as a 
peak in the APS centered on $\ell^\star=2\pi/\Delta\theta$. This angular scale 
decreases as we go deeper in redshift, since each copy of the MS-II cube 
covers smaller and smaller angles. This implies that the periodicity-induced 
peak will be located at a different multipole for each shell. Once the 
contributions from all shells are added up, this effect is largely averaged 
out, and the total APS is free from any evident features 
\citep[see Fig. 12 of][]{Zavala:2009zr}. Nevertheless, the spurious angular 
periodicity, in addition to the fundamental angular correlation associated 
with $\Delta\theta$, introduces smaller scale harmonics that affect multipoles 
larger than $\ell^\star$. 
Although these additional peaks are much smaller than the fundamental one, we 
decided to reduce this spurious effect by randomly rotating and translating 
every single replica within the past light cone instead of doing so only for 
every concentric shell.

\begin{figure}
\includegraphics[width=0.45\textwidth]{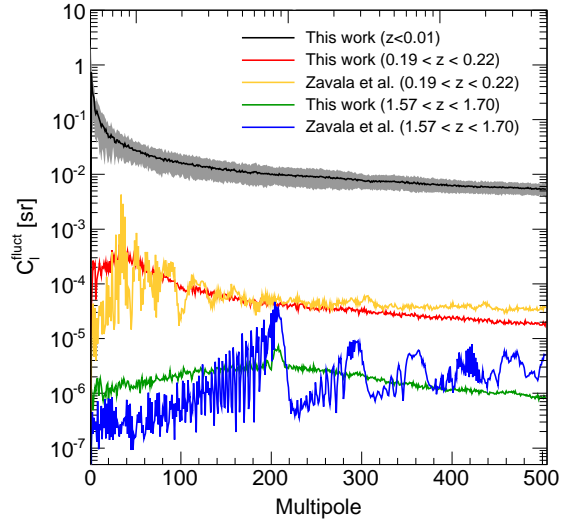}
\caption{\label{fig:old_vs_new_code} Fluctuation APS of anisotropies of the gamma-ray intensity produced by DM annihilating in extragalactic (sub)halos resolved in the MS-II, and located in a shell corresponding to $z<0.01$ (black line), $0.19<z<0.22$ (red and yellow lines) and $1.57<z<1.70$ (green and blue lines). Yellow and blue lines refer to the map-making algorithm presented in \citet{Zavala:2009zr}, while the black, red and green lines correspond to our improved algorithm; see text for details. The grey band around the black line indicates the $1\sigma$ standard deviation among 10 different realizations of the first shell ($z<0.01$).}
\end{figure}

The improvement of the new method becomes evident in Fig. 
\ref{fig:old_vs_new_code} where we show the comparison between the fluctuation 
APS\footnote{The APS will be formally defined in Sec. \ref{sec:APS}.} (for 
individual shells) computed with our map-making code (red and green lines) 
and the original one by \citet{Zavala:2009zr} (yellow and blue lines). The 
yellow and red lines refer to the shell with $z=0.21$, while the blue and 
green lines are for $z=1.63$. The small-scale spurious harmonics essentially 
disappear in the new method and the fundamental mode, although still present, 
is greatly reduced relative to the previous method.

The map-making code produces realizations of the distribution of DM halos 
around the observer through random rotations and translations of the MS-II 
boxes that fill the volume of the past-light cone. 
In order to quantify the effect of this random component in the simulated 
signal, we generate 10 different realizations of the first shell 
(corresponding to $z<0.01$) and compute the fluctuation APS for each of them. 
We only consider the effect of having different random rotations for the 
first shell since it is expected to be more important for nearby resolved 
structures, while shells at larger redshifts are less affected. In 
Fig. \ref{fig:old_vs_new_code} we plot the average APS over these 10 
realizations (black line) as well as the $1\sigma$ fluctuation (grey band). 
We can see that the effect induced on the APS is relatively small (at least 
compared to the other sources of uncertainties introduced later) and we 
neglect it from now on.

All halos up to $z=2$ are considered when computing the extragalactic signal. 
By this redshift, the cumulative emission has already reached $\gtrsim 80\%$ 
of the total signal (in the case of prompt emission $\sim90\%$ of the signal 
actually comes from $z<1$; see Fig. 9 of \citealt{Profumo:2009uf} and Fig. 11 
of \citealt{Zavala:2009zr}). The first shell of the extragalactic map starts at 
a distance of $R_{\rm min}=583$~kpc, corresponding to approximately twice the 
virial radius of the galactic halo. The volume within this distance is filled 
with the data from the AQ simulation (see Sec. \ref{sec:Galactic}).

\begin{figure*}
\includegraphics[width=0.45\textwidth]{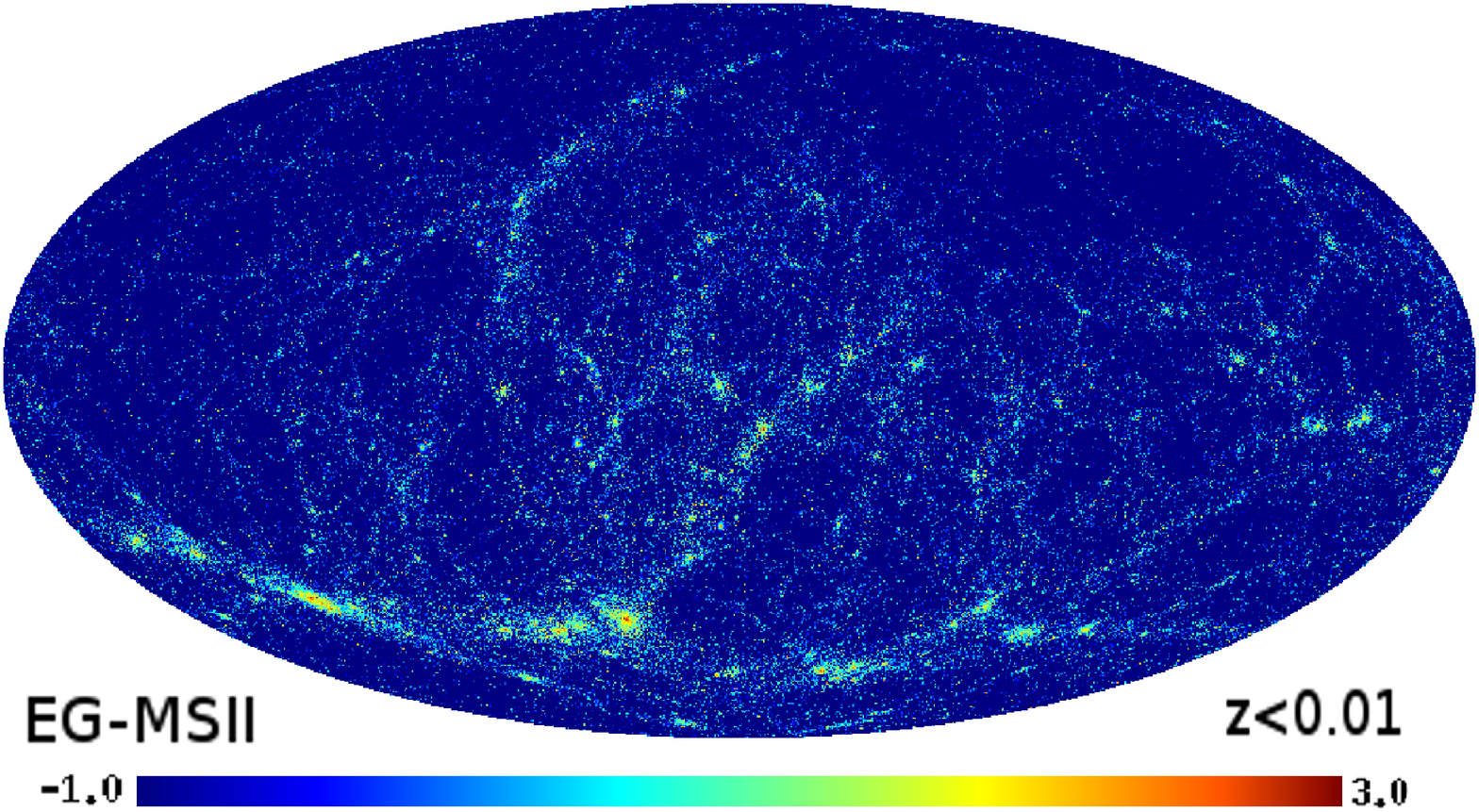}
\includegraphics[width=0.45\textwidth]{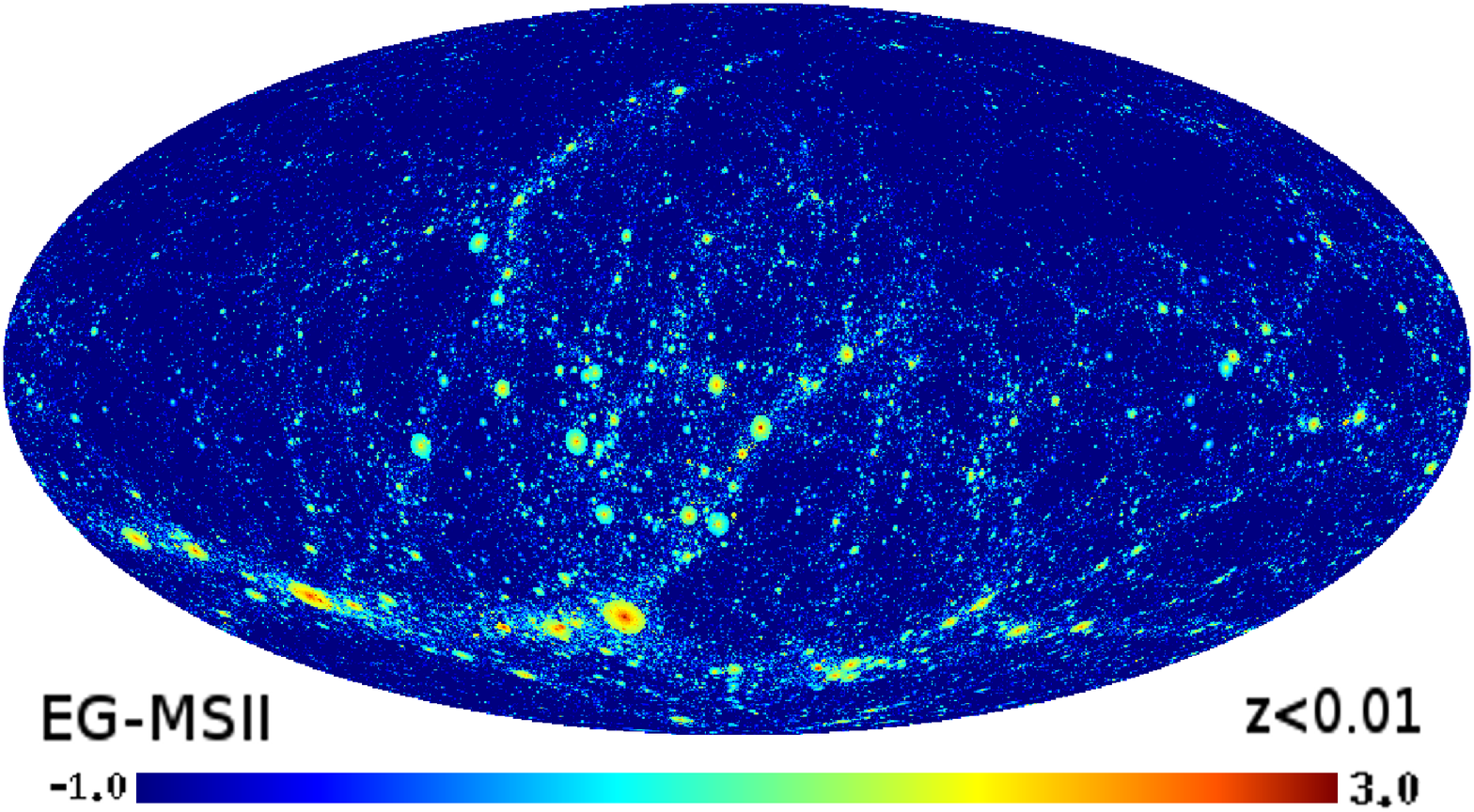}
\includegraphics[width=0.45\textwidth]{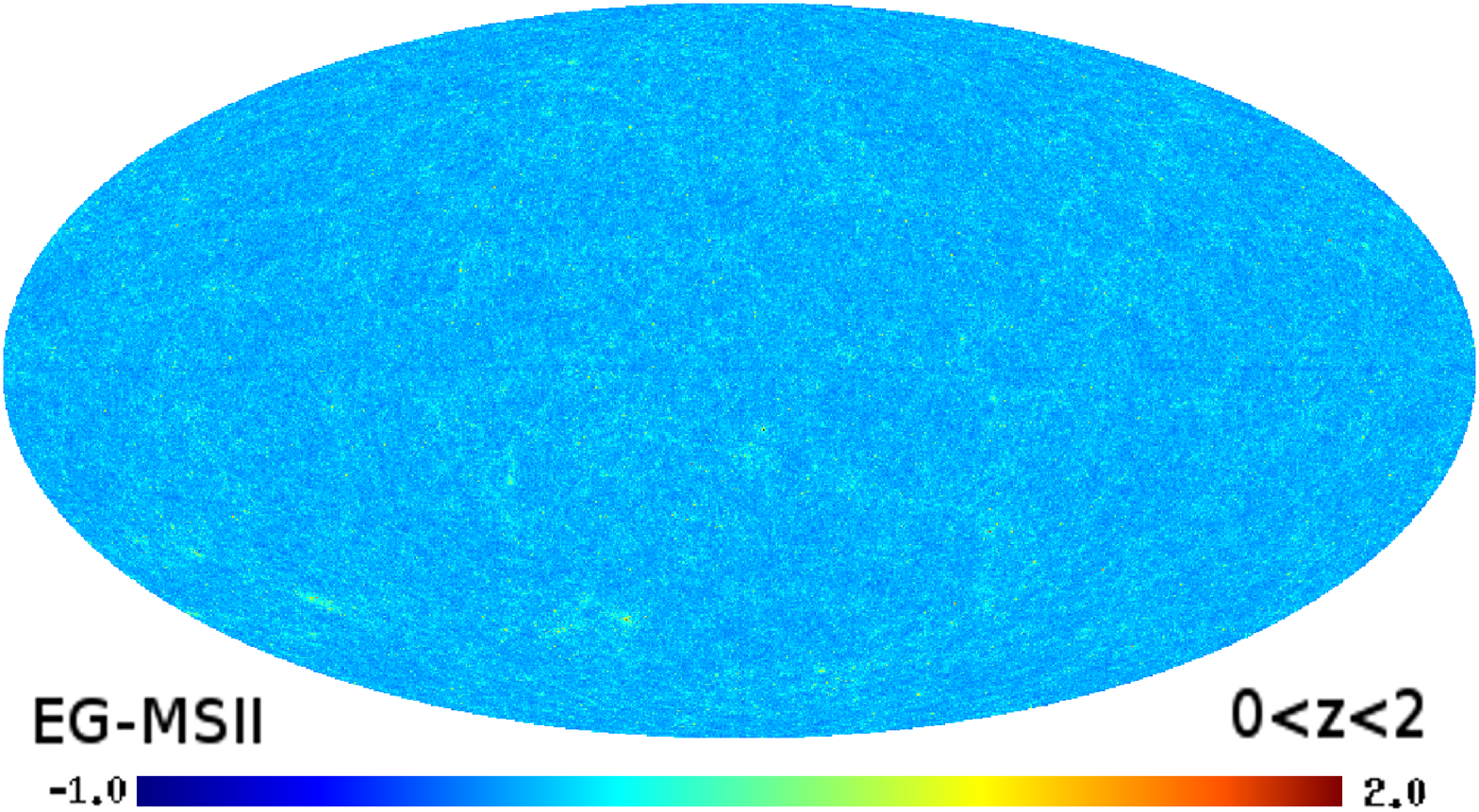}
\includegraphics[width=0.45\textwidth]{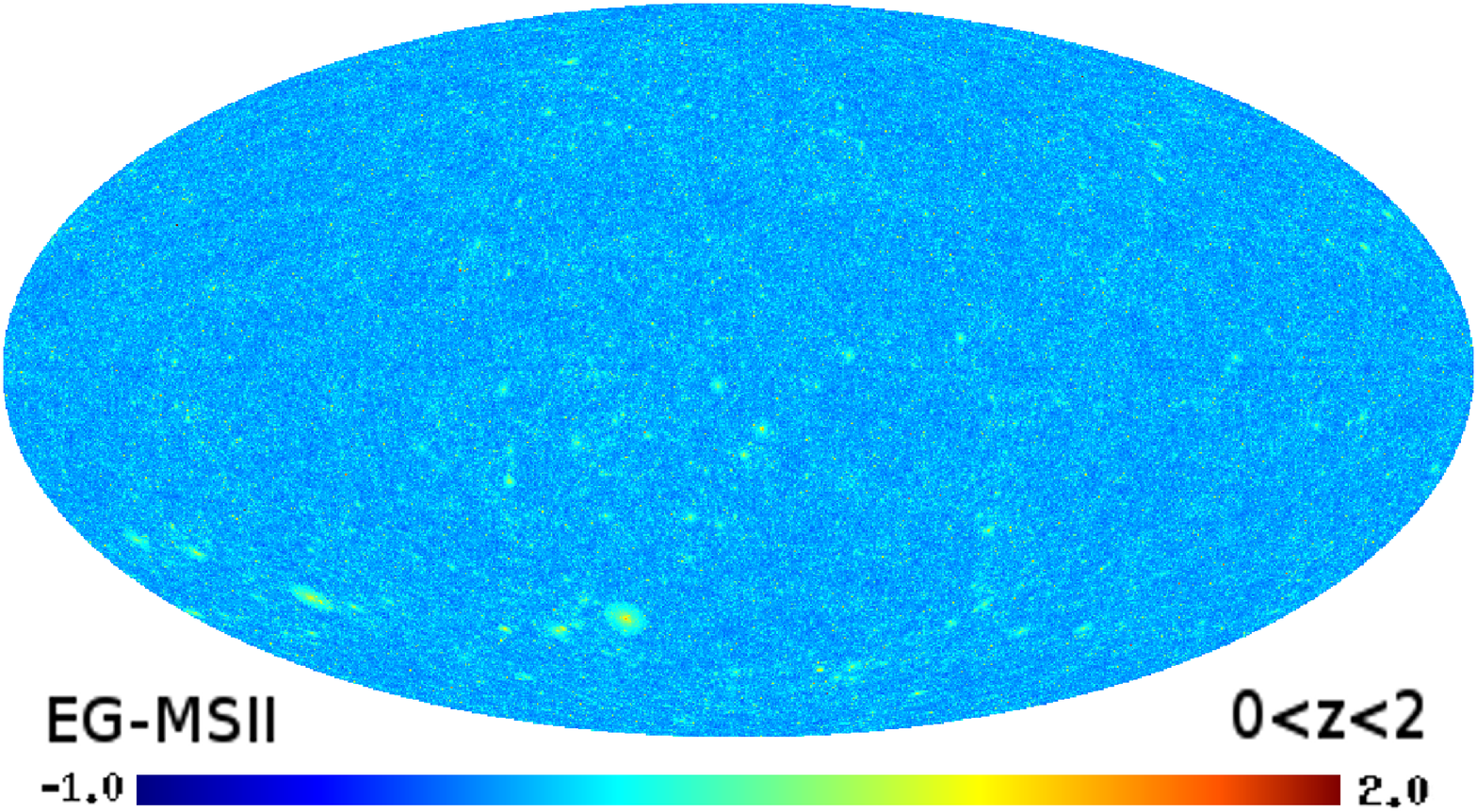}
\includegraphics[width=0.45\textwidth]{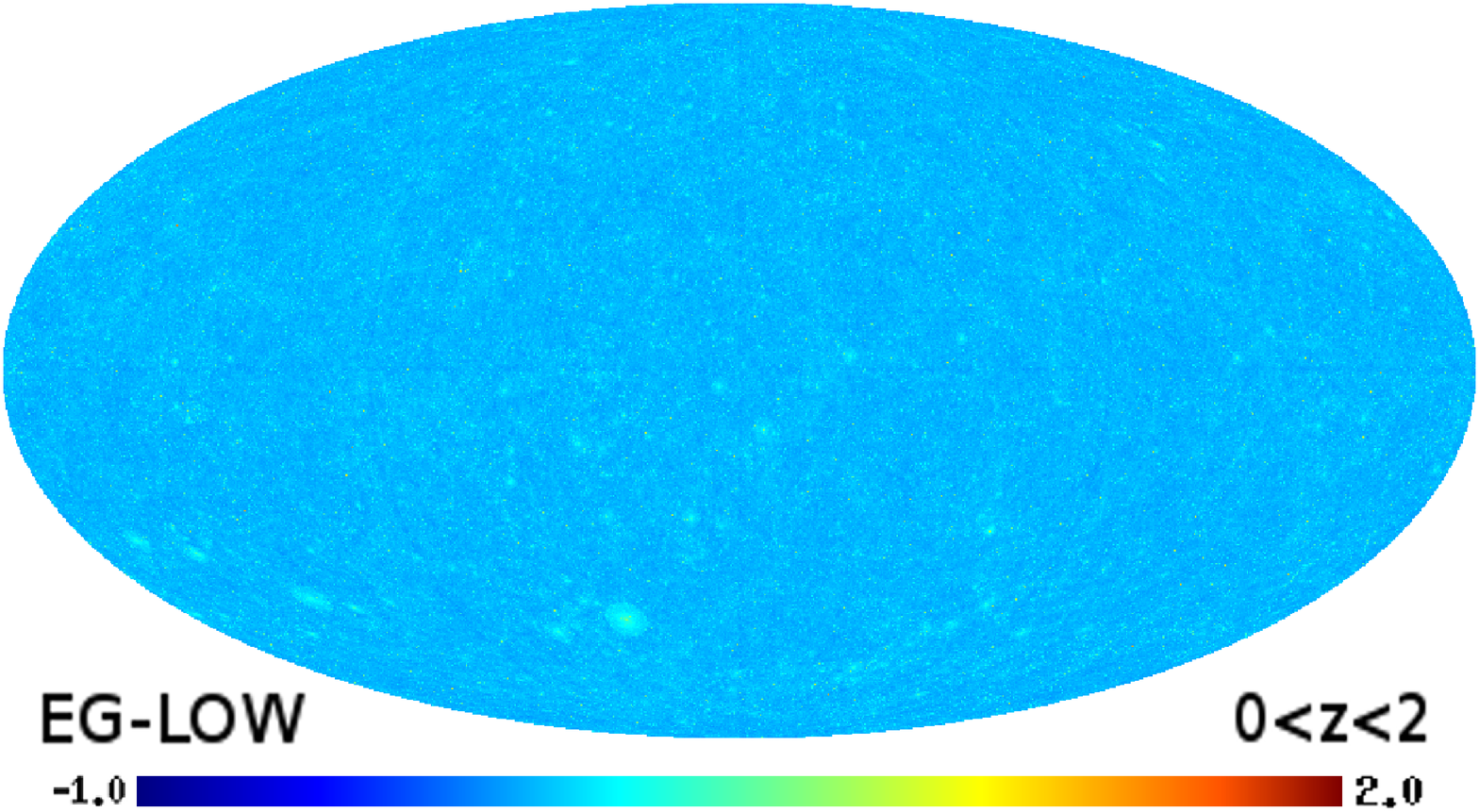}
\includegraphics[width=0.45\textwidth]{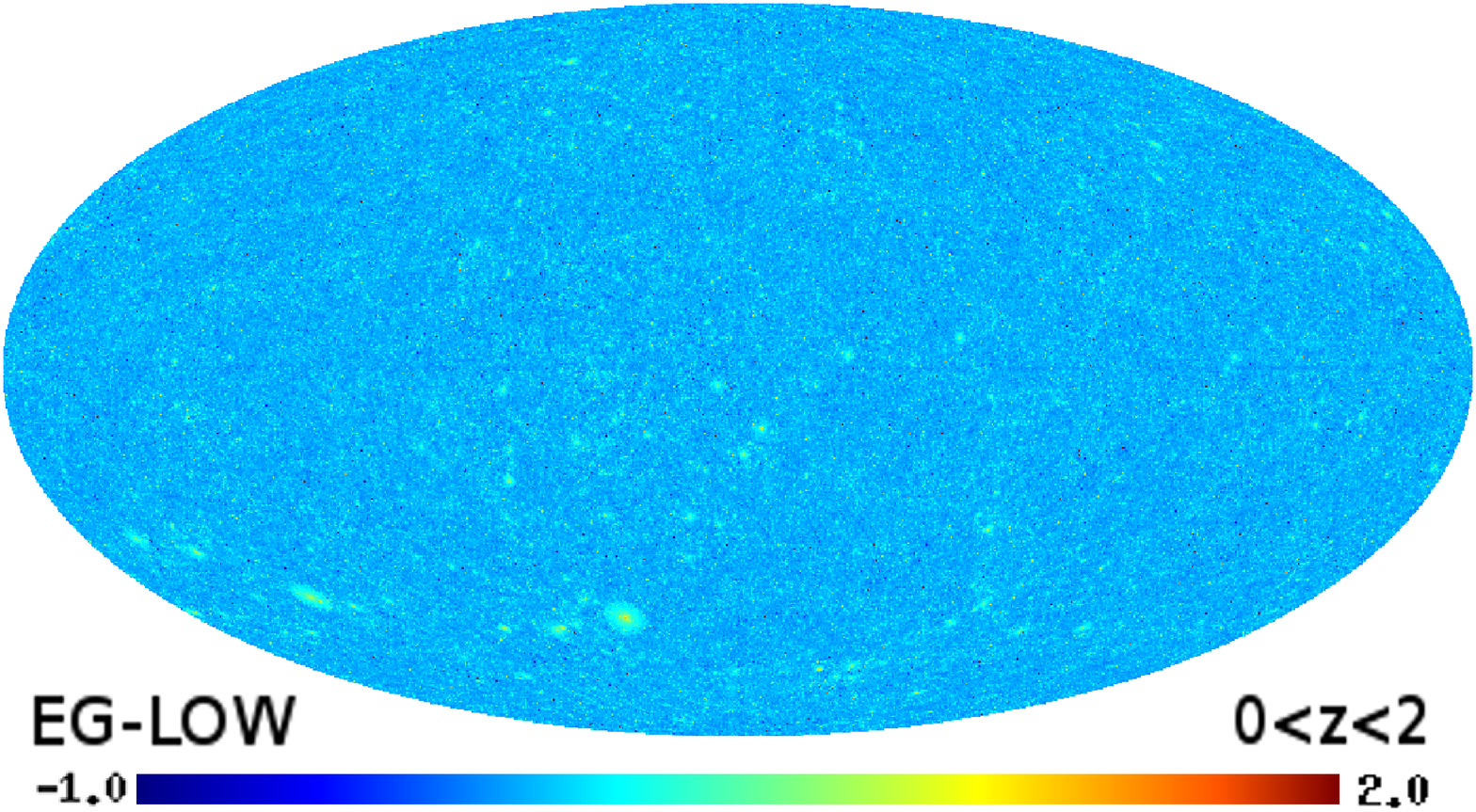}
\caption{\label{fig:MillenniumII_maps} All-sky maps of the gamma-ray intensity (in units of cm$^{-2}$s$^{-1}$sr$^{-1}$GeV$^{-1}$) at 4 GeV from DM annihilation (left panels) and DM decay (right panels). The figure shows the emission of all DM (sub)halos down to the resolution limit of the MS-II (EG-MSII component). In the upper row only nearby structures ($z<0.01$) are considered, while in the second row the emission up to $z=2$ is considered. In the last row we plot the emission from all extragalactic (sub)halos (resolved and unresolved) down to $M_{\rm min}=10^{-6} M_\odot/h$ with the LOW subhalo boost (see text for details). In all cases, annihilation or decay into $b$ quarks is assumed: for annihilating DM, $m_{\chi}=200$~GeV with a cross section of $3 \times 10^{-26}$cm$^3$s$^{-1}$, while for decaying DM, $m_{\chi}=2$~TeV with a lifetime of $2 \times 10^{27}$s. The photon yield receives contributions from prompt emission and IC off the CMB photons (see Sec. \ref{sec:particle_physics}). In each map we subtract the all-sky average intensity of that component, after moving to a logarithmic scale. Note the different scales in the first row.}
\end{figure*}

In the upper panels of Fig. \ref{fig:MillenniumII_maps} we show the gamma-ray 
intensity of the EG-MSII component at 4 GeV for the first snapshot ($z<0.01$), 
in the case of annihilating DM (left panels, $m_{\chi}=200$~GeV, 
$(\sigma_{\rm ann} v)=3\times10^{-26}{\rm cm}^3{\rm s}^{-1}$ and $B_b=1$) 
and decaying DM (right panels, $m_{\chi}=2$~TeV, $\tau=2\times10^{27}$s and 
$B_b=1$). The characteristic filaments of the cosmic web and individual DM 
halos are clearly visible, as well as (at least for the case of decaying DM) 
some subhalos hosted in large DM clumps. In the second row of this figure, 
we show the intensity up to $z=2$: the map is much more isotropic, even 
if some of the prominent, closest structures can still be seen.

\subsection{Unresolved main halos (EG-UNRESMain)}
\label{sec:unresolved_halos}
We describe now how we model the contribution of unresolved main 
halos, (i.e. those with masses below $M_{\rm res}$), a contribution that we call 
EG-UNRESMain. Since this is a regime which goes below the MS-II mass 
resolution, we are forced to resort to some assumptions concerning both the 
distribution and the individual properties of DM halos. Our approach is 
similar to the one of \citet{Zavala:2009zr}: we use main halos in the MS-II to 
perform an analytic fit to the single-halo luminosity (i.e. $L(M)$ defined 
in Eqs. \ref{eqn:annihilation_luminosity} and \ref{eqn:decay_luminosity}) as 
well as to the following function: 
\begin{equation}
F(M) = \frac{\sum L(M)}{\bar{M} \Delta \log M} 
\approx \ln10 L(M) \frac{\Delta n(\bar{M})}{\Delta M},
\label{eqn:F}
\end{equation}
which is the total luminosity of main halos with a mass in the logarithmic
mass range ${\rm log M} \pm \Delta {\rm log} M/2$, divided by its mean value 
$\bar{M}$ and the width of the logarithmic mass bin. The second equality
shows how $F(M)$ depends on the halo mass function $\Delta n/\Delta M$ in
the bin considered. 
By extrapolating the fit obtained for $F(M)$ above $M_{\rm res}$, it is possible 
to estimate the gamma-ray intensity due to DM main halos below $M_{\rm res}$ 
down to different values of $M_{\rm min}$. 
In \citet{Zavala:2009zr} (see their Figs. 5 and 6), the authors compared the 
predictions of such an extrapolation, in the case of annihilation, with the 
result of an analytical model based on the formalism of \citet{Sheth:1999su} 
for the halo mass function and \citet{Eke:2000av} for the concentration-mass 
relation. The total flux in unresolved main halos with a mass between 
$M_{\rm min}=10^{-6} M_\odot/h$ and $M_{\rm res}$ agrees within a factor of 5 
between the two approaches.

This missing flux is then added to the emission of resolved halos with mass 
between $1.39 \times 10^8$ and $6.89 \times 10^9 M_\odot/h$ (halos with 
particle number between 20 and 100). The decision to boost up only these halos 
is equivalent to assuming that halos smaller than $M_{\rm res}$ share the same 
spatial clustering as those within that mass range. This assumption is 
motivated by the fact that the two-point correlation function of halos 
approaches an asymptotic value already at these masses (see Fig. 7 of 
\citealt{BoylanKolchin:2009nc}). 

Nevertheless, the actual clustering of low-mass main halos is unknown and, 
even if they trace the distribution of more massive objects, treating their 
contribution simply as a boost factor for the halos with lowest masses in the 
MS-II may overestimate their true clustering. Assuming, instead, that they 
are distributed more isotropically would reduce their contribution to the 
total APS (especially at low multipoles), although it is difficult to estimate 
precisely by how much. 
In what follows we assume that the uncertainty in the clustering of unresolved 
main halos is small and can be ignored.

The same procedure described above is also applied for the case of DM decay. 

\subsection{Unresolved subhalos}
\label{sec:unresolved_subhalos}
In this section we describe how we account for the emission from unresolved 
subhalos, i.e.: $i)$ subhalos with masses below $M_{\rm res}$ that are hosted 
by main halos in the MS-II catalogs, and $ii)$ subhalos of unresolved main 
halos. We do not consider sub-subhalos since their contribution is likely 
negligible in comparison \citep[see, e.g.,][]{Martinez:2009jh}. Note also 
that, at least at low redshifts, most of these subhalos have been removed by 
tidal stripping \citep{Springel:2008by}.  

For the extragalactic emission, the impact of unresolved subhalos on the
intensity and angular anisotropy spectra is essentially to boost the 
luminosity of the host by a certain amount. Thus, in principle, any method 
that provides boosts within the range of what has been found previously in 
the literature is a reasonable one. The method we use here has the advantage 
of having a single parameter ($k$ in Eq. \ref{eqn:fs_original}) that 
controls the abundance of substructure, which can be easily adjusted to obtain 
the subhalo boosts that have been reported in the past.

\citet{Kamionkowski:2008vw} and \citet{Kamionkowski:2010mi} propose a method
to compute the subhalo boost factor for the annihilation rate of a MW-like DM 
halo, providing an expression for the total boost factor 
$B_{\rm ann}(M_{\rm MW})$, as well as the differential profile 
$B_{\rm ann}(M_{\rm MW},r)$ (expressing the boost factor at a distance $r$ from 
the center of the halo). This prescription was calibrated with the Via Lactea 
II simulation \citep{Diemand:2008in}. 
The distribution of particles in this simulation is used to derive the 
probability $P(\rho,r)$ of having a value of the DM density between $\rho$ 
and $\rho+d\rho$ at a distance $r$ from the center of the main halo. 
Two different components contribute to $P(\rho,r)$: the first one is Gaussian 
and corresponds to the smooth DM halo, while for higher values of $\rho$, the 
probability is characterized by a power-law tail due to the presence of 
subhalos (see Fig. 1 of \citealt{Kamionkowski:2010mi}). 
The fraction of the halo volume that is filled with substructures is well 
fitted by\footnote{Since current simulations are many orders of magnitude 
away from resolving the whole subhalo population down to $M_{\rm min}$, 
$f_{s}(r)$ is known with limited precision and represents one of the implicit 
uncertainties of our predictions.}:
\begin{equation}
1-f_s(r) = 
k \left( \frac{\rho_{sm}(r)}{\rho_{sm}(r=100 \mbox{ kpc})} \right)^{-0.26},
\label{eqn:fs_original}
\end{equation}
with $k=7 \times 10^{-3}$. 
$P(\rho,r)$ is then used to derive an expression for the boost factor 
$B_{\rm ann}(r)$ in the case of annihilating DM: 
$B_{\rm ann}(M,r) = \int_0^{\rho_{\rm max}} d\rho P(\rho,r) \rho^2 / \rho_{sm}^2(r)$,
where $\rho_{\rm max}$ is a maximum density, which is of the order of
the density of the earliest collapsing subhalos (see below) and $\rho_{sm}$
is the density of the smooth component.

\citet{SanchezConde:2011ap} extended the previous method to halos of all 
sizes, adopting a slight modification to the definition of $f_s(r)$:
\begin{equation}
1-f_s(r) = k \left( \frac{\rho_{sm}(r)}{\rho_{sm}(r=3.56 \times r_s)} 
\right)^{-0.26},
\label{eqn:fs}
\end{equation}
where $r_s$ is the scale radius of the host halo given in kpc\footnote{The 
value of 3.56 is chosen so that, for the MW halo in Via Lactea II, 
Eqs. \ref{eqn:fs} and \ref{eqn:fs_original} are identical.}. We note that 
this implies that halos of all masses have the same radial dependence of 
$f_{s}$, only rescaling it to the particular size of the halo. 
This is partially supported by the mass-independent radial distribution of 
subhalos found in simulations (e.g. \citealt{Angulo:2008xq}).
Using Eq.~\ref{eqn:fs}, \citet{SanchezConde:2011ap} found that $B_{\rm ann}<2$ 
for the MW dwarf spheroidals, while $B_{\rm ann}\sim 30-60$ for galaxy clusters 
(integrating up to the tidal and virial radius, respectively). In both cases, 
the morphology of the total gamma-ray emission coming from the halo is 
modified since the subhalo contribution makes the brightness profile 
flatter and more extended. 

For the case of annihilating DM, we account for the contribution of unresolved 
subhalos by implementing the procedure of \citet{SanchezConde:2011ap} in two 
different ways: 
\begin{itemize}
\item for the subhalos of unresolved main halos we integrate 
$F_{\rm ann}(M)B_{\rm ann}(M)$ to compute the total luminosity from $M_{\rm min}$ 
to $M_{\rm res}$. The result of this integral is then used to boost up the 
emission of main halos in the MS-II with a mass between $1.39 \times 10^8$ 
and $6.89 \times 10^9 M_\odot/h$.
\item for subhalos belonging to main halos that are resolved in the simulation 
we boost up the luminosity of each halo by the mass-dependent boost 
$B_{\rm ann}(M)$ (i.e. the integral of $B_{\rm ann}(M,r)$ up to the virial 
radius). If the halo is extended, in addition to a total luminosity boost, we 
assume a surface brightness profile as given by $B_{\rm ann}(M,r)$. We need to 
apply a correction to this procedure since these equations account for 
subhalos from a minimum mass $M_{\rm min}$ up to the mass of the main halo $M$, 
whereas subhalos with masses above $M_{\rm res}$ are resolved and already 
accounted for in the simulation (they belong to the EG-MSII component). To 
correct for this double-counting, we simply compute (and subtract) the 
emission due to subhalos down to a minimal mass equal to $M_{\rm min}=M_{\rm res}$.
\end{itemize}

We note that changing $M_{\rm min}$ corresponds to changing $\rho_{\rm max}$.
From \citet{Kamionkowski:2010mi}, the maximum density in a halo is the 
density that its smallest subhalo had at the moment this subhalo formed:
\begin{equation}
\rho_{\rm max}(M_{\rm min}) = 
\frac{200}{12} \frac{c_{200}^3(M_{\rm min},z_F)}{f(c_{200}(M_{\rm min},z_F))} 
\rho_{\rm crit}(z_F),
\label{eqn:rho_max}
\end{equation}
where $f(x)=\ln(1+x)-x/(1+x)$. 
The epoch of collapse $z_F$ as a function of halo mass can be computed using 
the spherical collapse model of DM halo formation and evolution (see, e.g., 
\citealt{SanchezConde:2006wu} and references therein), which shows that for 
low masses, up to $\sim1$ $M_\odot/h$, all halos collapse approximately at the 
same redshift, $z_F=40$. The initial concentrations are set by the formation 
epoch, which means that halos that collapse at roughly the same $z_F$ will have 
similar $c_{200}(z_F)$. Thus, according to Eq. \ref{eqn:rho_max}, all low-mass 
subhalos will be characterized roughly by the same 
$\rho_{\rm max} \sim 2.51 \times 10^9 M_\odot/$kpc$^{3}$ (for 
$M_{\rm min}<1 M_\odot/h$), after fixing $c_{200}(z_F)$ to a constant value of 3.5 
as suggested by simulations 
\citep[e.g.][]{Diemand:2006ey,Zhao:2008wd}\footnote{Here, a matter power 
spectrum parametrized as in \citet{Bardeen:1985tr} was used to compute $z_F$, 
with the most recent values of the cosmological parameters and with no 
exponential cut-off at the minimal mass of DM halos.}. 
We compute $\rho_{\rm max}$ for a set of reference values of $M_{\rm min}$
(see also Sec. \ref{sec:Mmin}, noting that $z_F>5$ for 
$M_{\rm min} \lesssim 10^9 M_\odot/h$, which implied that we can safely assume
$c_{200}(z_F)=3.5$ \citep{Zhao:2008wd}. Note also that, by using the case of 
$M_{\rm min}=M_{\rm res}=6.89 \times 10^9 M_\odot/h$ we can correct for the 
aforementioned problem of double-counting the subhalos with masses above the 
MS-II mass resolution. 

Recently, \citet{Pinzke:2011ek} and \citet{Gao:2011rf} also estimated the 
substructure boost for DM halos of mass ranging from those of dwarf 
spheroidals to those of galaxy clusters. They point to substantially larger 
boost factors than those found by \citet{SanchezConde:2011ap} for the same 
mass range. 
This is mainly a consequence of the different methodologies. In the former 
cases, the subhalo mass function and the concentration-mass relation are 
power-laws calibrated at the resolved masses and extrapolated to lower 
unresolved masses. On the contrary, in the method by 
\cite{Kamionkowski:2010mi} (with the modification implemented in 
\citealt{SanchezConde:2011ap}), the dependence on $M_{\rm min}$ is flatter 
towards lower masses due to the limit on the natal concentrations.

Nevertheless, using the procedure described in the previous paragraphs, we 
can obtain similar subhalo boosts to those given by \citet{Pinzke:2011ek} and 
\citet{Gao:2011rf} if we {\it substantially} increase the parameter that 
controls the abundance of substructure in Eq.~\ref{eqn:fs} to $k=0.15$. Both 
cases ($k=7 \times 10^{-3}$ as in \citealt{SanchezConde:2011ap}, and $k=0.15$ 
to reproduce the results of \citealt{Pinzke:2011ek} and \citealt{Gao:2011rf}) 
are considered in this paper as representative of scenarios with a small and 
a large subhalo boost and are referred in the following as the LOW and HIGH 
scenarios, respectively. These two cases represent the extreme values 
reported in the literature for the contribution of unresolved subhalos. By 
obtaining predictions for the total DM-induced emission for these extrema, 
we aim at estimating how large is the uncertainty associated with the 
unresolved subhalo population. Parameterizing such uncertainty in this way
represents a ``hybrid'' approach, since it does not rely completely either on 
a direct extrapolation of the results of simulations \citep{Zavala:2009zr} or 
on analytical estimates such as the stable clustering hypothesis 
\citep{Afshordi:2009hn}.

Up to now, the discussion of how to model unresolved subhalos refers only to 
the case of annihilating DM. For decaying DM, there is no need to model this 
contribution since these subhalos are too small to be detected by the subhalo 
finder and their mass is already accounted for in the mass of the host halo. 
Since for decaying DM the total luminosity of a halo is proportional to its 
mass, the unresolved subhalos contribute to what we call the 
``smooth component''\footnote{For the case of DM annihilation note that, 
although the mass of unresolved subhalos is also accounted for as part of 
the ``smooth component'', this does not imply that their contribution to the 
gamma-ray intensity is already considered since the annihilation rate is not 
proportional to the DM density, but to the density squared.}.
This is strictly valid only if we consider the total halo luminosity. If the
intensity profile is needed, we should consider that the true spatial 
distribution of unresolved subhalos is expected to be different from that of 
the smooth component. In the case of the extragalactic emission we neglect 
this effect since only the halos that are close by appear extended in the 
maps, while the vast majority appear as point sources. For the case of the 
galactic emission we comment on this issue on Sec. \ref{sec:Aquarius}.

\section{The gamma-ray emission from the Milky-Way halo}
\label{sec:Galactic}

\subsection{The smooth Milky-Way halo}
\label{sec:smooth_halo}
Our model for the emission from the smooth DM halo of our own galaxy is 
partially based on the results of the Aquarius project
\citep{Springel:2008cc,Navarro:2008kc}. With the goal of studying the 
evolution and structure of MW-size halos, the Aquarius project selected a
group of MS-II halos with properties similar to the MW halo and 
resimulated them at increasing levels of resolution. The different AQ
halos are characterized by virial masses between 0.95 and 
$2.2 \times 10^{12}M_{\odot}/h$ and have a variety of mass accretion histories 
\citep{BoylanKolchin:2009an}.
In this sense, they are not expected to be a perfect match to the dynamical 
properties of our own MW halo, but rather to be a representative sample of 
MW-size halos within the context of the CDM paradigm.
We consider here the halo dubbed Aq-A-1, containing more than one billion 
particles within $r_{200}$ and having a mass resolution of $1250 \, M_\odot/h$. 
\begin{figure*}
\includegraphics[width=0.45\textwidth]{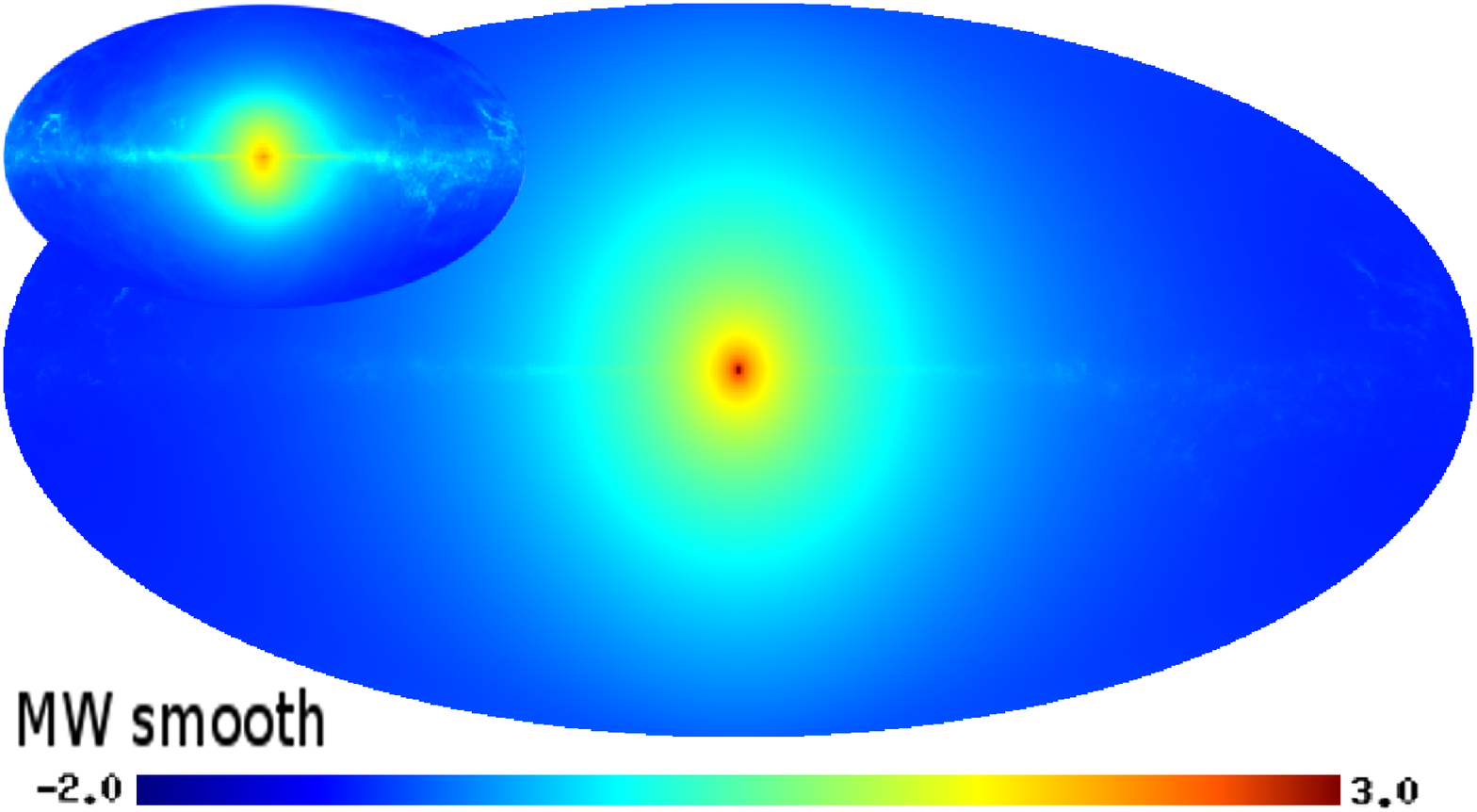}
\includegraphics[width=0.45\textwidth]{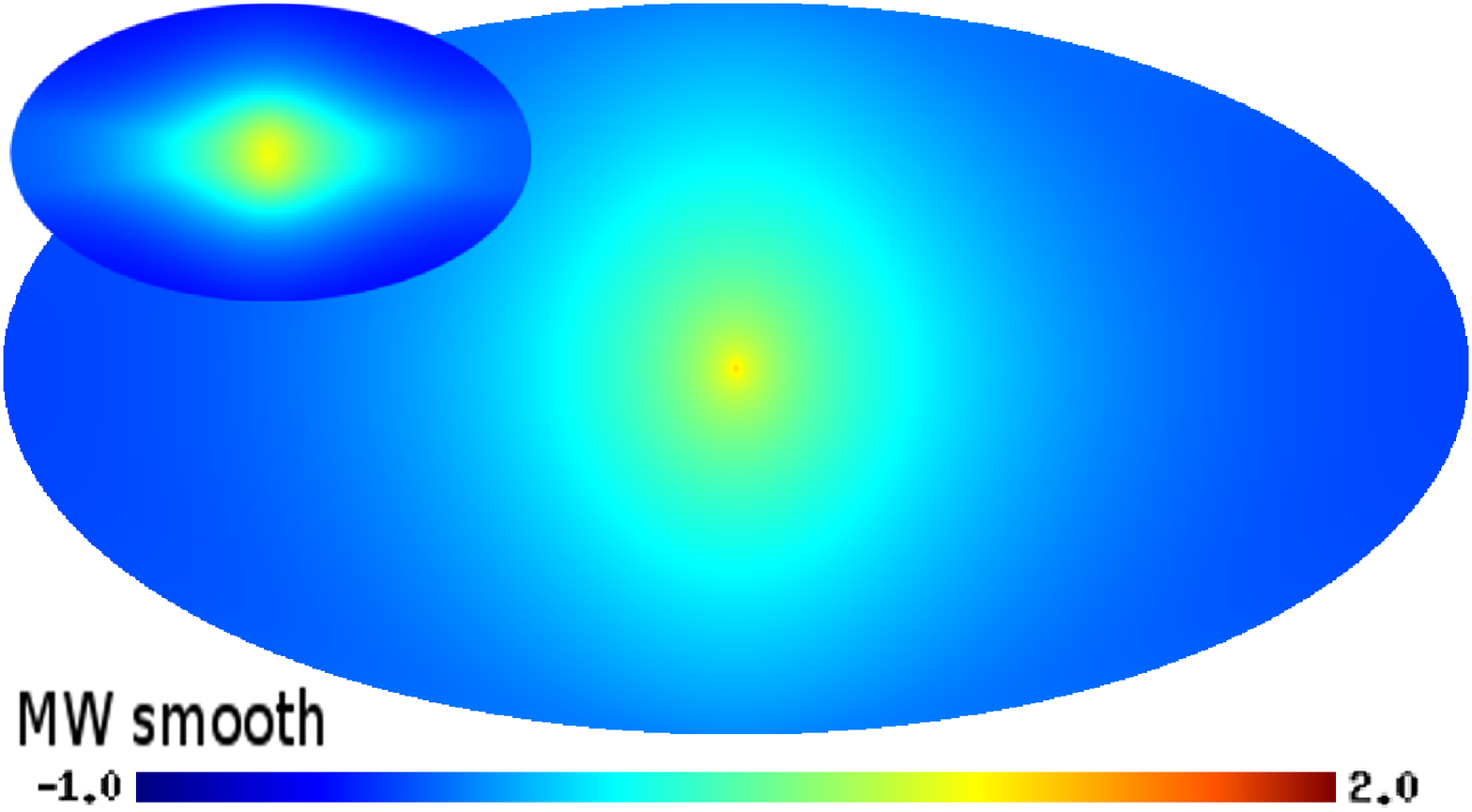}
\includegraphics[width=0.45\textwidth]{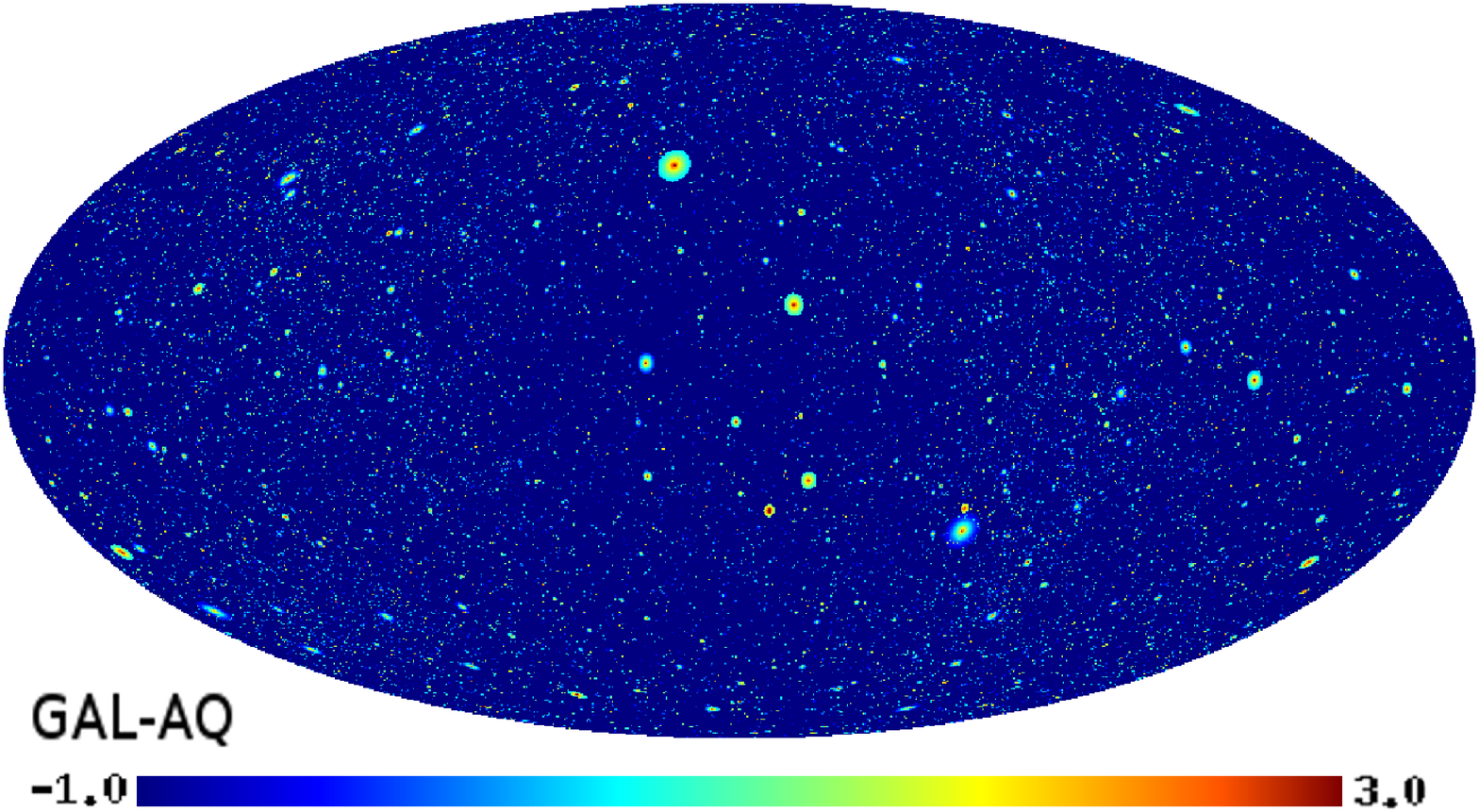}
\includegraphics[width=0.45\textwidth]{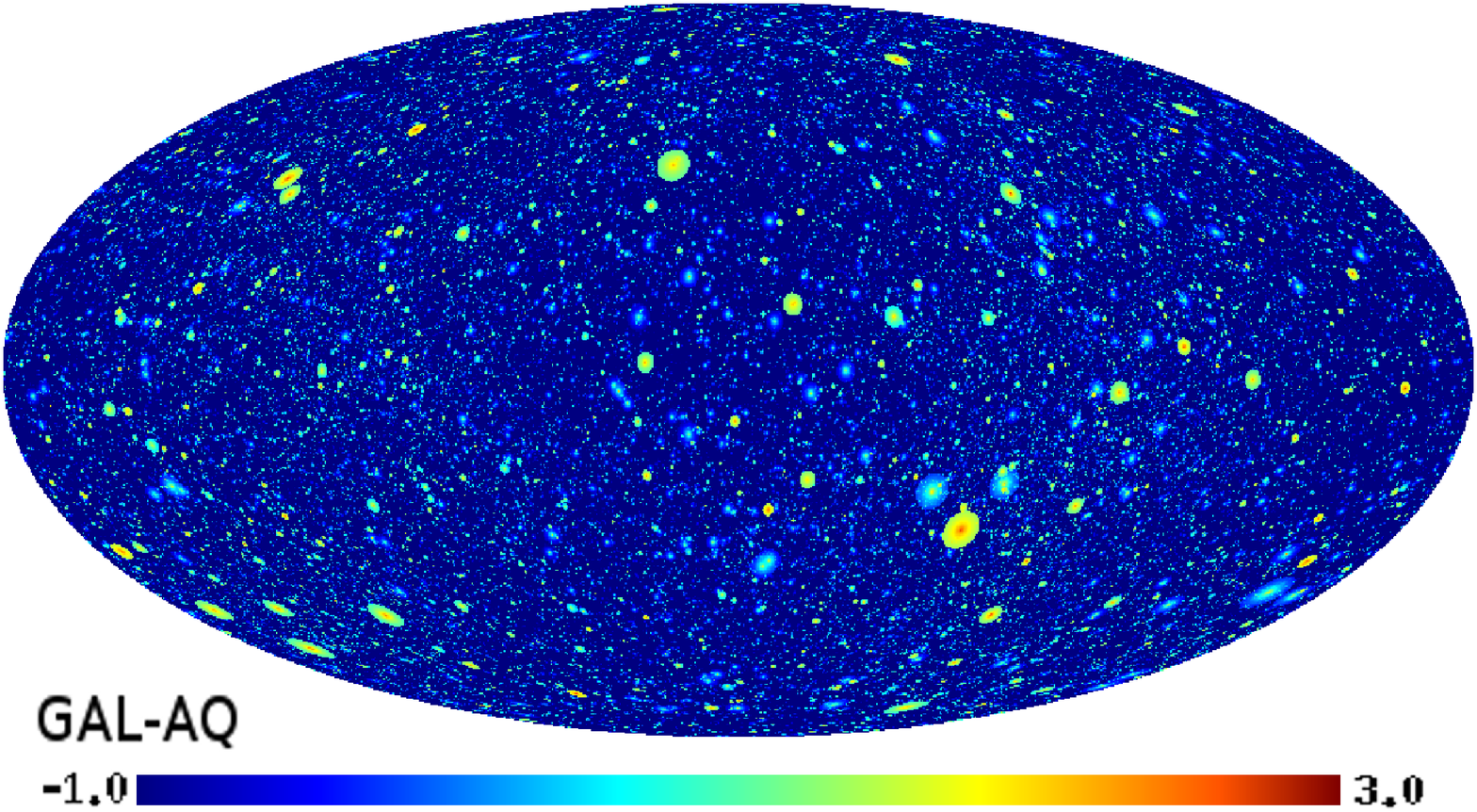}
\includegraphics[width=0.45\textwidth]{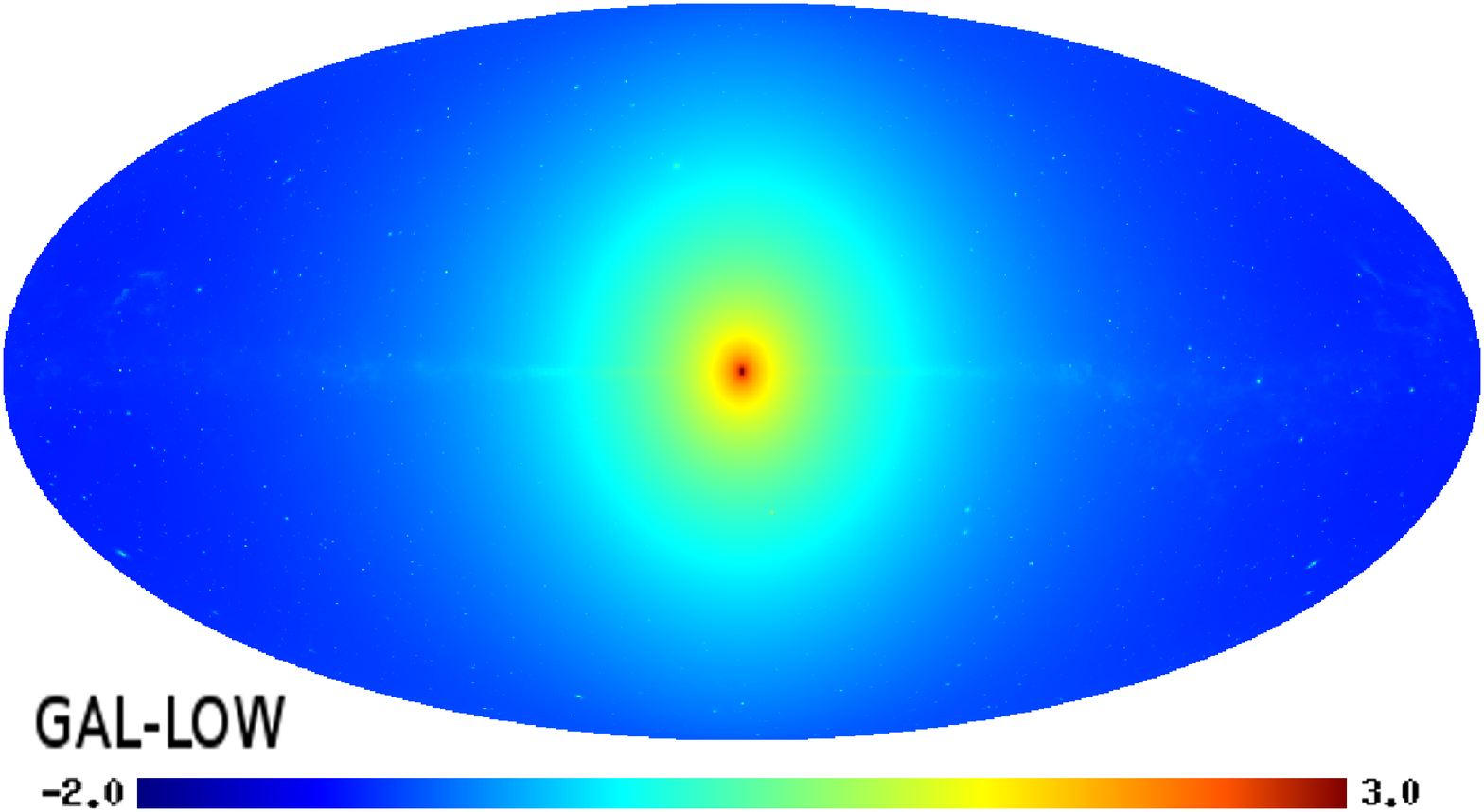}
\includegraphics[width=0.45\textwidth]{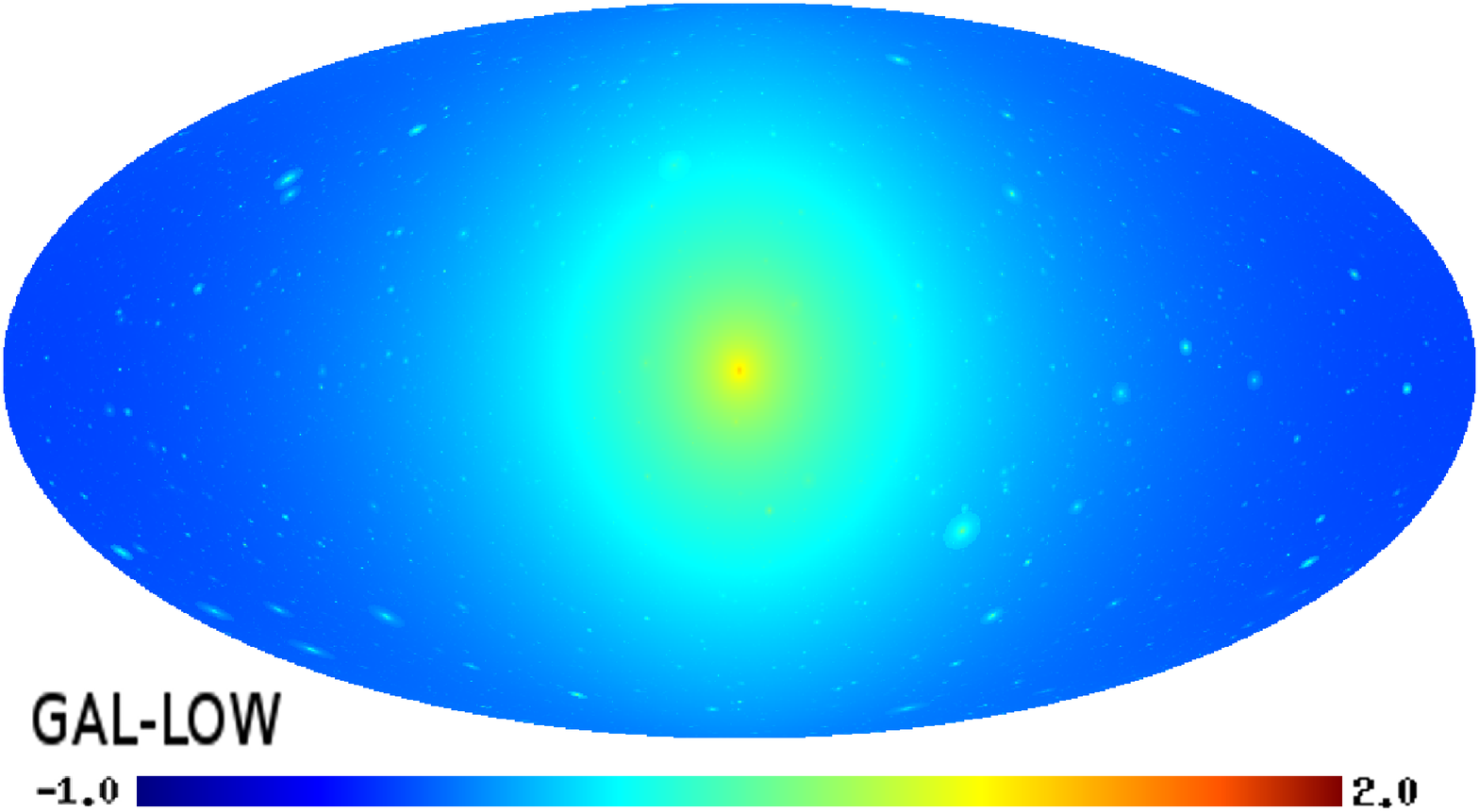}
\caption{\label{fig:Aquarius_maps} All-sky map of the galactic gamma-ray intensity (in units of cm$^{-2}$s$^{-1}$sr$^{-1}$GeV$^{-1}$) at 4 GeV from DM annihilation (left panels) and decay (right panels). In the first row, we show the emission from the smooth MW halo, while the contribution of resolved subhalos in the Aquarius Aq-A-1 halo (GAL-AQ component) is shown in the second row. The maps on the last row indicate the total galactic emission accounting for the MW smooth halo and its (resolved and unresolved) subhalos down to $M_{\rm min}=10^{-6} M_\odot/h$ (for the LOW subhalo boost). As in Fig. \ref{fig:MillenniumII_maps}, $m_\chi=200$ GeV, the cross section is $3 \times 10^{-26}$cm$^3$s$^{-1}$ and $B_b=1$ for the left panels, while $m_\chi=2$ TeV with a lifetime of $2 \times 10^{27}$s and $B_b=1$ for the right ones. The intensity includes contributions from prompt emission and IC with the CMB photons (see Sec. \ref{sec:particle_physics}). For the emission of the MW smooth halo we also consider IC with the complete ISRF, as well as hadronic emission. The non-prompt emission alone is shown in the smaller panels overlapping with the maps of the first row. In each map we subtract the all-sky average intensity of that component, after moving to a logarithmic scale. Note the different scale in the different panels.}
\end{figure*}
A careful analysis of the density profile of the smooth component of the 
Aq-A-1 halo performed by \citet{Navarro:2008kc} shows that the simulation 
data is best fitted by an Einasto profile (preferred over an NFW profile):
\begin{equation}
\ln \left( \frac{\rho(r)}{\rho_{-2}} \right) =
\left( \frac{-2}{\alpha} \right) 
\left[ \left( \frac{r}{r_{-2}} \right)^{\alpha} - 1\right],
\label{eqn:Einasto}
\end{equation}
with $r_{-2} \sim 15.14$~kpc, $\rho_{-2}=3.98 \times 10^6 M_\odot$/kpc$^{3}$ and 
$\alpha \sim 0.170$. 

Stellar dynamics and microlensing observations can be used to constrain the 
absolute value of the DM density at the position of the Earth, $\rho_{\rm loc}$. 
Different results point towards a range of values between 0.2 and 
0.5 GeV/cm$^{3}$ \citep{Prada:2004pi,Catena:2009mf,Pato:2010yq,Salucci:2010qr,
Iocco:2011jz}. Noting that a different value for the local DM density would 
shift up or down our predictions for the intensity of the emission from DM 
annihilation/decay in the MW smooth halo proportionally to $\rho_{\rm loc}^2$ 
and $\rho_{\rm loc}$, respectively, we decide to renormalize the value of 
$\rho_{-2}$ of Aq-A-1 in order to reproduce a reference value of 
$\rho_{\rm loc}=0.3$ GeV/cm$^{3}$ (a similar approach was used in 
\citealt{Pieri:2009je}).

To build our template map for the smooth MW halo, we assume that the observer 
is located at the solar circle at a distance of 8.5~kpc from the
GC and we integrate the DM-induced emission along the line of sight up to a 
distance of 583~kpc ($\sim 2.5~r_{200}$ of Aq-A-1). This distance marks the 
transition between our galactic and extragalactic regimes and it is selected 
because the Aq-A-1 halo is still simulated with high resolution up to this 
radius, and it therefore provides a better representation of the outermost 
region of the MW halo than the MS-II. For the smooth component, in addition 
to the prompt emission and secondary emission from IC scattering with the CMB
photons, we also consider the emission due to IC scattering with the complete 
InterStellar Radiation Field (ISRF) provided in \citet{Moskalenko:2005ng} as 
well as hadronic emission from interactions with the interstellar gas (see 
Appendices \ref{sec:IC_emission} and \ref{sec:hadronic_emission} for details). 
The first row in Fig.~\ref{fig:Aquarius_maps} shows the gamma-ray emission 
from DM annihilation (left panel) and decay (right panel) in the smooth MW
halo. The secondary emission correlated with the MW ISRF and the interstellar
gas can be seen along the galactic plane and is plotted independently in
the small panels overlapping with the maps of the first row.

\subsection{The Milky Way subhalos (GAL-AQ and GAL-UNRES)}
\label{sec:Aquarius}
This section focuses on the contribution of galactic subhalos, dealing with 
$i)$ subhalos that are resolved in the Aq-A-1 halo, (which we refer to
as the GAL-AQ component) and $ii)$ subhalos with masses below the mass 
resolution of AQ (which we call the GAL-UNRES component). As we did in 
Sec. \ref{sec:Millennium_II}, we use the subhalo catalog to compute the 
luminosity of each object from its $V_{\rm max}$ and $r_{\rm max}$ 
values\footnote{As in the case of extragalactic (sub)halos, we correct the 
values of $V_{\rm max}$ and $r_{\rm max}$ for numerical effects (see
Sec. \ref{sec:Millennium_II}).}. Only subhalos with more than 100 particles 
are considered, resulting in an ``effective'' AQ mass resolution of 
$1.71 \times 10^5 M_\odot$. The gamma-ray intensity in a given direction $\Psi$ 
is then obtained by summing up the contribution from all subhalos 
encountered along the line of sight, up to a distance of $583$~kpc. The GAL-AQ 
component is shown in the second row of Fig. \ref{fig:Aquarius_maps} in the 
case of annihilation (left) and decay (right).

For an annihilating DM candidate, the contribution of unresolved galactic 
subhalos is accounted for using the same procedure as for unresolved 
extragalactic subhalos described in Sec. \ref{sec:unresolved_subhalos}, 
introducing the LOW and HIGH cases as representatives of scenarios with a small 
and a large subhalo annihilation boost. The LOW boost is taken again directly 
from \citet{Kamionkowski:2010mi} and \citet{SanchezConde:2011ap} (which 
assumes $k=7 \times 10^{-3}$), while the HIGH boost is tuned to reproduce the 
results of \citet{Springel:2008by} who estimated a total subhalo boost of 232 
(integrating up to $r_{200}$ for the Aq-A-1 halo and for 
$M_{\rm min}=10^{-6}M_\odot/h$); we reproduce this result using 
$k=0.2$\footnote{The formalism by \citet{Kamionkowski:2010mi} overestimates 
the subhalo abundance in the inner region of the MW-like halo, namely within 
20 kpc. To correct for this, we assume that within this radius, the spatial 
distribution of unresolved subhalos follows the AQ distribution, being 
well fitted by an Einasto profile with $\alpha=0.678$ and $r_{-2}=199$ kpc.}.

In the case of decaying DM, we note that the mass contained in resolved 
subhalos is $2.7 \times 10^{11} M_\odot$ ($\sim15\%$ of $M_{200}$ for the 
Aq-A-1 halo, if we consider subhalos down to $1.71 \times 10^5 M_\odot$, see 
Eq. 5 of \citet{Springel:2008cc}. 
This goes up to $3.9 \times 10^{11} M_\odot$ if we extrapolate the subhalo mass 
function down to $M_{\rm min}=10^{-6} M_\odot/h$, which implies that unresolved 
subhalos contribute to the halo mass (and hence to the total decay luminosity)
slightly less than resolved ones (see end of 
Sec. \ref{sec:unresolved_subhalos}). Thus, an upper limit to the gamma-ray 
intensity from DM decay coming from these unresolved subhalos can be obtained 
by considering the flux coming from resolved subhalos, which is less than 1\% 
of the flux coming from the smooth component. Hence, we decide to ignore the 
contribution of unresolved subhalos to the amplitude of the galactic DM-decay 
emission. 

Regarding the contribution to the APS from unresolved subhalos, we note that
although subhalos just below the mass resolution of Aq-A-1 ($\sim 10^5 M_\odot$) 
might still contribute to the anisotropies, mainly through a Poisson-like 
APS, their abundance is so large (the subhalo mass function grows as 
$\propto M^{-1.9}$) that the intrinsic anisotropies of the gamma-ray intensity
produced by them would be very small. Because of this, the APS at multipoles 
above $l\sim100$ is likely dominated by subhalos with masses above $10^5M_\odot$ 
(see Sec. \ref{sec:APS} and the top panel of Fig. 8 of \citealt{Ando:2009fp}),
allowing us to neglect the contribution of subhalos with lower masses. 
We have verified this is indeed the case using the analytical model of
\citet{Ando:2009fp} (see discussion in Sec. \ref{sec:Galactic_APS} and
Appendix \ref{sec:Galactic_unresolved_subhalos}).

We do not take into account annihilation boosts due to fine-grained 
phase-space structures like streams and caustics. For a standard DM model 
without specific boost mechanisms (e.g. Sommerfeld enhancement) these effects 
are subdominant \citep{Vogelsberger:2007ny,White:2008as,Vogelsberger:2009bn,
Vogelsberger:2010gd}. If a mechanism like the Sommerfeld enhancement is 
invoked, fine-grained streams increase significantly the main halo 
annihilation, but their contribution is typically still less than that from 
subhalos \citep{Zavala:2011tt}.

Finally, in Tab. \ref{tab:nomenclature} we summarize the nomenclature used to 
identify the different components of DM-induced extragalactic and galactic 
emission introduced in the present section and in the previous one.

\begin{table}
\begin{center}
\begin{tabular}{cc}
{\bf Name} & {\bf Description} \\
\hline
 & DM halos and subhalos in MS-II catalogs \\
EG-MSII & with more than 100 particles (i.e. with \\
 & a mass larger than $M_{\rm res}=6.89 \times 10^9 M_\odot/h$) \\
\hline
EG-UNRESMain & extragalactic DM (main) halos with a mass \\
 & between $M_{\rm min}$ and $M_{\rm res}=6.89 \times 10^8 M_\odot/h$ \\
\hline
 & resolved and unresolved (sub)halos down to \\
 & $M_{\rm min}$. The unresolved subhalos are simulated \\
EG-LOW & following \citet{SanchezConde:2011ap} with \\
 & $k=7 \times 10^{-3}$ \\
 & (includes EG-MSII and EG-UNRESMain) \\
\hline
 & resolved and unresolved (sub)halos down to \\
 & $M_{\rm min}$. The unresolved subhalos are simulated \\
EG-HIGH & following \citet{SanchezConde:2011ap} with \\
 & $k=0.15$ \\
 & (includes EG-MSII and EG-UNRESMain) \\
\hline
\hline
 & smooth MW DM halo, parametrized by an \\
MW smooth & Einasto profile as in \citet{Navarro:2008kc}, \\
 & and normalized to a local DM density of \\ 
 & 0.3 GeV/cm$^3$ \\
\hline
 & DM subhalos in the AQ catalogs \\
GAL-AQ & with more than 100 particles (i.e. with a \\
 & mass larger than $1.71 \times 10^{5} M_\odot$) \\
\hline
 & DM subhalos with a mass between $M_{\rm min}$ \\
GAL-UNRES (LOW) & and $1.71 \times 10^5 M_\odot$, simulated \\
 & following \citet{SanchezConde:2011ap} \\
 & with $k=7 \times 10^{-3}$ \\
\hline
 & DM subhalos with a mass between $M_{\rm min}$ \\
GAL-UNRES (HIGH) & and $1.71 \times 10^5 M_\odot$, simulated \\
 & following \citet{SanchezConde:2011ap} \\
 & with $k=0.2$ \\
\hline
GAL-LOW & MW smooth + GAL-AQ + \\
 & GAL-UNRES (LOW) \\
\hline
GAL-HIGH & MW smooth + GAL-AQ + \\
 & GAL-UNRES (HIGH) \\
\end{tabular}
\caption{\label{tab:nomenclature} Summary table of the nomenclature used in the paper to identify the different components of the DM-induced emission.}
\end{center}
\end{table}

\section{Energy and angular power spectra of the Dark-Matter-induced gamma-Ray emission}
\label{sec:energy_APS_spectra}
Before showing the analysis of our simulated maps, we note that changing 
the particle physics scenario (i.e. considering a different value for $m_\chi$ 
and/or selecting a different annihilation/decay channel) would require, in 
principle, re-running our map-making code for the extragalactic intensity, 
since the photon emission spectrum is redshifted along the line of 
sight. This is a computationally expensive task given that one complete 
realization takes approximately 50000 CPU hours. 
However, this is not necessary since it is possible, given a reference 
all-sky map obtained for a particular particle physics model, to derive the 
corresponding map for a different model simply applying a set of 
re-normalization factors for different redshifts. Such prescription is 
described in detail in Appendix \ref{sec:Particle_physics_method}.

\subsection{Analysis of the energy spectrum}
\label{sec:energy_spectrum}

\subsubsection{Extragalactic emission}
\begin{figure*}
\includegraphics[width=0.49\textwidth]{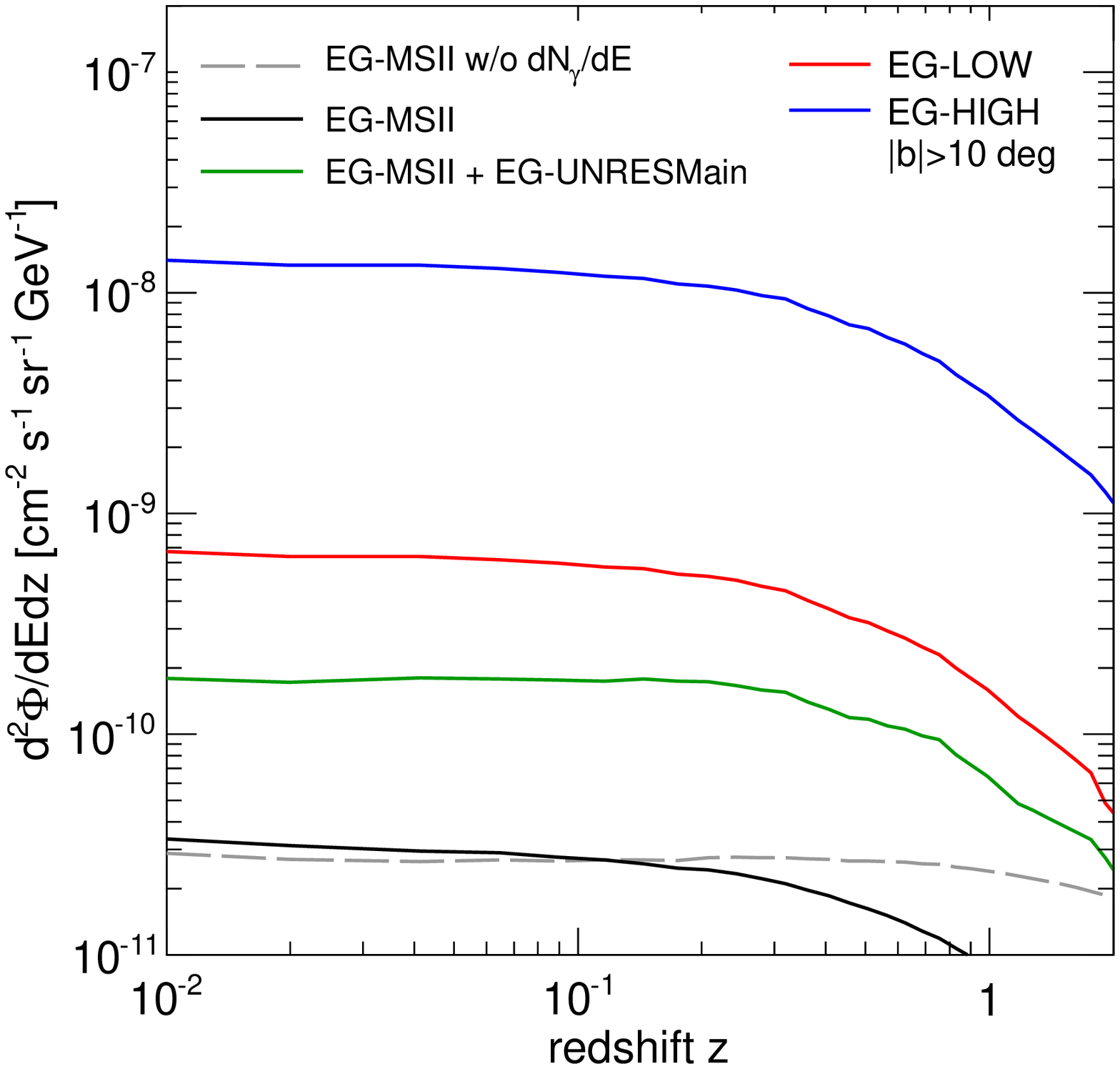}
\includegraphics[width=0.49\textwidth]{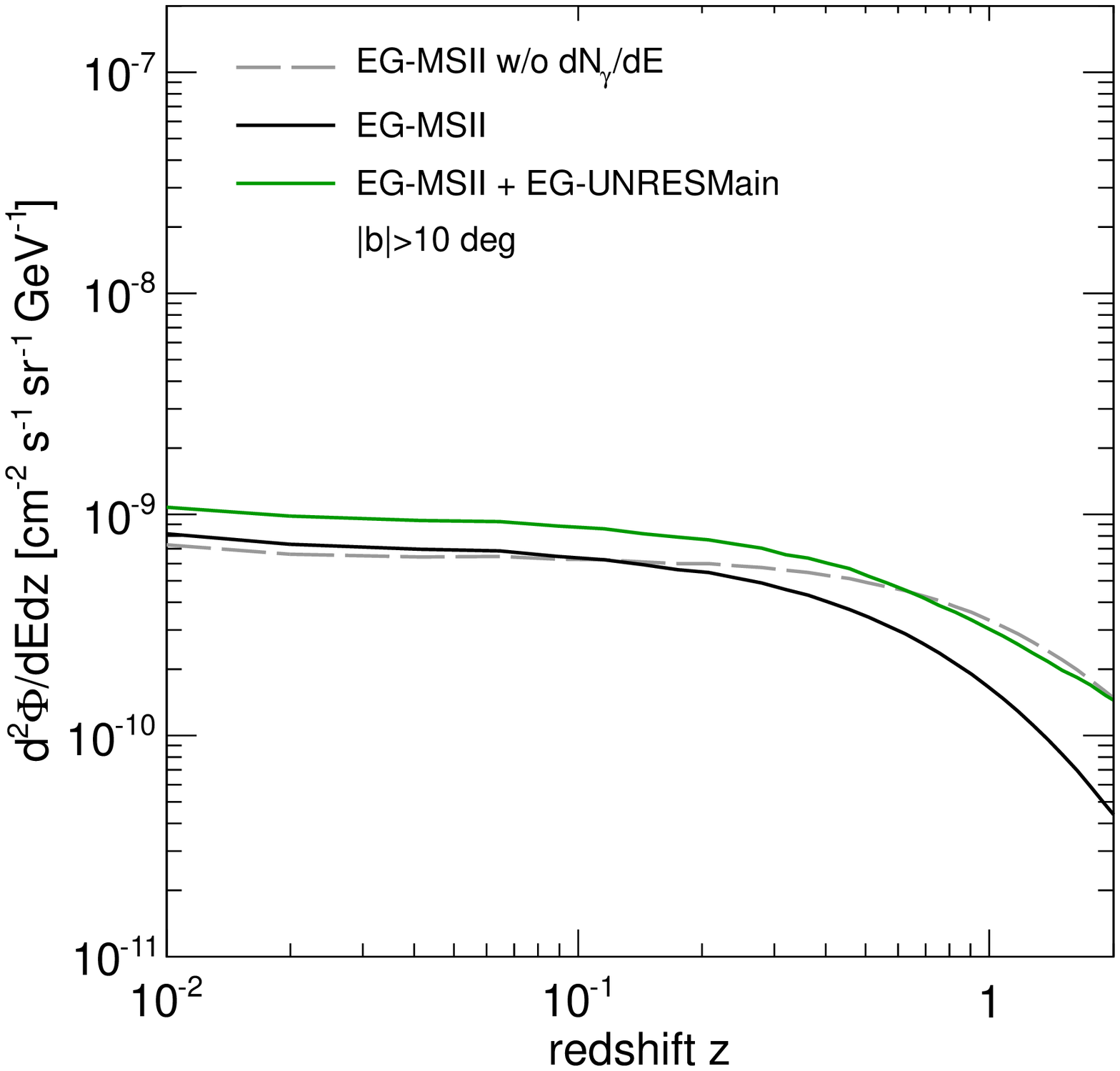}
\caption{\label{fig:IGRB_flux_redshift} Average of the extragalactic gamma-ray intensity per unit of redshift as a function of redshift at 4 GeV from DM annihilation (left panel) and DM decay (right panel) for $|b|>10^\circ$. Solid black lines correspond to the contribution from resolved (sub)halos in the MS-II (EG-MSII), while the solid green lines include in addition the boost from unresolved main halos (EG-UNRESMain, see Section \ref{sec:unresolved_halos}). The solid red and blue lines include all the previous components and the emission from unresolved subhalos down to a minimum mass $M_{\rm min}=10^{-6}M_{\odot}$ according to the method described in Section \ref{sec:unresolved_subhalos} for the LOW and HIGH case, respectively. In all cases, annihilation or decay into bottom quarks is assumed: for annihilating DM, $m_{\chi}=200$~GeV and $(\sigma_{\rm ann} v)=3\times10^{−26}{\rm cm}^3{\rm s}^{-1}$, while for decaying DM, $m_{\chi}=2$~TeV and $\tau=2\times10^{27}{\rm s}$. The photon yield receives contributions from prompt emission and IC of the CMB photons. The dashed grey line shows the ``astrophysical'' part of the signal (with an arbitrary normalization) for the EG-MSII component, by neglecting the $dN_\gamma/dE$ factor in Eqs.~\ref{eqn:annihilation_flux} and \ref{eqn:decay_flux}.}
\end{figure*}

Fig. \ref{fig:IGRB_flux_redshift} shows the average DM-induced gamma-ray 
intensity per unit redshift of our simulated extragalactic maps as a function 
of redshift (left and right panels for DM annihilation and DM decay, 
respectively), for an energy of 4 GeV. The average is computed over the whole 
sky except for a strip of $10^\circ$ along the galactic plane, since this is 
the region used in \citet{Abdo:2010nz} to determine the Fermi-LAT IGRB energy 
spectrum. Note that the intensity in each concentric shell filling up the 
volume of the past light cone is divided by the width of the particular shell 
in redshift space $\Delta z$: this is roughly equivalent to computing the 
average of the integrand of Eqs.~\ref{eqn:annihilation_flux} and 
\ref{eqn:decay_flux} over the redshift interval of each shell. 
The intensity from extragalactic resolved (sub)halos in the MS-II (EG-MSII) 
is shown with a solid black line. This same contribution is shown with a 
dashed grey line once the photon yield $dN_\gamma/dE$ is removed from the 
intensity (arbitrary normalization) in Eqs.~\ref{eqn:annihilation_flux} and 
\ref{eqn:decay_flux}, leaving only the ``astrophysical'' part of the signal. 
In the case of annihilation, the grey line is essentially flat, with all 
redshifts contributing equally to the gamma-ray intensity (see also Fig.~1 of 
\citealt{Abdo:2010dk}). Note that, in principle, the EBL attenuation should be 
visible in the shape of the grey dashed line, but at 4 GeV its effect is 
negligible and the line only depends on how the DM distribution changes with 
$z$. In the case of decaying DM, the astrophysical part of the signal drops 
more quickly with redshift since it is proportional to the DM density 
(which in average grows as $\propto(1+z)^3$) instead of to the density 
squared. Once the modulation of the photon yield $dN_\gamma/dE$ is included, 
we see that the majority of the signal comes from low redshifts (more so for 
decaying DM): in order to contribute to the emission at 4 GeV, photons coming 
from higher redshifts need to be more energetic, and their intensity is damped 
due to a lower photon yield. For the benchmark shown in 
Fig. \ref{fig:IGRB_flux_redshift}, the signal drops by a factor of $\sim3-5$ 
from $z=0$ to $z=1$.

Once the EG-MSII component is boosted up to include the contribution of 
unresolved main halos (EG-UNRESMain) with masses down to 
$M_{\rm min}=10^{-6} M_\odot/h$, the signal increases by a factor of $\sim$ 7
($\sim 1.5$) in the case of DM annihilation (decay). The contribution of 
unresolved main halos is given by integrating $F_{\rm ann}(M)$ and 
$F_{\rm decay}(M)$ in Eq.~\ref{eqn:F} from $M_{\rm min}$ to $M_{\rm res}$. These 
cumulative luminosities are ultimately connected to the halo mass function 
and the single-halo luminosities $L_{\rm ann}(M)$ and $L_{\rm decay}(M)$ in 
Eqs. \ref{eqn:annihilation_luminosity} and \ref{eqn:decay_luminosity}. 
Interestingly, they combine to produce a mass-dependent contribution that 
diverges towards lower masses in the case of DM annihilation 
($F_{\rm ann}(M)\propto M^{-1.04}$), but converges in the case of DM decay 
($F_{\rm decay}(M)\propto M^{-0.92}$). This is the reason why the EG-UNRESMain 
component is much larger than the resolved component in the case of 
annihilating DM, while the two remain rather similar for decaying DM. This 
implies that for the case of decay, the signal is essentially independent of 
$M_{\rm min}$, as long as $M_{\rm min}$ is low enough (see below).
\begin{figure*}
\includegraphics[width=0.49\textwidth]{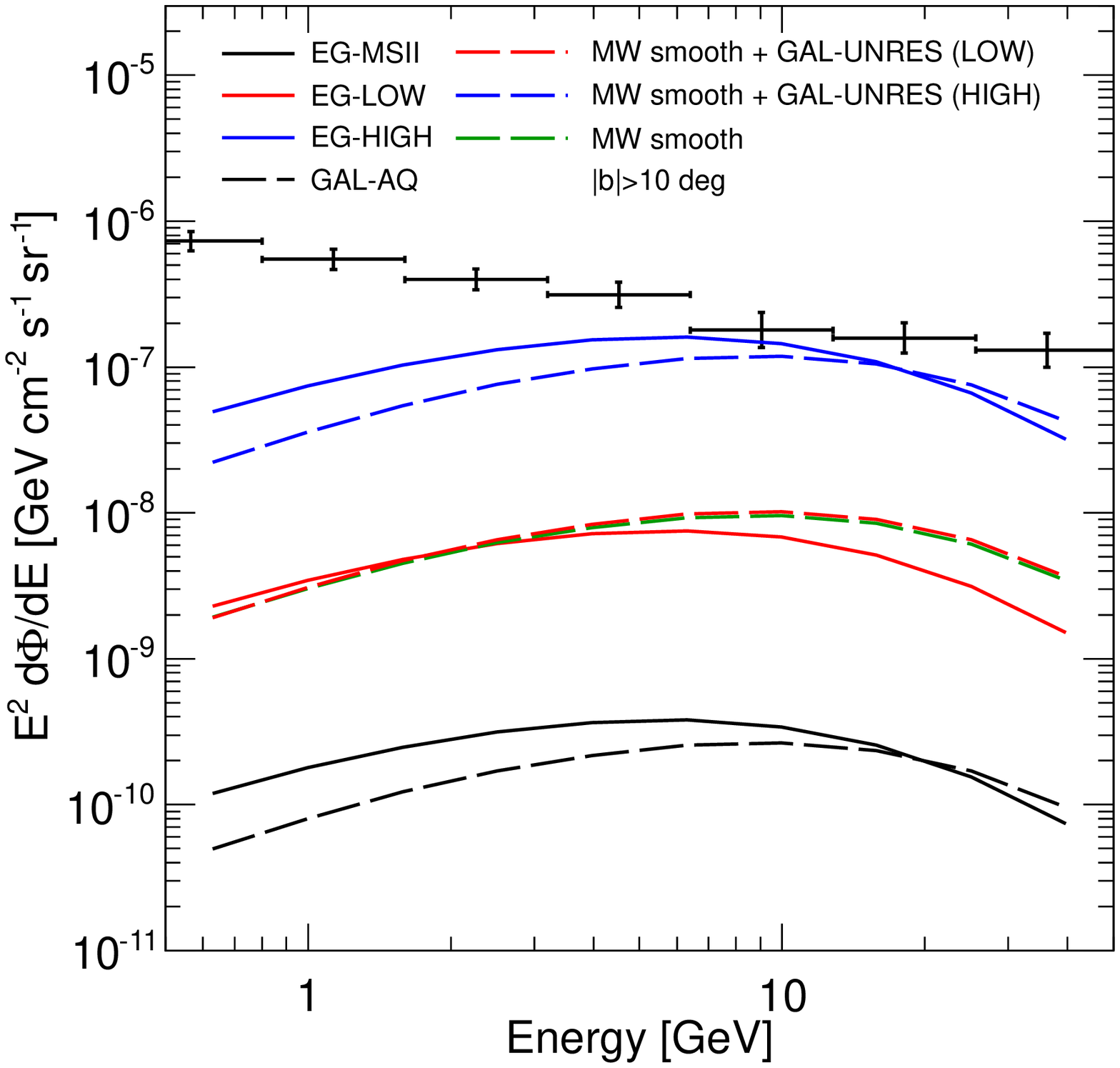}
\includegraphics[width=0.49\textwidth]{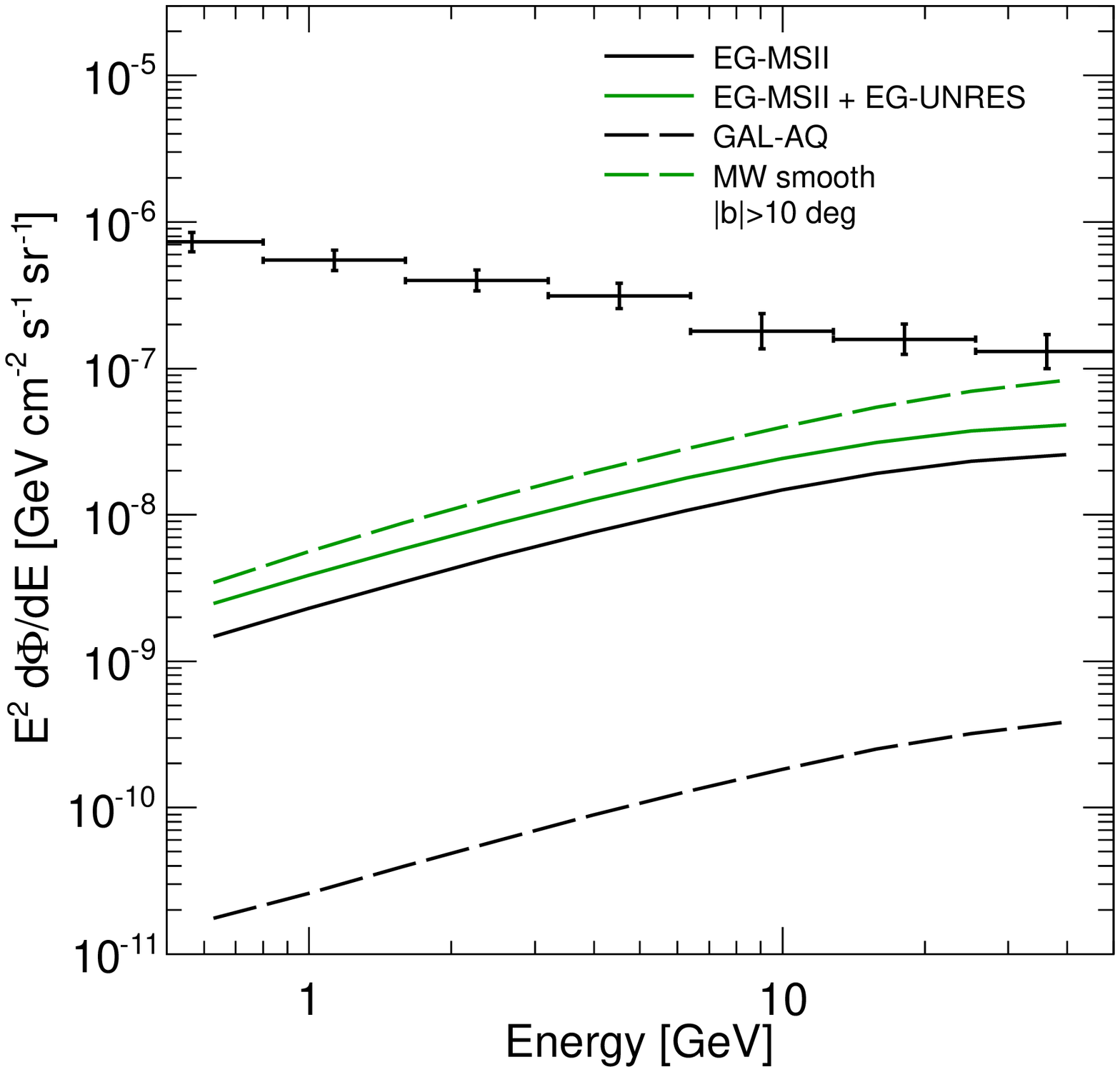}
\caption{\label{fig:IGRB_flux} Average of the gamma-ray intensity coming from DM annihilation (left) and DM decay (right) as a function of observed energy for $|b|>10^\circ$. Solid lines are for the extragalactic contribution, while dashed lines are for the galactic one. The color coding for the solid lines is the same as in Fig.~\ref{fig:IGRB_flux_redshift}, while for the dashed lines, the green one indicates the contribution of the smooth MW halo, the black one is for resolved subhalos (GAL-AQ) and the red and blue lines indicate the emission from the MW smooth halo and its unresolved subhalos (GAL-UNRES) in the LOW and HIGH case respectively (only for the left panel). The observational data points with error bars refer to the measurement of the IGRB as given in \citet{Abdo:2010nz}.}
\end{figure*}

The total emission is obtained by summing the previous components and the 
contribution of unresolved subhalos down to $M_{\rm min}$. The LOW (red line) 
and HIGH (blue line) scenarios in the left panel bracket the uncertainty 
associated with the subhalo contribution, for a fixed value of 
$M_{\rm min}=10^{-6} M_\odot/h$. We can see that unresolved (sub)halos boost the 
signal by a factor between 25 and 400 compared to the EG-MSII component. As 
noted before, such uncertainty is not present in the case of decaying DM, 
since the contribution of unresolved (sub)halos is essentially negligible. 
Note that the subhalo boost is smaller at high redshifts since the number of 
massive resolved main halos decreases with redshift and hence, the overall 
subhalo boost decreases as well.

Fig. \ref{fig:IGRB_flux} shows the energy spectrum of the average amplitude 
of the extragalactic (solid lines) DM-induced gamma-ray intensity. We only 
consider an energy range between 0.5 GeV and 50 GeV, approximately the same 
range where the IGRB Fermi-LAT data are available \citep{Abdo:2010nz}. As in 
Fig. \ref{fig:IGRB_flux_redshift}, the average is computed in the region with 
$|b|>10^\circ$. The left (right) panel is for annihilating (decaying) DM. The 
color-coding of the solid lines is the same as in 
Fig.~\ref{fig:IGRB_flux_redshift}. The full extragalactic signal, including 
resolved and unresolved (sub)halos, is expected to lie between the solid red 
and blue lines. 

In the case of DM annihilation, the extragalactic contribution is dominated by 
unresolved (sub)halos. This prediction agrees well with those from previous 
works. For instance, the grey band in Fig.~2 of \citet{Zavala:2011tt} can be 
compared with our ``uncertainty'' range bracketed by the red and blue 
lines\footnote{Note, however, that although the DM particle mass and the 
annihilation channel are the same, the annihilation cross section in Fig.~2 
of \citet{Zavala:2011tt} is a factor of 5 lower than the one we use in 
Fig. \ref{fig:IGRB_flux}.}. To be precise, the methodology implemented in the 
present paper and the one in \citet{Zavala:2011tt} are not identical, since
the emission of unresolved subhalos is accounted for in a different way.
Nevertheless, we find that the range covered between our LOW and HIGH subhalo
boosts is similar to those reported in Fig.~2 of \citet{Zavala:2011tt} (see 
also \citealt{Abdo:2010dk}).

\subsubsection{Galactic emission}
In Fig. \ref{fig:IGRB_flux} we also show the galactic DM gamma-ray intensity,
receiving contributions from the resolved subhalos of the Aq-A-1 halo 
(GAL-AQ, black dashed line), the smooth MW halo (green dashed line), and from 
unresolved subhalos (down to $M_{\rm min}=10^{-6}M_{\odot}/h$, red and blue 
dashed lines, for annihilating DM). We see that the emission from resolved 
galactic subhalos is essentially negligible, indeed being roughly two orders 
of magnitude smaller than the one from the smooth component (both for 
annihilating and decaying DM). The effect of unresolved subhalos is important 
only for DM annihilation and it is estimated to be between less than a factor 
of 2 (LOW, dashed red) and 10 (HIGH, dashed blue) times more than the smooth 
component. 
This represents an important difference with respect to what is found for the 
extragalactic case, where the subhalo boost can be even larger than two 
orders of magnitude. It can, however, be understood by noting that for the 
extragalactic case a given main halo and its subhalos are located essentially 
at the same distance from the observer, while for the galactic case, the 
observer is located much closer to the GC than to the bulk of the subhalo 
emission (on the outskirts of the halo). This is something that has already 
been noted by \citet{Springel:2008by}, where the subhalo boost to the smooth 
component of the Aq-A-1 halo (down to $M_{\rm min}=10^{-6}M_{\odot}/h$) was 
estimated to be 1.9, whereas for a distant observer it was 232. The value of 
1.9 is smaller than what we find for the HIGH case, even if the total boost 
of 232 for the case of a distant observer is compatible with our value. This 
is due to the slightly different radial distribution of the unresolved 
subhalos in the HIGH scenario, compared to what is found in 
\citet{Springel:2008by}.

In the case of decaying DM, the gamma-ray intensity is dominated by the smooth
component (approximately compatible with the results of 
\citealt{Ibarra:2009nw}).

Comparing the total galactic and extragalactic contributions, we see that they
are of the same order for the energy range and annihilation/decay channel 
explored in Fig.~\ref{fig:IGRB_flux}\footnote{Notice the slightly different 
shapes of the energy spectra of the extragalactic and galactic components due 
to redshifting and photon absorption at high energies in the case of 
extragalactic objects.}. This is roughly consistent with what has been reported 
previously (e.g, see Fig. 3 of \citealt{Abdo:2010dk}, and also Figs. 1 and 2 
of \citealt{Hutsi:2010ai}). 

For the particular DM candidates explored in Fig. \ref{fig:IGRB_flux}, the 
total DM-induced emission is able to account for the observed IGRB intensity if
the HIGH subhalo boost is considered (at least in one energy bin).

\subsection{Analysis of the angular power spectrum of anisotropies}
\label{sec:APS}
We consider now the statistical properties of the anisotropies of our 
simulated maps, which is the main objective of the present paper. Two 
slightly different definitions of the APS will be used: $i)$ the so-called 
``intensity APS'' ($C_\ell$), defined from the decomposition in spherical 
harmonics of the two-dimensional sky map after subtracting the average 
value of the intensity over the sky region considered:
\begin{eqnarray}
\Delta_{\rm flux}(\Psi) &=& 
\frac{d\Phi}{dE}(\Psi)-\left\langle\frac{d\Phi}{dE}\right\rangle
=\sum_{l=0}^\infty\sum_{m=-l}^{m=l}a_{lm}Y^\ast_{lm}(\Psi), \nonumber \\
C_\ell &=& \frac{1}{2\ell+1}\left(\sum_{|m| \geq \ell}\vert  a_{\ell m}\vert^2\right),
\label{eqn:intensity_APS}
\end{eqnarray}
and $ii)$ the so-called ``fluctuation APS'' ($C_\ell^{\rm fluct}$), which is 
dimensionless and is obtained from the decomposition of the {\it relative} 
fluctuations of an all-sky map. The fluctuation APS can be obtained from the 
intensity one simply dividing by $\langle d\Phi/dE \rangle^{2}$.

The intensity APS has the advantage of being an additive quantity, meaning
that the intensity APS of a sum of maps is the sum of the intensity APS of 
each individual component (assuming that the maps are uncorrelated, otherwise 
their cross-correlations should also be taken into account). On the other 
hand, the fluctuation APS of multiple components can be summed only after 
multiplying by the square of the relative emission of each component with
respect to the total:
\begin{equation}
\label{eqn:sum_APS}
C_{\ell}^{\rm fluct} \equiv \langle \frac{d\Phi}{dE} \rangle^{-2} C_\ell =
\sum_i \frac{\langle d\Phi^i/dE \rangle^{2}}{\langle d\Phi/dE \rangle^{2}} 
C_{\ell,i}^{\rm fluct}  = \sum_i f_i^2 C_{\ell,i}^{\rm fluct}.
\end{equation}

In order to compare directly the APS from our maps with the Fermi-LAT APS 
measurement, it would be necessary to consider the same target region as in
\cite{Ackermann:2012uf}, masking out the known point sources and the region 
along the galactic plane ($|b| \leq 30^\circ$), where the contamination due to 
the galactic foreground emission is larger. In this work we only present the 
APS as obtained directly from our maps and leave the comparison to the 
Fermi-LAT APS data for future work. 

We use HEALPix to compute the APS of our template maps, and note that the APS 
is conventionally plotted once multiplied by $\ell(\ell+1)/2\pi$, which for 
large multipoles is proportional to the variance of $\Delta_{\rm flux}$ (see 
Eq. 35 of \citealt{Zavala:2009zr}).

\begin{figure*}
\includegraphics[width=0.49\textwidth]{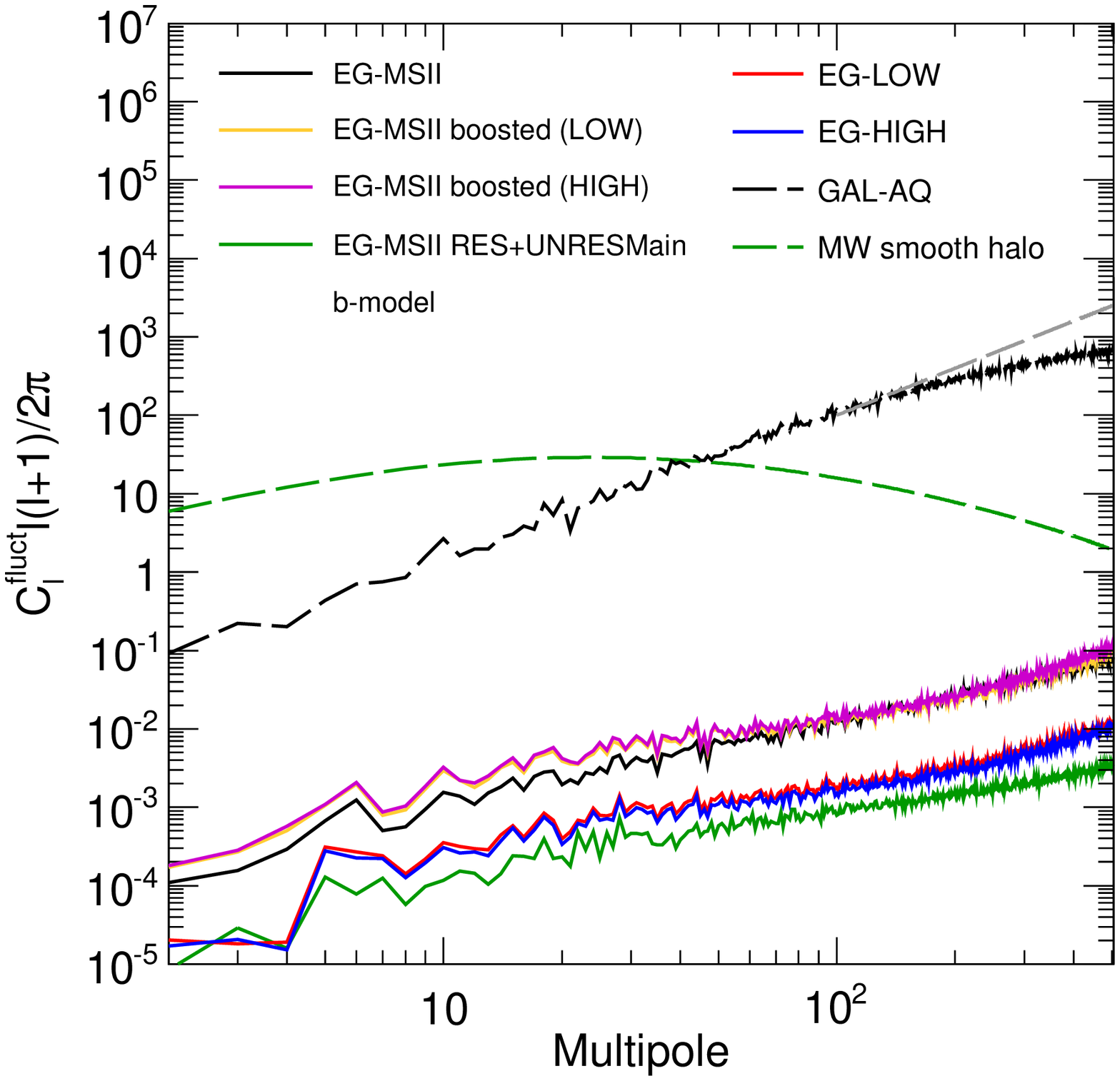}
\includegraphics[width=0.49\textwidth]{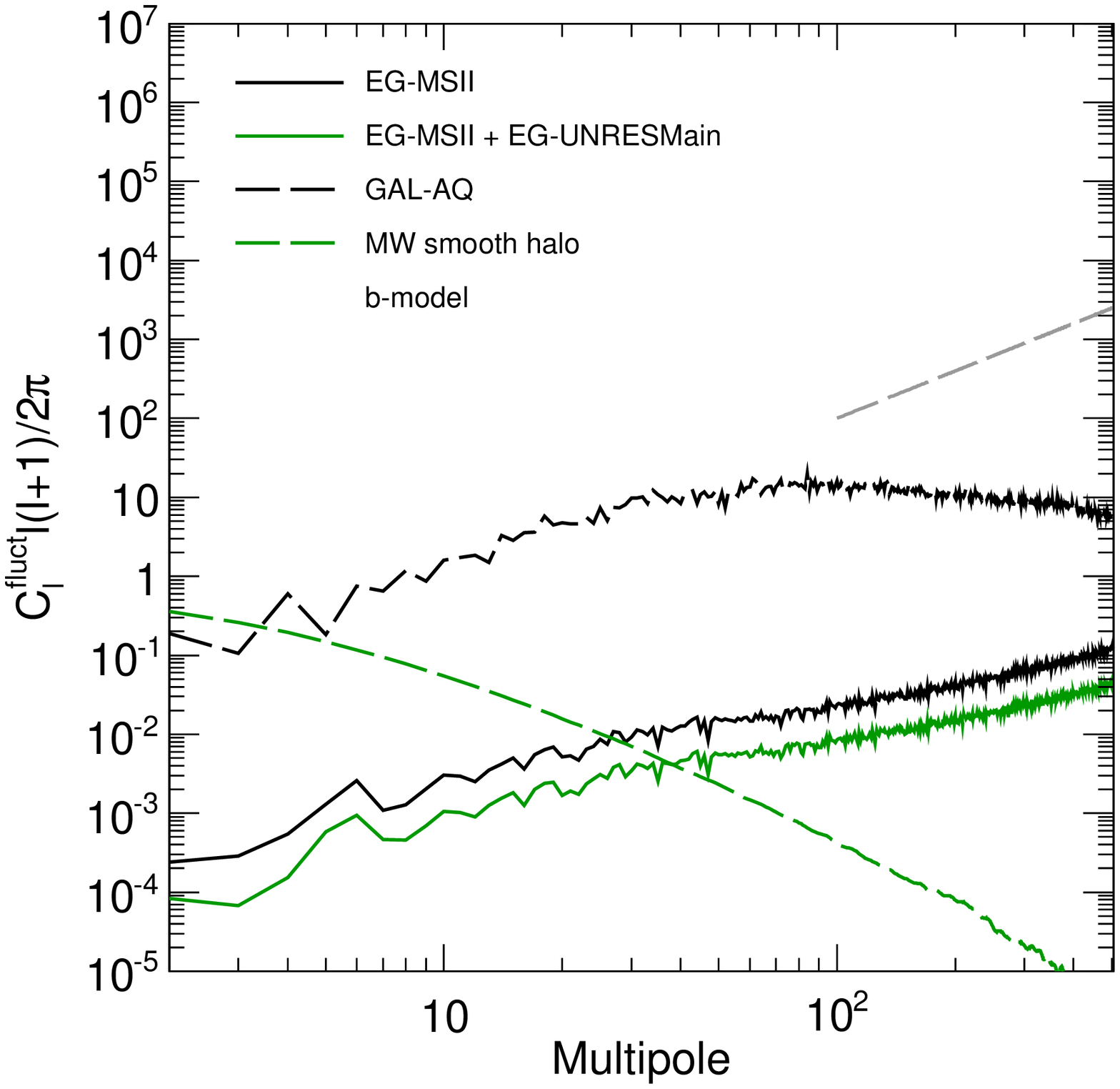}
\includegraphics[width=0.49\textwidth]{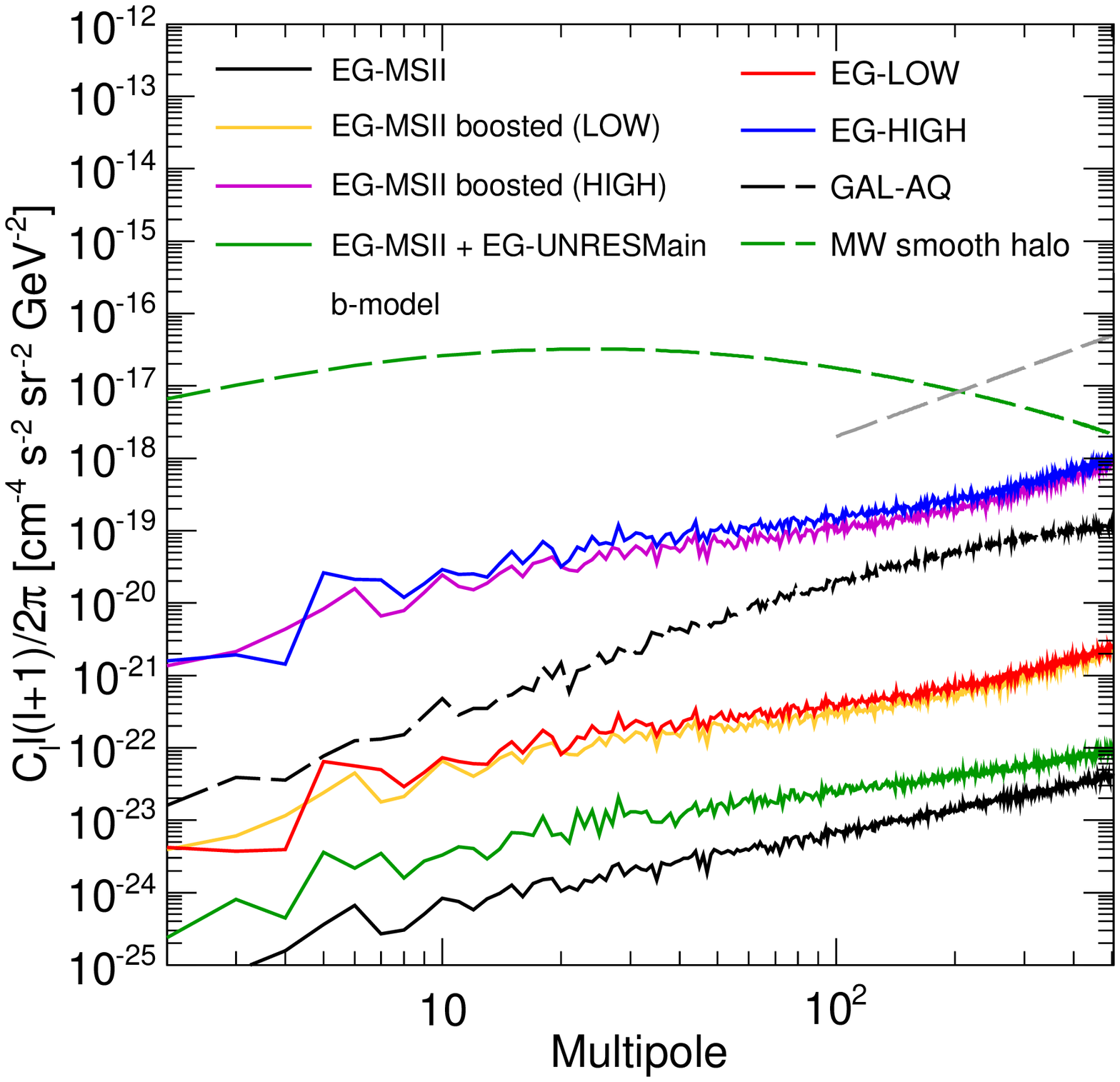}
\includegraphics[width=0.49\textwidth]{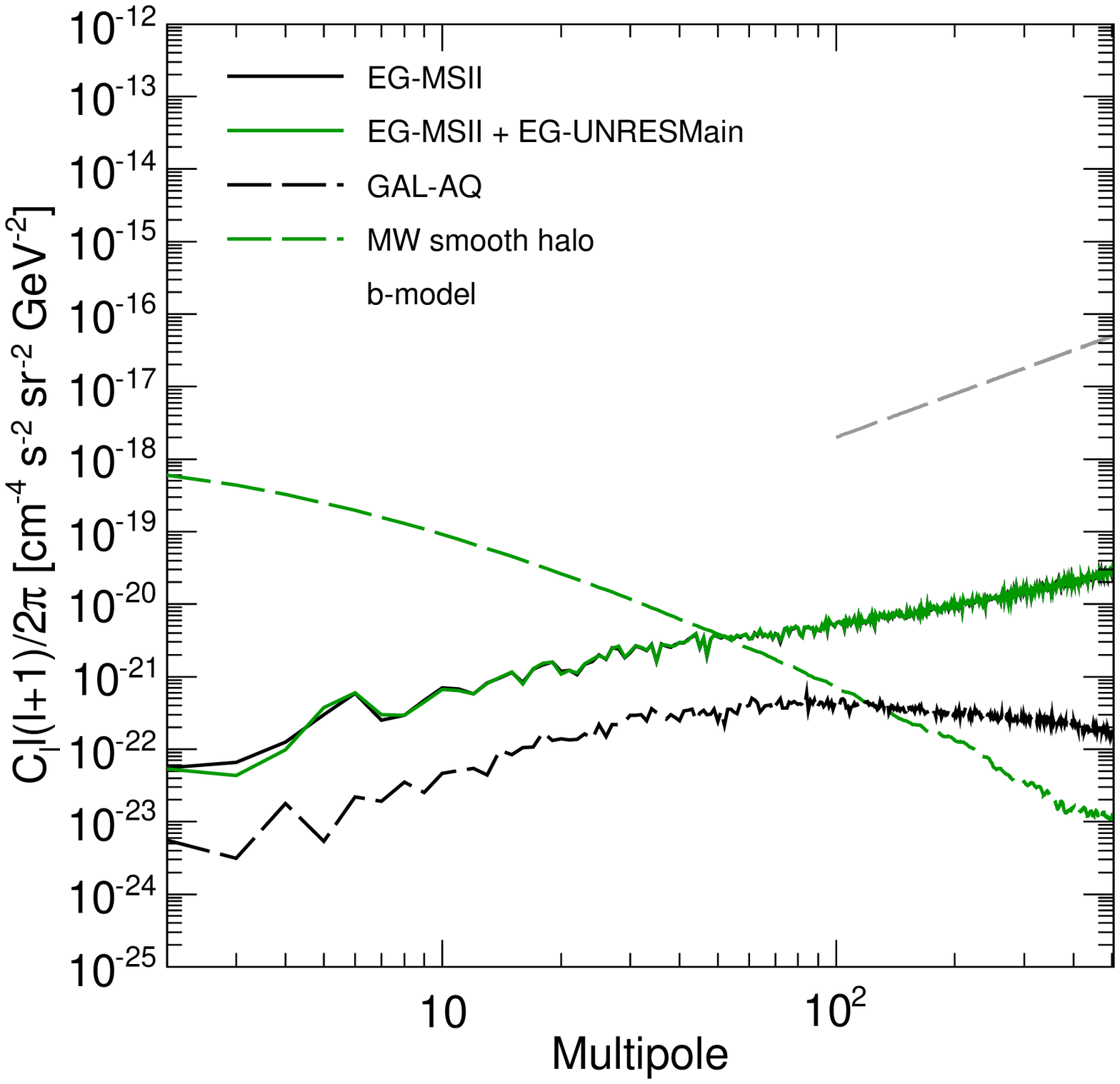}
\caption{\label{fig:APS} Upper panels: Fluctuation APS of the template gamma-ray maps at an observed energy of 4~GeV for annihilating DM (left) and decaying DM (right). The particle physics parameters (including $M_{\rm min}$) as well as the color coding are the same as those in Figs. \ref{fig:IGRB_flux_redshift} and \ref{fig:IGRB_flux}. Solid (dashed) lines indicate the extragalactic (galactic) emission. Bottom panels: The same as the upper panels but for the intensity APS (see Eq. \ref{eqn:intensity_APS}). The upper panels give a measure of the relative anisotropies of the different components, whereas the bottom panels are an absolute measurement of the anisotropies and clearly show which components dominate the APS. The grey dashed line (with arbitrary normalization) indicates a Poissonian APS independent on multipole.}
\end{figure*}

\subsubsection{Extragalactic APS}
The upper panels of Fig.~\ref{fig:APS} show the fluctuation APS of our 
template maps at an observed energy of 4~GeV for the case of annihilating DM 
(left panel) and decaying DM (right panel), using the same particle physics 
benchmark models used in Figs.~\ref{fig:IGRB_flux_redshift} and 
\ref{fig:IGRB_flux} (defined in Sec. \ref{sec:particle_physics}). 
The color-coding is also the same as Fig. \ref{fig:IGRB_flux_redshift}: solid
lines indicate extragalactic components, while dashed ones stand for galactic
ones. The minimal halo mass is assumed to be $M_{\rm min}=10^{-6}M_{\odot}/h$. 

The fluctuation APS (upper panels) illustrates clearly the difference in the 
intrinsic anisotropies pattern of the different components which can be
summarized as follows\footnote{We remind the reader that the extragalactic 
APS is affected by a deficit of power at large angular scales due to finite 
size of the MS-II box.}: 

\noindent
{\it Resolved (sub)halos in MS-II (EG-MSII):} in the 
case of DM annihilation, the extragalactic signal from the resolved (sub)halos 
(solid black line) is less steep than a pure shot-noise power spectrum, 
characteristic of perfectly unclustered sources and, not surprisingly, it is 
in agreement with the results found by \citet{Zavala:2009zr} (see the black 
solid line of their Fig.~12). At large multipoles, this component is 
approximately compatible also with the top right panel of Fig. 2 of 
\citet{Ando:2006cr}.

\noindent
{\it Unresolved subhalos of MS-II main halos:} the solid yellow and 
purple lines correspond to the case in which the emission of resolved main 
halos is boosted up by the contribution of unresolved subhalos, for the
LOW and HIGH subhalo boosts, respectively. We see that at large angular 
scales, where the APS is related to the clustering of main halos, the yellow 
and purple lines have a larger normalization than the black one, although 
their shapes are approximately the same. This is because subhalos give a 
larger boost to the most massive halos, which are also more clustered 
(biased). At intermediate scales, from $\ell=30$ to 100, the APS gets 
shallower reflecting the internal distribution of subhalos within the largest 
halos, which is considerably less peaked than their smooth density profiles. 
Finally, at larger multipoles ($\ell>100$), the emission is dominated by 
low-mass main halos and thus the yellow and violet solid lines are 
essentially on top of the solid black line.

\noindent
{\it Unresolved main halos (EG-UNRESMain):} on the other hand, the 
solid green line indicates the case in which the contribution from unresolved 
main halos is added to the resolved component. The fluctuation 
APS of the EG-UNRESMain component alone is characterized by a lower 
normalization than the solid black line, since we assume that unresolved main 
halos have the same distribution of the least massive halos in MS-II (see 
Sec. \ref{sec:unresolved_halos}). Moreover, these are mainly point sources 
(and very numerous), thus their APS is less steep than the case of the 
EG-MSII component, being mainly sensitive to what is called the 
``2-halo term'', i.e., to correlations between points in different halos 
\citep[e.g.][]{Ando:2005xg}. The green line can be compared with the dashed 
line in Fig. 12 of \citet{Zavala:2009zr}: we note a significant difference 
for $\ell>40$, where the APS in \citet{Zavala:2009zr} is closer to a pure 
shot-noise behaviour. 
This difference already appears in Fig. \ref{fig:old_vs_new_code} where the 
APS obtained with the code used in \citet{Zavala:2009zr} exhibits more power 
at large multipoles than what we find with our improved map-making code. We 
speculate that the steep APS of the dashed line in Fig. 12 of 
\cite{Zavala:2009zr} is a consequence of the spurious features that can be 
seen in Fig. \ref{fig:old_vs_new_code} and that we have reduced in the present 
work.

\noindent
{\it Total extragalactic emission:} once the unresolved subhalo boost 
is applied to halos below and above the MS-II mass resolution, we obtain the 
full extragalactic emission, for either the LOW (solid red line) or HIGH 
(solid blue line) subhalo cases. The contribution of unresolved halos (even 
with the subhalo boost) to the fluctuation APS is subdominant and the shape 
of the solid red and blue lines is exactly the same as the solid yellow and 
purple lines, respectively. 
The decrease in the normalization is due to the fact that the resolved 
structures generate anisotropies that only contribute to a small fraction of 
the total emission (the $f_i$ factor in Eq. \ref{eqn:sum_APS}).

In the lower panels of Fig.~\ref{fig:APS} we show the intensity APS, which 
allow us to estimate the absolute contribution of the different components. 
Large values of the intensity APS can be obtained from a particularly 
anisotropic component or from a very bright one. The angular dependence for
all components is the same as in the fluctuation APS, but now, due to a very 
small average intensity, the EG-MSII component has the lowest intensity APS 
(black solid line), followed by the solid green line, corresponding to the sum 
of the EG-MSII and EG-UNRESMain components (even if the fluctuation APS 
is larger for the former than for the latter). Once the full extragalactic 
emission is considered (solid red and blue lines), the intensity APS is 
between a factor of $100$ and $5 \times 10^4$ larger than the intensity APS of 
EG-MSII, depending on the subhalo boost used.
Notice that the solid yellow and purple lines (that only include resolved 
(sub)halos and the subhalo boost to the resolved main halos) have essentially 
the same intensity APS as the solid red and blue lines, which implies that 
the total intensity APS of the DM annihilation signal is dominated by the 
extragalactic unresolved subhalos of the massive main halos.

In the case of DM decay (right panels), we can see that the 
fluctuation APS of the EG-MSII component (solid black line) has the same
shape as the solid green line (which adds the contribution of EG-UNRESMain), 
but a higher normalization. This is because the signal is dominated by the 
massive resolved (sub)halos. We also see this in the case of the intensity 
APS (bottom right panel), where the contribution of low-mass halos to the 
intensity APS is essentially negligible (the solid green line overlaps with 
the solid black line).

\begin{figure}
\includegraphics[width=0.49\textwidth]{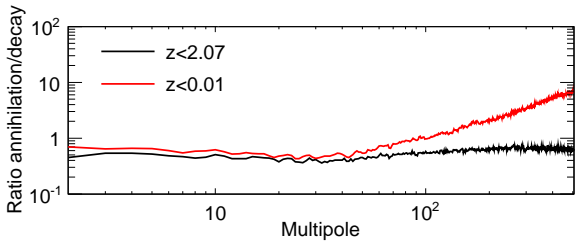}
\caption{\label{fig:APS_ratio} Ratio of the fluctuation APS of the extragalactic maps (resolved (sub)halos, EG-MSII) between the case of annihilating and decaying DM (for the same particle physics models as in previous figures). The black line corresponds to the DM-induced emission up to $z<2.07$ while the red line only accounts for the emission in the first shell ($z<0.01$).}
\end{figure}

The fluctuation APS of the extragalactic maps for the case of DM annihilation 
and DM decay are very similar. This can be seen more clearly in 
Fig.~\ref{fig:APS_ratio}, where we plot the ratio of the fluctuation APS of 
the EG-MSII component in the case of annihilating and decaying DM. The black 
line corresponds to the APS of the past-light cone up to $z<2.07$, while for 
the red line we only consider the first concentric shell ($z<0.01$). The red 
line shows that the annihilation and decay cases are different mainly at 
large multipoles ($\ell>50-60$) where the APS is sensitive to the inner halo 
profile: the more extended the surface brightness profile is, the less steep 
the APS is. Thus, we expect the APS to be steeper (at large multipoles) for the 
case of annihilating DM than for decaying DM. However, this effect is only 
evident for the objects that are closer to us: (sub)halos that are further 
away appear point-like (for the angular resolution of the maps) and, in that 
case, the signal from annihilation and decay becomes indistinguishable, as is 
shown by the black line in Fig.~\ref{fig:APS_ratio}.

\begin{figure*}
\includegraphics[width=0.49\textwidth]{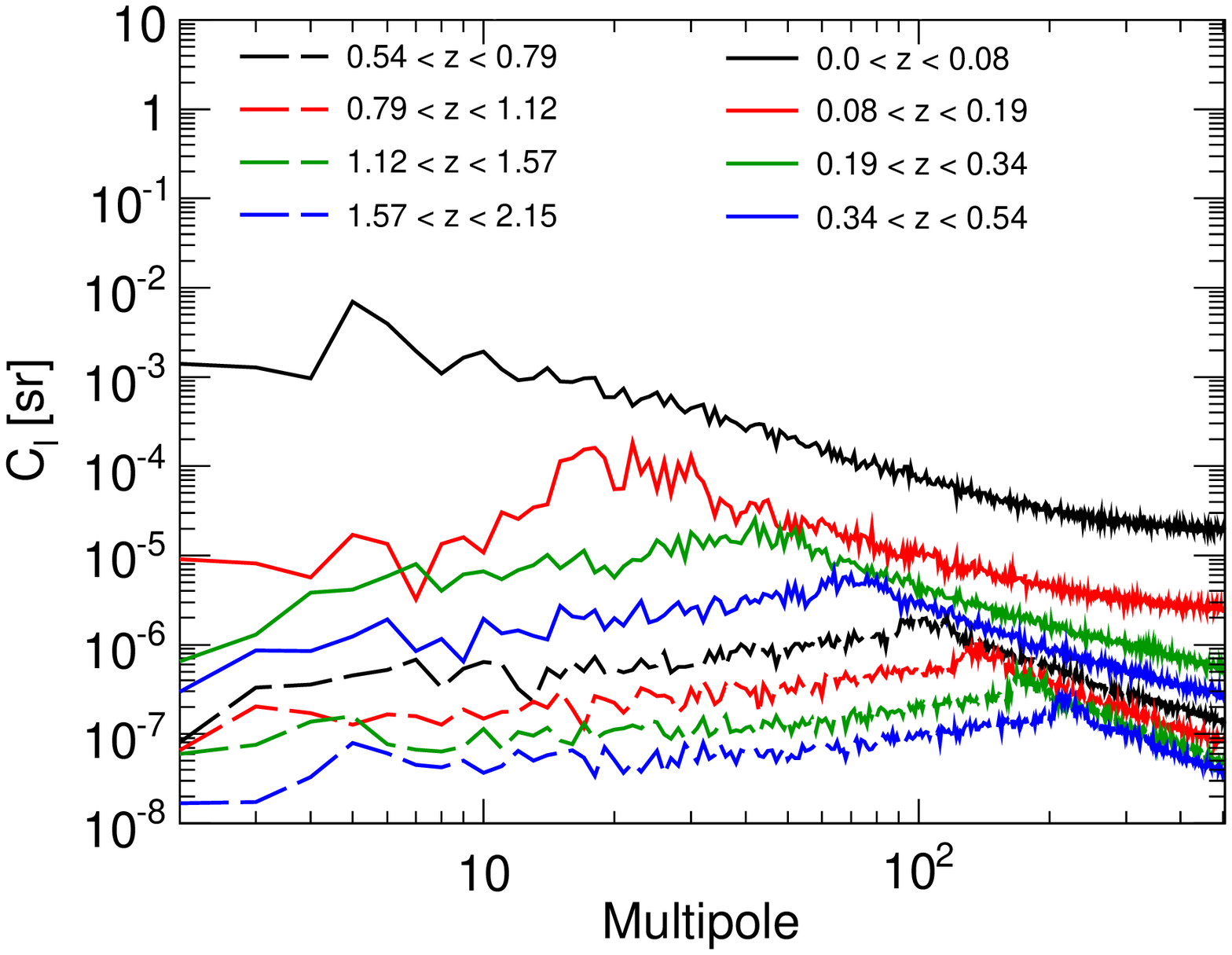}
\includegraphics[width=0.49\textwidth]{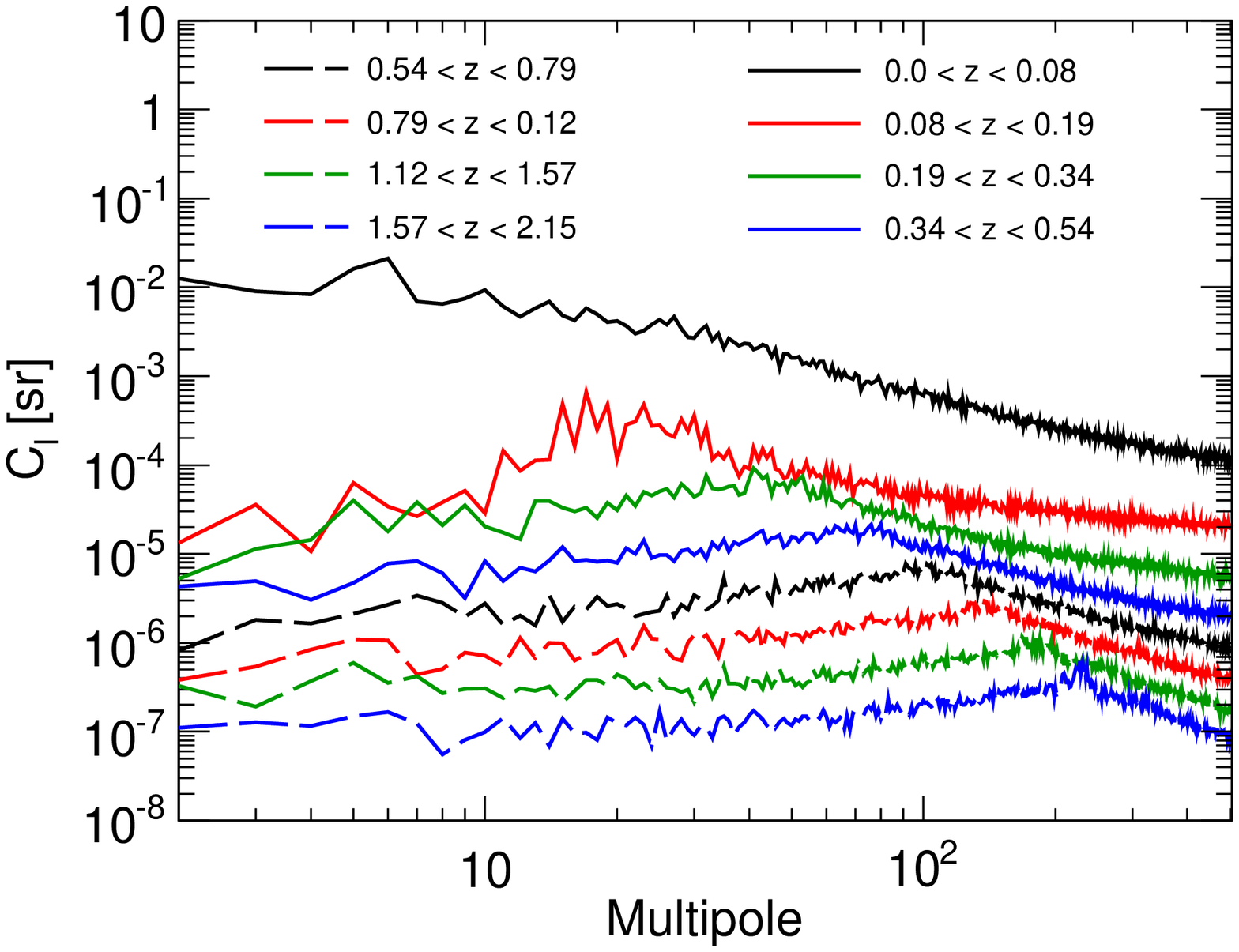}
\caption{\label{fig:APS_redshift} Fluctuation APS for the extragalactic gamma-ray intensity coming from DM annihilation (left) and DM decay (right) for different redshift bins. The APS is computed at an energy of 4 GeV and for the LOW subhalo boost (in the case of annihilation) with $M_{\rm min}=10^{-6}M_\odot/h$. The $b$-model is assumed.}
\end{figure*}

\subsubsection{Galactic APS}
\label{sec:Galactic_APS}
The dashed lines in Fig. \ref{fig:APS} indicate our results for the APS
of the galactic components. They can be summarized as follows:

\noindent
{\it MW smooth halo:} since the position of the observer is offset with 
respect to the GC, the DM-induced emission associated with our own galaxy is 
larger when looking towards the GC. 
This creates a large scale dipole\footnote{Strictly speaking the effect of
having an emission peaking towards one particular direction does not affect 
only the APS at $\ell=1$, as a real dipole, but extends to much larger 
multipoles.}, that can be seen in the APS of the smooth component (dashed 
green lines in Fig.~\ref{fig:APS}), which decreases more rapidly for the case 
of decay than for annihilation since the luminosity profile is more centrally 
concentrated in the latter. 

\noindent
{\it Resolved AQ subhalos (GAL-AQ):} in contrast to the 
previous case, the emission from the resolved subhalos (GAL-AQ, dashed black 
lines) is much more anisotropic at larger multipoles, being rather similar in 
shape to the extragalactic one. 
The exact shape of the GAL-AQ contribution can be affected by the position of 
the observer relative to the local subhalo population: if a subhalo is very 
close to the observer, it would appear as a very extended source in the sky 
map increasing the power at low multipoles, making the APS steeper.
In order to quantify this effect, we constructed 100 different sky-maps of the 
GAL-AQ component, randomly changing the position of the observer in the 
surface of a sphere centered in the GC with a radius of 8.5 kpc. We find that 
the first and third quartiles of the distribution (at $\ell=200$) are located 
only a factor of 2 below and above the median, respectively. 

\noindent
{\it Unresolved galactic subhalos (GAL-UNRES):} to evaluate the 
contribution of unresolved galactic subhalos to the APS we follow the method 
presented in \citet{Ando:2009fp}, which uses analytical relations to 
calculate the APS from galactic substructures for a specified subhalo 
distribution and luminosity function. The details of our implementation are 
described in Appendix \ref{sec:Galactic_unresolved_subhalos}. Basing our 
subhalo models on the results of \citet{Springel:2008cc} (both for a LOW 
and HIGH boost), the contribution of unresolved galactic substructures to the 
intensity APS is small: for annihilation, the contribution to the APS is less 
than $\sim 10\%$ of that from resolved subhalos, while for decay their 
contribution is at most a few percent of that from resolved subhalos. We 
therefore choose to not include this contribution to the APS.

Overall, considering the galactic and extragalactic contributions, the APS
signal is clearly dominated by the smooth halo component in the case of DM 
annihilation, although the extragalactic emission could be important at very 
large multipoles ($\ell\gtrsim300$) if subhalos give a large boost. 
On the contrary, for DM decay, the extragalactic emission dominates already 
from $\ell\gtrsim20$ and it is only at the very large scales that the 
anisotropy of the smooth halo dominates the signal. However, if a mask is
introduced along the galactic plane (as in \citealt{Ackermann:2012uf}), we 
expect that the  balance between galactic and extragalactic components will 
change, reducing significantly the impact of all the components characterized 
by a large emission around the GC (see Sec. \ref{sec:discussion}).

\section{Discussion}
\label{sec:discussion}

\subsection{Redshift dependence of the extragalactic APS}
\label{sec:APS_redshift}
In Fig.~\ref{fig:APS_redshift}, we divide the extragalactic gamma-ray emission 
in redshift bins (each bin including 4 MS-II snapshots) and compute the 
fluctuation APS for the EG-LOW component (i.e. the total emission, including 
all (sub)halos, down to $M_{\rm min}=10^{-6} M_\odot/h$ and with a LOW subhalo 
boost) in each bin. The APS is computed at an energy of 4 GeV. We can see 
that for both, DM annihilation (left panel) and DM decay (right panel), the 
lower redshifts are characterized by a larger anisotropy. This is due to the 
fact that the volume of the past light cone grows with redshift, as well as 
the number of gamma-ray emitting (sub)halos. Thus, the first snapshots are 
those characterized by the lowest number of (sub)halos and are more affected 
by their discrete distribution. Moreover, the clustering of DM (sub)halos 
is larger at lower redshifts.
The peaks that move towards higher multipoles with increasing redshift
are a remnant of the spurious effect related to the periodicity of the 
MS-II box discussed in Sec. \ref{sec:Millennium_II}. For a particular 
redshift, the peaks indicate the angular size of the MS-II box at 
that redshift (what we called $\ell^\star$ in Sec. \ref{sec:Millennium_II}): 
multipoles smaller than $\ell^\star$ are affected by a loss of power due to 
the missing modes at wavelenghts larger than the MS-II box, and therefore we 
cannot trust our predictions below $\ell^\star$. This fact is, however, not 
relevant for a comparison with the Fermi-LAT APS data, since we are mainly 
interested in the multipole range between $\ell=155$ and 500, where the 
extragalactic APS is dominated by the first redshifts, for which 
$\ell^\star<20-30$.

\subsection{Energy dependence of the APS}
The extragalactic fluctuation APS increases with increasing energy, a fact 
already pointed out in the past 
\citep{Ando:2005xg,Zavala:2009zr,Ibarra:2009nw} and related to the redshift 
dependence discussed in the previous section: following Eq. \ref{eqn:sum_APS}, 
the total fluctuation APS at a particular energy $E_\gamma$ can be written as 
the sum $\sum_i f_i^2(E_\gamma) C_i^{\rm fluct}$ over the fluctuation APS of each 
concentric shell $C_i^{\rm fluct}$ normalized by the square of the relative 
emission in the $i$-th shell with respect to the total.
Since individual shells are thin in redshift space, each single 
$C_i^{\rm fluct}$ does not depend on energy, and thus changing the energy only 
has the effect of modifying the $f_i$ factors that determine the balance 
among the APS of the different shells. These $f_i$ factors depend on the 
annihilation/decay channel selected for the particular DM candidate, as well 
as on how much the DM density changes with $z$ within a particular shell 
(see Fig. \ref{fig:IGRB_flux_redshift}). For high energies, the shells that 
contribute the most to the signal, i.e. those with the largest $f_i$ factors, 
are the first shells, which are characterized by the largest APS. Thus, the 
total fluctuation APS increases as energy increases.

Note, however, that in the cases where the fluctuation APS is dominated by 
the galactic emission, the fluctuation anisotropy will not change with energy, 
neither in normalization nor in shape. 

\subsection{Inner density profile of DM (sub)halos}
\label{sec:DM_profiles}
When dealing with the extragalactic emission, we have assumed that (sub)halos 
have a smooth NFW density profile, which is characterized by a slope that 
tends asymptotically to $-1$ for small radii, and lies between the steeper 
Moore \citep{Moore:1999gc} and cored Burkert \citep{Burkert:1995yz} profiles. 
Current high resolution $N$-body simulations have demonstrated that the 
Einasto profile (Eq. \ref{eqn:Einasto}) produces an even better fit than NFW. 
The slope of the Einasto profile decreases as a power-law as the distance 
from the center decreases, and it is shallower than NFW at small radii. 

It is also important to note that the process of galaxy formation within DM 
halos has an impact on the DM distribution in the central regions where it 
is believed that the halo is adiabatically contracted resulting in a more 
concentrated DM distribution 
\citep[e.g.][]{Mo:1997vb,Gnedin:2004cx,Ahn:2007ty}. However, recent 
hydrodynamical simulations with strong supernovae feedback claim that 
including the effect of baryons can actually result in the development of a 
central DM core in intermediate mass halos 
\citep{Pontzen:2011ty,Maccio':2011eh}.

For the extragalactic emission, the uncertainty on the inner DM density 
profile has very limited effect on the APS since only a small fraction of 
(sub)halos cover more than one pixel in our maps, and also because, even 
if the object is characterized by extended emission, the difference 
between a cuspy or cored profile is only noticeable at very small projected 
radii. On the other hand, we do expect a change in the total intensity of the 
DM-induced emission: for instance, if an Einasto profile is used instead of a 
NFW, the annihilation rate per halo will increase by $50\%$ 
\citep{Zavala:2009zr}. Considering the extreme cases of a Burkert and a 
Moore profile, the difference is roughly an order of magnitude 
\citep{Profumo:2009uf}. For decaying DM the total luminosity of a halo is 
directly proportional to its mass, independently of the DM profile assumed.

For the galactic emission, the reasoning above applies to the resolved 
subhalos. On the other hand, assuming a different profile for the smooth 
halo may have a stronger impact on the APS, since this represents the largest 
contribution (at least at low multipoles, and particularly for the case of 
annihilating DM). The effect of assuming a NFW\footnote{Taken from 
\citealt{Prada:2004pi} and normalized to the same local density than the
Einasto profile introduced in Sec. \ref{sec:smooth_halo}.} rather than an 
Einasto profile is evident at low multipoles (with the APS of the former 
being smaller than the APS of the latter), but the difference becomes smaller 
at larger multipoles. This can be explained by noting that the emission 
towards the GC is larger with respect to the anti-center in the case of an 
Einasto profile\footnote{If the two profiles are normalized to the same 
density at 8.5 kpc, the NFW will have a larger intensity within $\sim 10$ pc, 
but in the region between $\sim 1$ kpc and $10$ kpc from the GC (where the 
majority of the emission comes from) an Einasto profile is characterized by 
a larger density.}, resulting in a more anisotropic APS. The differences are 
less evident for the case of decaying DM.

Finally, it is important to remember that any uncertainty in the inner
MW density profile will be reduced if the region around the GC is masked.
For instance, for $|b|>30^\circ$, the different reasonable DM 
profiles are practically indistinguishable \citep{Bertone:2008xr}.

\begin{figure}
\includegraphics[width=0.49\textwidth]{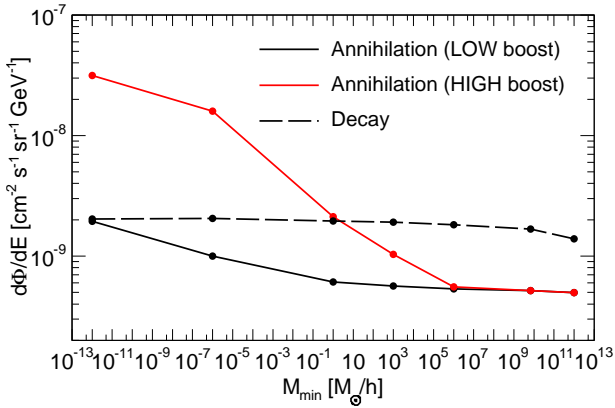}
\caption{\label{fig:Mmin_flux} Total gamma-ray intensity from DM annihilation (solid lines) and DM decay (dashed lines) at 4 GeV as a function of the minimal halo mass $M_{\rm min}$. The LOW and HIGH subhalo boosts (see Secs. \ref{sec:unresolved_subhalos} and \ref{sec:Aquarius}) are shown with black (red) lines, respectively. The emission has been computed only for the values of $M_{\rm min}$ indicated by the full dots, while the lines are obtained by interpolation.}
\end{figure}

\begin{figure*}
\includegraphics[width=0.49\textwidth]{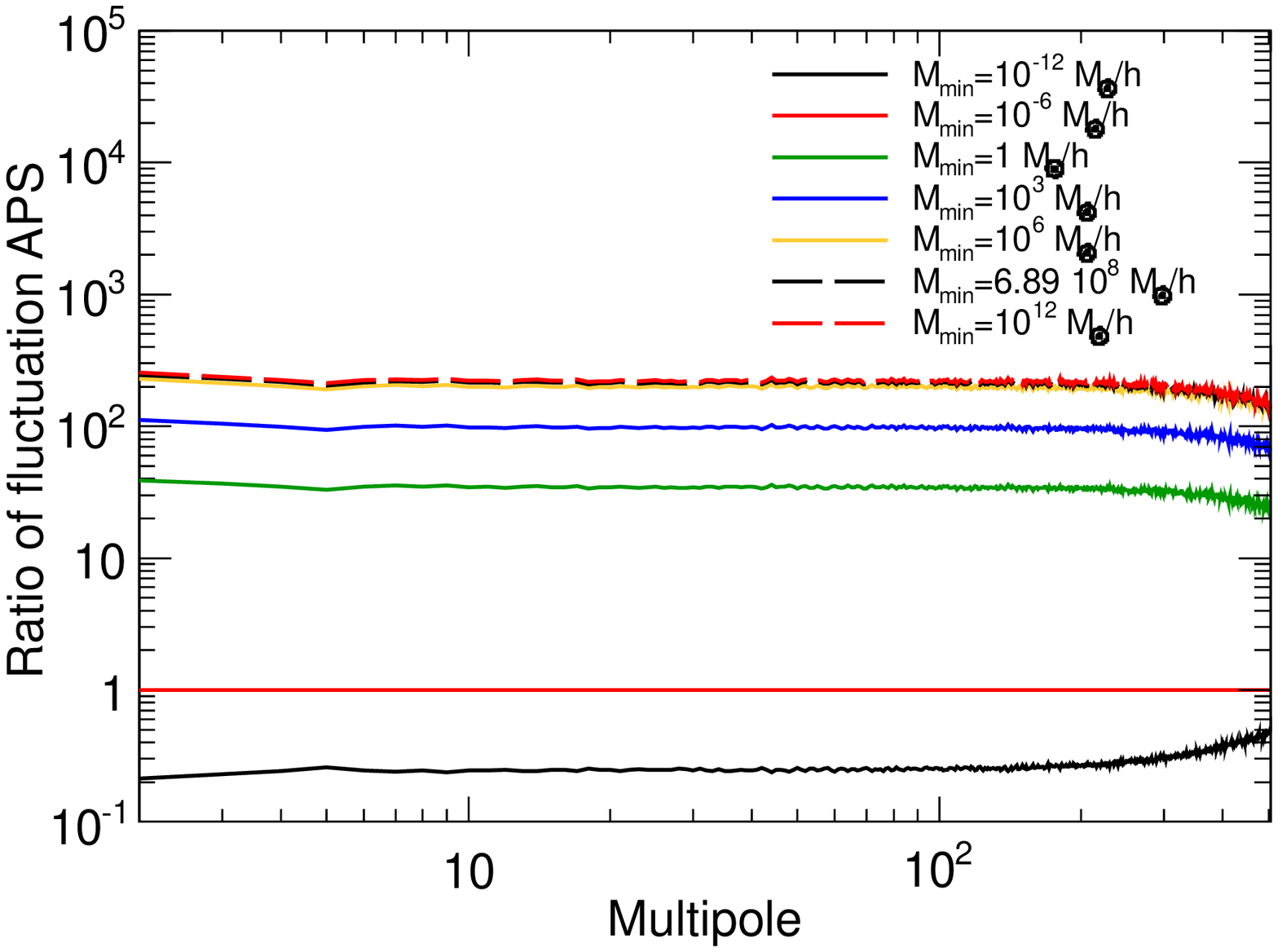}
\includegraphics[width=0.49\textwidth]{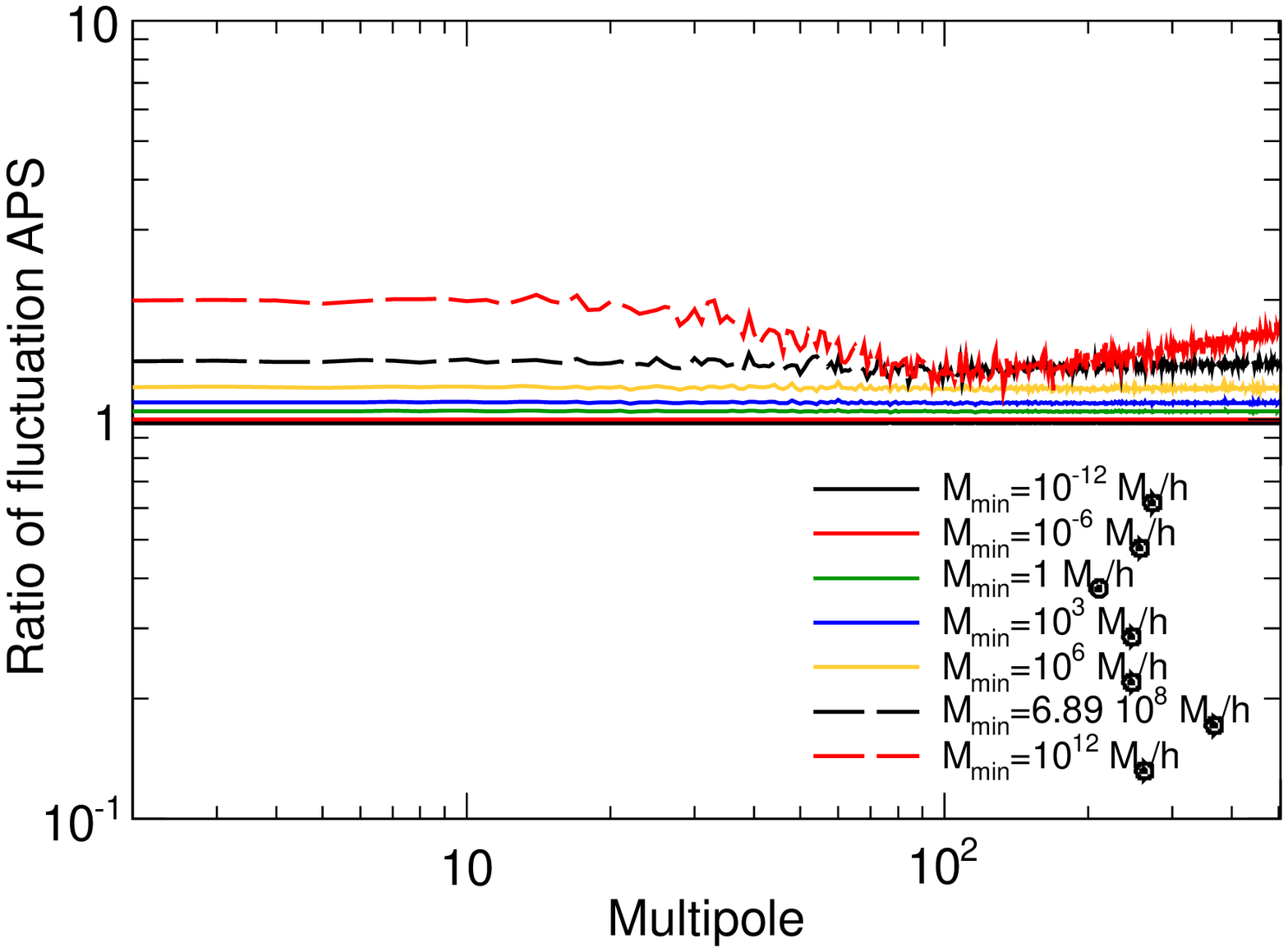}
\caption{\label{fig:Mmin_APS} Ratio of the total fluctuation APS (galactic and extragalactic) for different values of $M_{\rm min}$ with respect to the reference case $M_{\rm min}=10^{-6} M_\odot/h$. The APS is computed at 4 GeV. The left panel refers to the case of annihilating DM with the HIGH subhalo boost, while the right panel is for decaying DM.}
\end{figure*}

\subsection{The minimum self-bound halo mass $M_{\rm min}$} 
\label{sec:Mmin}
The particle nature of DM determines the small-scale cutoff in the matter 
power spectrum of density fluctuations, and hence, the value of $M_{\rm min}$. 
For neutralinos, the most common WIMP DM candidates, typical values for 
$M_{\rm min}$ go from $10^{-11}M_\odot/h$ to $10^{-3}M_\odot/h$ 
\citep[e.g.][]{Bringmann:2009vf}. 
Although this range can be considered as a reference for all WIMP candidates, 
a particular scenario might lie outside this range. In order to investigate 
the impact of different values of $M_{\rm min}$ in our predictions, we generate 
template maps for $M_{\rm min}$ equal to $10^{-12}M_\odot/h$ and 1 $M_\odot/h$. 
We also consider a few larger values (namely $M_{\rm min}=10^{3}$, 
$10^{6}$, $6.89 \times 10^{8}$ and $10^{12}M_\odot/h$) that, although clearly far 
above the expected mass range for WIMP models, are discussed in order to 
understand how halos of different masses contribute to the gamma-ray 
intensity and APS.

In terms of the mean gamma-ray intensity, we can see the impact of changing 
$M_{\rm min}$ in Fig. \ref{fig:Mmin_flux}; solid (dashed) lines for the case of 
DM annihilation (decay), while black and red lines refer to the LOW and HIGH
subhalo boosts, respectively. For annihilation, the mean flux decreases only 
by a factor of $\sim5$ between $M_{\rm min}=10^{-12}M_\odot/h$ and 1 $M_\odot/h$, 
for the LOW case, while the difference is one order of magnitude for
the HIGH case. For even higher values of $M_{\rm min}$, the intensity stays 
essentially constant for the LOW case, while it decreases further for the HIGH 
case until reaching a plateau. In both cases, the point where the intensity 
reaches an approximately constant value marks the region where the total 
intensity passes from being dominated by the extragalactic component (for
low $M_{\rm min}$) to being dominated by the emission in the MW (for larger
$M_{\rm min}$). The transition happens at the larger values of $M_{\rm min}$
for the HIGH subhalo boost, because (see Sec. \ref{sec:energy_spectrum}) 
increasing the subhalo abundance produces more significant effects for the 
extragalactic emission than for the galactic one.

In the case of DM decay (as discussed before), the bulk of the emission is 
dominated by large mass halos and, in particular, is already accounted for in 
(sub)halos with masses larger than $10^8-10^9 M_\odot/h$, with smaller objects 
contributing only marginally. 

The effect of $M_{\rm min}$ on the total APS is shown in 
Fig. \ref{fig:Mmin_APS}: the two panels indicate the ratio of the fluctuation 
APS at 4 GeV for 7 values of $M_{\rm min}$ with respect to the reference case 
of $M_{\rm min}=10^{-6} M_\odot/h$. The panel on the left shows the case of an 
annihilating DM candidate with a HIGH subhalo boost: looking at 
Fig. \ref{fig:APS}, for $M_{\rm min}=10^{-6} M_\odot/h$, the total intensity APS 
is dominated by the MW smooth halo, while the contribution of extragalactic 
(sub)halos plays a role only at large multipoles. Now, going from 
$M_{\rm min}=10^{-6}$ to $10^{-12} M_\odot/h$ does not have a strong impact on the 
galactic component but it makes the total extragalactic emission increase by a factor of a few (see Fig. \ref{fig:Mmin_flux}). The net effect, following 
Eq. \ref{eqn:sum_APS}, is that the total fluctuation APS decreases because 
less intensity is associated with the component that dominates the intensity 
APS (i.e. the galactic one). This is also the reason why the total fluctuation 
APS increases from $M_{\rm min}=10^{-6}$ to 1, $10^3$ and $10^6 M_\odot/h$. When the
total emission starts to be dominated by the MW smooth halo (i.e. above
approximately $\sim 10^{6} M_\odot/h$), there is essentially no change to the 
APS due to variations in $M_{\rm min}$.

The same features appears in the case of a LOW subhalo boost (the figure is
not present), even if this case is characterized by a smaller relative 
difference (all the lines are within one order of magnitude), and, since the 
emission of the DM smooth halo starts to dominate already at 
$M_{\rm min}=1$ $M_\odot/h$, the APS does not change for $M_{\rm min}$ larger 
than that value.

The right panel in Fig. \ref{fig:Mmin_APS} is for decaying DM: the different 
lines follow the same behaviour as for annihilating DM but the effect of 
changing $M_{\rm min}$ is highly reduced. The only important deviation is for 
the largest value of $M_{\rm min}$: at low multipoles the APS is still dominated 
by the smooth MW halo and, thus, we expect only a different normalization. 
However, when the extragalactic component becomes relevant, the dashed red 
line decreases because the extragalactic fluctuation APS for 
$M_{\rm min}=10^{12} M_\odot/h$ is smaller than the case at 
$M_{\rm min}=10^{-6} M_\odot/h$ since it is sensitive, at these multipoles, to 
the inner DM profile of the largest objects.

\begin{figure}
\includegraphics[width=0.49\textwidth]{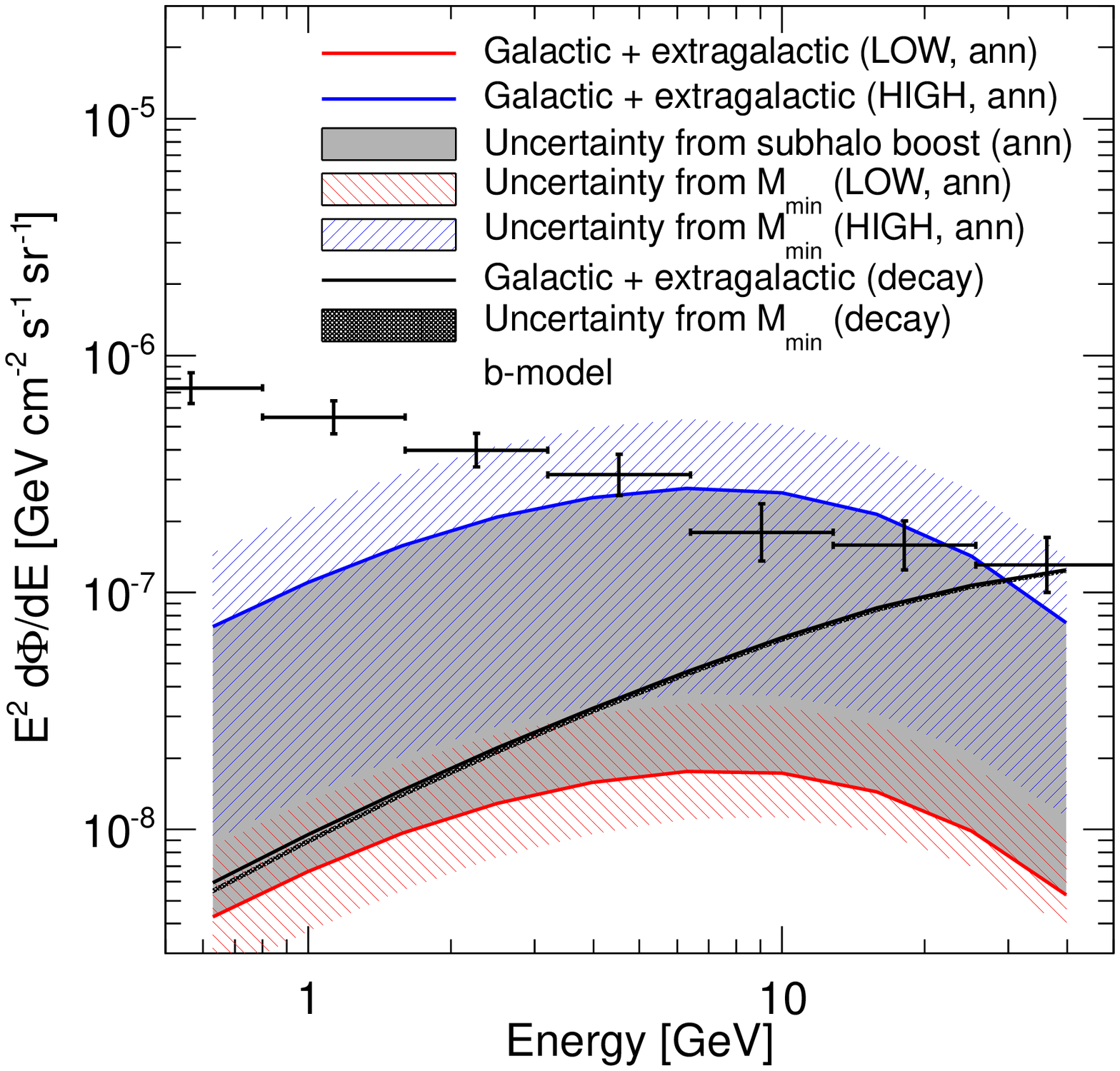}
\caption{\label{fig:Flux_bands} Energy spectrum of the average gamma-ray intensity from DM annihilation (color lines) or decay (black line) from extragalactic and galactic (sub)halos. The blue and red lines correspond to the LOW and HIGH subhalo boosts, respectively, so that the filled grey area between them corresponds to the uncertainty due to the subhalo boost, for a fixed value of $M_{\rm min}$. The red (blue) shaded area around the red (blue) solid line indicates the uncertainty in changing the value of $M_{\rm min}$ from $10^{-12}$ to 1 $M_\odot/h$, for the LOW (HIGH) scenario boost. The solid black line shows the prediction for a decaying DM candidate and the black shaded area (appearing as a thickening of the solid black like) indicates the uncertainty in changing $M_{\rm min}$ from $10^{-12}$ to 1 $M_\odot/h$. The observational data points with error bars refer to the measurement of the IGRB as given in \citet{Abdo:2010nz}. Only the emission with $|b|>10^\circ$ is considered. The DM candidates are described in Sec. \ref{sec:particle_physics}.}
\end{figure}

\subsection{Theoretical uncertainty bands}
In the current section we summarize our predictions for the energy and 
angular power spectra of the DM contribution to the IGRB emission. We also 
present ``theoretical error bands'' that bracket the uncertainties discussed 
in the previous sections. These predictions are given only for a fixed 
particle physics scenario (the $b$-model, see 
Sec. \ref{sec:particle_physics}), while the analysis of different DM 
candidates, (i.e., changing $m_\chi$, the annihilation cross section, decay 
life time and annihilation/decay channels), will be discussed in a
follow-up paper. 

The energy spectrum of the DM-induced signal (averaged over the region with 
$|b|>10^\circ$) is shown in Fig. \ref{fig:Flux_bands}. 
The grey area between the red (LOW subhalo boost) and blue line (HIGH subhalo
boost) spans approximately a factor of 50 and quantifies the uncertainty 
associated with the unknown subhalo boost, for a fixed value of 
$M_{\rm min}=10^{-6} M_\odot/h$. The additional red and blue shaded areas indicate 
the uncertainties introduced by changing the value of $M_{\rm min}$ between 
$10^{-12} M_\odot/h$ and 1 $M_\odot/h$.
For the case of decaying DM, our predictions are completely determined by 
massive (sub)halos so the theoretical uncertainties are much smaller than for 
the case of DM annihilation. The Fermi-LAT data from \citet{Abdo:2010nz} are 
also plotted with error bars.

\begin{figure*}
\includegraphics[width=0.49\textwidth]{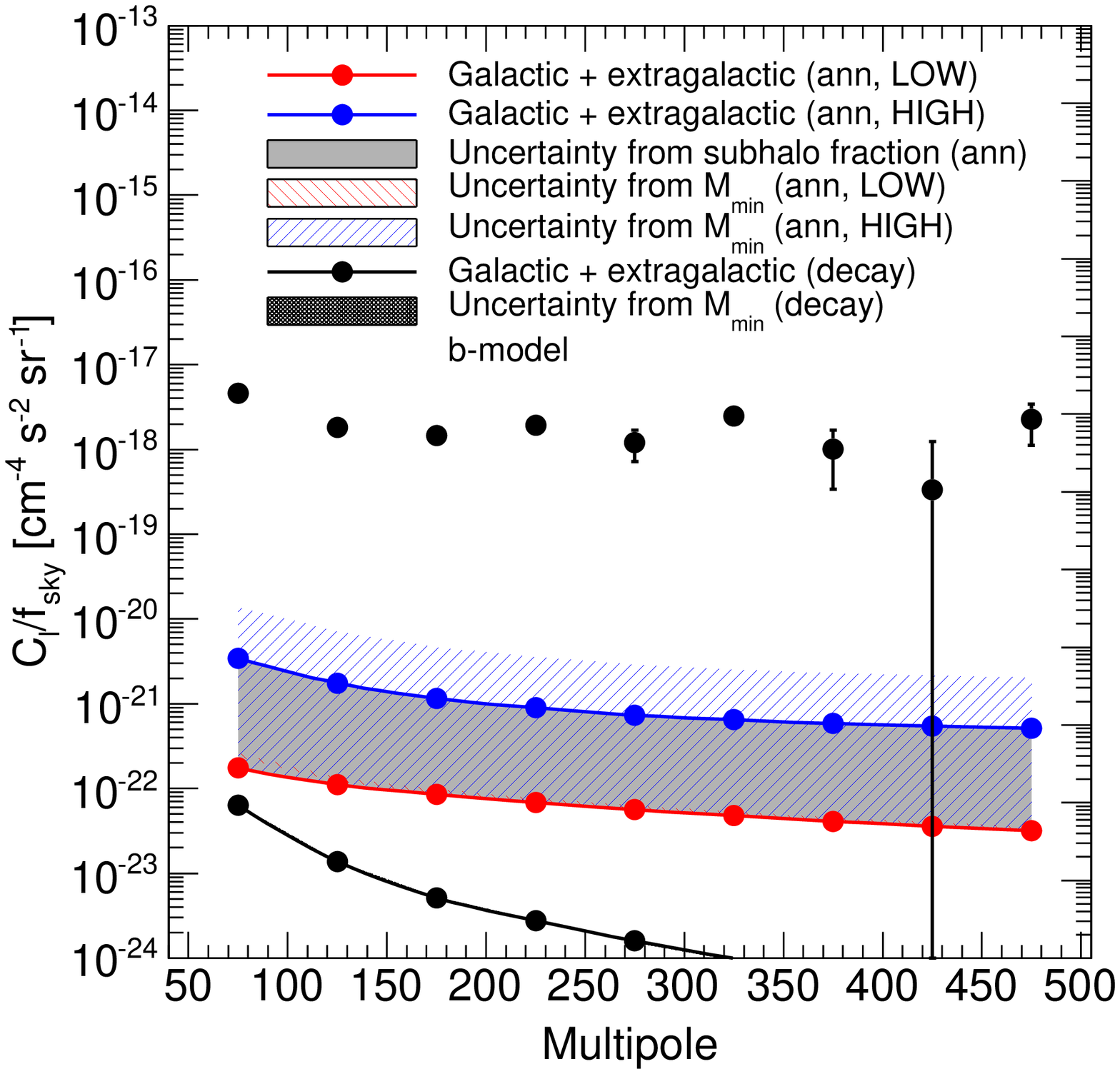}
\includegraphics[width=0.49\textwidth]{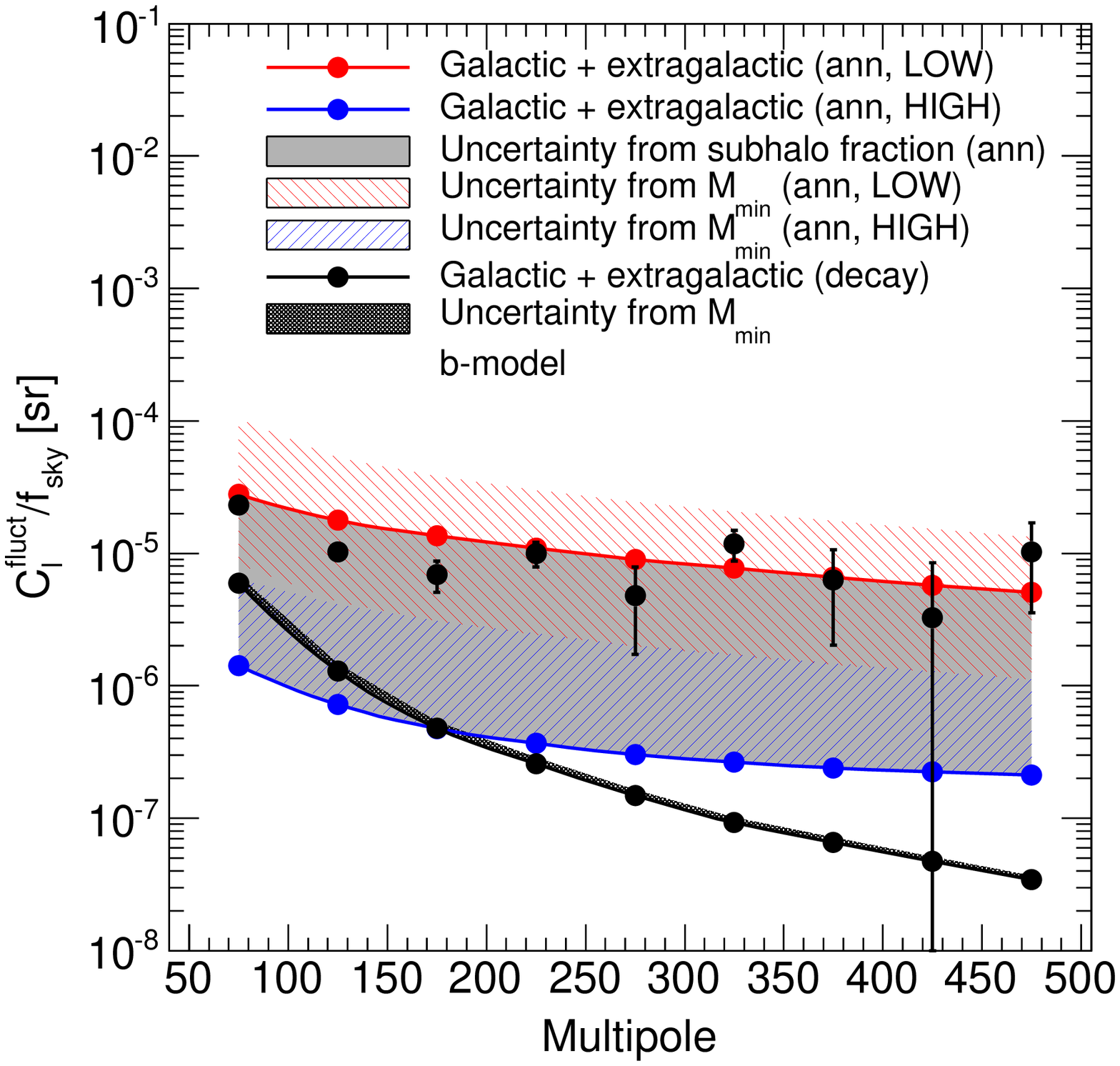}
\caption{\label{fig:APS_bands} Total intensity APS of the gamma-ray emission from DM annihilation (color lines) or decay (black line) in extragalactic and galactic (sub)halos. The blue and red lines correspond to the LOW and HIGH subhalo boosts, respectively, so that the filled grey area between them corresponds to the uncertainty due to the subhalo boost, for a fixed value of $M_{\rm min}$. The red (blue) shaded area around the red (blue) solid line indicates the uncertainty in changing the value of $M_{\rm min}$ from $10^{-12}$ to 1 $M_\odot/h$, for the LOW (HIGH) case. The solid black line shows the prediction for a decaying DM candidate and the small black shaded area, appearing as a thickening of the solid black line, indicating the uncertainty in changing $M_{\rm min}$ from $10^{-12}$ to 1 $M_\odot/h$. The APS is measured in the energy bin between 2 to 5 GeV. The observational data points with error bars refer to the measurement of the APS as given in \citet{Ackermann:2012uf}. A region of $30^\circ$ around the galactic plane has been masked and the APS has been binned with a binsize of $\Delta\ell=50$. The DM candidates are described in Sec. \ref{sec:particle_physics}.}
\end{figure*}

Fig. \ref{fig:APS_bands} summarizes our predictions for the DM-induced APS 
(intensity APS in the left panel and fluctuation APS in the right panel). 
Contrary to the plots presented in the previous sections, the APS is now 
computed after having integrated the gamma-ray emission between 2 and 5 GeV. 
Moreover, the APS has been averaged in bins of $\Delta\ell=50$ starting from 
$\ell=5$, and we introduce a mask covering the region with $|b|<30^\circ$. We
approximately correct for the effect of the mask by dividing the raw APS by
the fraction of the sky $f_{\rm sky}$ left unmasked, as it was done in 
\citet{Ackermann:2012uf}. All of this is for comparison purposes with the 
Fermi-LAT APS data in the same energy bin, taken from 
\citealt{Ackermann:2012uf}\footnote{We do not mask the point sources in the 
1-year catalog, as in \citet{Ackermann:2012uf}, so that our $f_{\rm sky}$ is 
0.5.}.
The inclusion of the mask has strong effects both on the average emission of 
the smooth MW halo and on its APS since we are masking the region where the 
signal peaks. On the other hand, it has a limited effect on the extragalactic 
emission. After masking, the total intensity APS for annihilating DM is 
dominated by the resolved galactic subhalos in the case of the LOW subhalo 
boost and by the extragalactic unresolved (sub)halos for the HIGH subhalo 
boost, i.e., contrary to what is shown in Fig. \ref{fig:APS}, the smooth MW 
halo only represent a subdominant contribution. For decaying DM, all these 
three components (extragalactic emission, resolved galactic subhalos and the 
smooth MW halo) have a comparable intensity APS.

In Fig. \ref{fig:APS_bands}, the red and blue lines indicate our predictions 
for an annihilating DM candidate in the LOW and HIGH scenario, respectively. 
Thus, the grey area indicates the uncertainty associated with the unknown 
subhalo boost. If we had plotted only the extragalactic intensity APS in the 
left panel, the LOW case would have been a factor 500 below the line for the 
HIGH case (as in Fig. \ref{fig:APS}). However, the resolved galactic subhalos 
increase the intensity APS for the LOW case, while having a less important 
role for the HIGH case. Thus, the red and blue lines are only one order of 
magnitude away from each other. Moreover, the uncertainty due to 
$M_{\rm min}$ is completely negligible in the LOW case since the APS is 
determined by the galactic resolved subhalos, and thus is not sensitive to 
changes in $M_{\rm min}$. The same is true for the case of decaying DM (black 
line), whose APS is determined by massive (sub)halos.

The right panel of Fig. \ref{fig:APS_bands} shows the fluctuation APS: the 
red line, corresponding to the LOW subhalo boost is now above the blue line, 
relative to the HIGH subhalo boost. This is because the galactic subhalos 
(the component that dominates the total APS in the former case) are 
associated with a larger intrinsic anisotropy than the extragalactic 
(sub)halos, which dominate the APS in the latter case.

We conclude this section with a comment on the comparison between our 
predictions for the DM-induced APS with the Fermi-LAT data shown in 
Fig. \ref{fig:APS_bands}. Although a rigorous comparison is left for future 
work we can already see that the fluctuation APS from DM annihilation 
is of the same order, and has a similar shape, as the data (at least in the 
case of the LOW subhalo boost). On the contrary, for a decaying DM candidate, 
the predictions are not compatible with a flat APS and they are also 
characterized by a normalization which is too low.

Nevertheless, even if the annihilating DM candidate we used here is able to 
reproduce the same level and shape of the fluctuation APS inferred from the 
data, it still does not represent a viable interpretation, since such a
candidate is characterized by a very low intensity APS (left panel).
Improvements in the analysis of the gamma-ray data are still possible both 
from the experimental side (e.g. increasing the statistics, especially at
high energies), and from the theoretical side (e.g. one can think of 
selecting, for each DM candidate, the energy bin that maximizes the DM-induced 
intensity APS).

\section{Summary and conclusions}
\label{sec:conclusion}
In the present paper we generated all-sky gamma-ray maps from the 
annihilation/decay of DM in extragalactic (sub)halos and in the halo and 
subhalos of the MW. Apart from the prompt gamma-ray emission, we also 
considered emission due to the IC scattering of $e^+/e^-$ produced in the 
annihilation or decay with CMB photons and, for the smooth MW halo, additional 
contributions from starlight (either directly or re-scattered by dust) and 
the so-called ``hadronic emission'' (see Appendices \ref{sec:IC_emission} and 
\ref{sec:hadronic_emission}) are also considered.

The DM distribution was modeled using state-of-the-art $N$-body simulations: 
Millennium-II for extragalactic (sub)halos and Aquarius (Aq-A-1) for the 
galactic halo and its subhalos. To compute the extragalactic emission, we 
improved the algorithm described in \citet{Zavala:2009zr} and simulated the 
past light cone up to $z=2$. The MS-II allows us to account for the emission of 
structures with a mass larger than $M_{\rm res}\sim 10^9 M_\odot/h$. We then 
considered the intensity from unresolved (sub)halos down to a minimum 
self-bound mass $M_{\rm min}$, by a hybrid method that combines an extrapolation 
of the behaviour of the least massive resolved halos in MS-II with the 
subhalo boost model introduced in 
\citet{Kamionkowski:2008vw} and \citet{Kamionkowski:2010mi} and refined in 
\citet{SanchezConde:2011ap}. On the other hand, the galactic emission was 
modeled assuming that the smooth halo of the MW is given by an Einasto 
profile, renormalized to a value of 0.3 GeV/cm$^3$ for the local DM 
density. Resolved galactic subhalos are taken directly from the Aquarius
simulation (down to a mass of $\approx 10^5 M_\odot$), while the contribution 
of unresolved galactic subhalos is estimated by means of the same procedure 
used for the extragalactic emission.

The template maps of the DM-induced emission were then used to derive the 
energy spectrum of the different components from 0.5 GeV and 50 GeV (see 
Fig. \ref{fig:IGRB_flux}). 
The main goal of the paper is the characterization of the anisotropies of the 
DM-induced emission, which was done in Sec. \ref{sec:APS}, where we computed 
the APS of the different components up to $\ell=500$ (see Fig. \ref{fig:APS}), 
which is the range covered by the recent Fermi-LAT analysis of the APS of the 
diffuse gamma-ray emission \citep{Ackermann:2012uf}. 

We also discuss the possible effects of modifying some of the assumptions in 
our modeling of the DM distribution. Most notably, we consider two different 
scenarios with a small and a large subhalo contribution (referred to as LOW 
and HIGH throughout the text). Additionally, we study how the energy spectrum 
and APS depend on the value of the minimal self-bound halo mass $M_{\rm min}$. 
A discussion on the effects of using different DM halo profiles is also given.
Quantifying the impact of these uncertainties helps us understanding which
are the ones that primarily affect the APS, as well as to associate a 
``theoretical uncertainty band'' to our predictions.

The main results of our study are:
\begin{itemize}
\item An improvement of the procedure used in \citet{Zavala:2009zr} to 
compute the extragalactic DM-induced intensity introducing independent 
rotations for each of the replicas of the simulation box. This notably reduces 
spurious features in the APS of the simulated maps due to residual 
correlations introduced by the periodicity of the MS-II box.
\item For annihilating DM, the total extragalactic emission (once all 
(sub)halos down to $M_{\rm min}=10^{-6}M_\odot/h$ are considered) is a factor of 
20 (500) larger than the emission produced in the (sub)halos resolved by 
the MS-II simulation. On the other hand, the extragalactic emission for 
decaying DM is dominated by the structures resolved in the simulation, with a 
total intensity that only increases by a factor of 2 once unresolved objects 
are taken into account.
\item The effect of including unresolved subhalos is less important for the 
galactic component, since these are mainly located in the outskirts of the MW
halo, far from the observer, contrary to the nearby GC that produces a 
significant contribution to the signal. Our prediction for the total galactic 
intensity (down to $M_{\rm min}=10^{-6} M_\odot/h$) is between a factor of 2 and 
10 times larger than the emission of the smooth MW (for annihilating DM). The 
contribution of unresolved subhalos is negligible in the case of DM decay.
\item The extragalactic intensity APS in the case of annihilating DM is 
dominated by unresolved (sub)halos. The intensity APS of the total emission
is between 100 and $5 \times 10^4$ times larger than if only the resolved 
MS-II (sub)halos are considered, even though its fluctuation APS is lower 
than the fluctuation APS of the resolved component. In the case of the 
galactic substructures, the intensity APS is dominated by the resolved 
subhalos (which have the largest intrinsic anisotropies of all components) in 
the Aquarius halo (down to $\sim 10^5M_{\odot}$), while unresolved 
subhalos are not expected to contribute.
The total intensity APS is dominated by the smooth DM halo of the MW, at least 
for low multipoles, while above $\ell=300$, the extragalactic contribution 
becomes important (if the HIGH subhalo boost is assumed).
\item The case of decaying DM is quite different: the APS of the smooth MW
halo decreases more rapidly, so that the total intensity APS is dominated 
by extragalactic halos around $\ell=20-30$. Galactic subhalos, on the other 
hand, are characterized by large anisotropies but their low intensity forces 
them to play only a minor role in the total intensity APS.
\item Both for annihilating and decaying DM, the total intensity APS depends 
mainly on structures in the local Universe, with objects located at $z>0.26$ 
contributing to less than 10\% of the total signal.
\item Changing the value of $M_{min}$ from 1 to $10^{-12} M_\odot/h$ has a very
small effect for decaying DM, while our predictions can change dramatically 
for annihilating DM, especially for a HIGH subhalo boost: the left panel of 
Fig. \ref{fig:APS_bands} shows that an uncertainty of almost two orders of 
magnitude is associated with the total intensity APS in this case.
\end{itemize}

In a future work the DM template maps produced here will be used to derive 
constraints on the particle physics nature of DM from a comparison with the 
Fermi-LAT data. 
In Fig. \ref{fig:APS_bands} we made a first comparison for a particular
DM candidate used in this work and find that even if the DM-induced 
fluctuation APS is of the same order of the Fermi-LAT data (for DM 
annihilation), this particular DM candidate is not able to account for the 
bulk of the signal detected by Fermi-LAT since its intensity APS is too low. A 
more rigorous comparison (coupled with a scan over a reasonable set of DM 
models and using a broader energy range) is still required in order to derive 
more conclusive statements.
Based on the energy spectra of the DM candidates considered here relative
to the measured IGRB (see Fig. \ref{fig:IGRB_flux}), the APS of the 2-5 GeV 
energy band shown in Fig. \ref{fig:APS_bands} is likely not the optimal choice 
for setting constraints, but it should be considered as an example for the
comparison between the Fermi-LAT data and our predictions.

It is also important to note that the majority of the IGRB emission is 
expected to be produced by standard astrophysical unresolved sources, such as 
blazars, star-forming galaxies and pulsars. Thus, a complete study of the 
IGRB emission can only be performed with a model that also includes these 
contributions.
In this case, the so-called ``energy anisotropy spectrum'', i.e. the 
fluctuation APS at a fixed multipole but as a function of the energy, is a
particularly useful observable since it has been shown that modulations in 
the energy anisotropy spectrum may mark transitions between regimes where
different classes of sources are responsible for the bulk of the IGRB 
intensity \citep{SiegalGaskins:2009ux}.

In any case, the study of the IGRB energy spectrum and of its anisotropies 
are not the only tools one can resort to for the study of the IGRB nature. 
For instance, in \citet{Xia:2011ax} the authors compute the cross-correlation 
of the Fermi-LAT data with the angular distribution of objects detected in 
different galaxy surveys. Assuming that these objects represent the detected 
counterparts of unresolved astrophysical sources contributing to the IGRB, 
they used the cross-correlation measurement to put constraints on the IGRB 
composition.
Moreover, \citet{Dodelson:2009ih}, \citet{Baxter:2010fr} and 
\citet{Malyshev:2011zi} showed that the analysis of the probability 
distribution of the photon counts can be used efficiently to distinguish a DM 
signal from a cumulative emission of astrophysical sources in the IGRB data.
In principle, the maps produced in the present paper represent unique tools
to extend the techniques exploited in \citet{Xia:2011ax} and 
\citet{Dodelson:2009ih} by including a possible DM contribution. 

\section*{Acknowledgments}
We thank Alberto Dominguez for providing us with tables for the EBL 
attenuation factor. JZ thanks Niayesh Afshordi for fruitful discussions.
We thank Alessandro Cuoco and Anne Green for useful comments and discussions. 
We also thank Chris Hirata and the support of the Consolider-Ingenio 2010 
Programme under grant MultiDark CSD2009-00064. JZ is supported by the 
University of Waterloo and the Perimeter Institute for Theoretical Physics. 
Research at Perimeter Institute is supported by the Government of Canada 
through Industry Canada and by the Province of Ontario through the Ministry of 
Research \& Innovation. 
JZ acknowledges financial support by a CITA National Fellowship. JSG 
acknowledges support from NASA through Einstein Postdoctoral Fellowship grant 
PF1-120089 awarded by the Chandra X-ray Center, which is operated by the 
Smithsonian Astrophysical Observatory for NASA under contract NAS8-03060. 
TD was supported by the Spanish MICINN’s Consolider-Ingenio 2010 Programme 
under grant CPAN CSD2007-00042. We also thank the support of the MICINN under 
grant FPA2009-08958, the Community of Madrid under grant 
HEPHACOS S2009/ESP-1473, and the European Union under the Marie Curie-ITN 
program PITN-GA-2009-237920. 
The calculations for this paper were performed on the ICC Cosmology Machine, 
which is part of the DiRAC Facility jointly funded by STFC, the Large 
Facilities Capital Fund of BIS, and Durham University, and the clusters
at the Max Planck Institute for Astrophysics. We acknowledge use of the 
facilities of the Shared Hierarchical Academic Research Computing Network 
(SHARCNET:www.sharcnet.ca) aCompute/Calcul Canada.

\appendix 
\section{Inverse Compton emission}
\label{sec:IC_emission}
The secondary IC emission has been described in detail in 
\citet{Blumenthal:1970gc}, where the authors also provide useful formul\ae~ 
to reproduce their calculation. This process consists in a transfer of 
momentum from a high energy cosmic ray (CR) electron or positron to a low 
energy photon of the ISRF.

A model for the ISRF provided by \citet{Moskalenko:2005ng} is publicly 
available on the GALPROP webpage\footnote{http://galprop.stanford.edu/}. 
In order to compute the IC emission using semi-analytical methods, it is 
convenient to fit the GALPROP model of the ISRF as a sum of five black-body 
spectra \citep[e.g.][]{Delahaye:2010ji}. One of these is the CMB, while the 
others come from a fit to the model and have less physical meaning, although 
they derive from dust and stellar emissions. In this procedure, it is 
necessary to assume a homogeneous ISRF which might impact on the morphology 
of the resulting gamma-ray emission, although it should be quite moderate 
since variations of the ISRF affect both the $e^+/e^-$ spatial density and 
gamma-ray emissivity in opposite directions.

Apart from the ISRF, one also needs to know the $e^+/e^-$ distribution and 
propagation in the galaxy in order to compute the IC emission. These processes 
are governed by the following diffusion-loss equation (neglecting convection 
and re-acceleration effects):
\begin{equation}
-\nabla \cdot (D(E,\mathbf{x}) \nabla f) - 
\frac{\partial}{\partial E} (b(E) f) = Q(E,\mathbf{x}),
\label{eqn:diffusion}
\end{equation}
where $f(E,\mathbf{x})$ is the $e^+/e^-$ number density per unit of energy at 
the point $\mathbf{x}$, $D(E,\mathbf{x})$ is the diffusion coefficient while 
$b(E)$ describes the energy losses (due to synchrotron and IC emissions). 
Finally $Q(E,\mathbf{x})$ indicates the source term which in our case is 
DM annihilation/decay.

Eq. \ref{eqn:diffusion} governs diffusion inside a so-called diffusion zone, 
outside of which electrons and positrons are not confined by magnetic fields 
and escape from the galaxy. The coefficients defining the different terms in 
Eq. \ref{eqn:diffusion} are constrained by the available observational data 
(mainly the boron-to-carbon CR ratio), but important uncertainties are still 
present (see, e.g. \citealt{Donato:2003xg} and their definition of the 
MIN/MED/MAX scenarios). We use here the semi-analytical methods described in 
\citet{Delahaye:2007fr} which take into account the full expression of the 
energy losses in the Klein-Nishina regime. 

As explained in \cite{Boehm:2010qt}, the morphology of the galactic IC
emission created by the $e^+/e^-$ produced by DM annihilation/decay is 
very sensitive to the choice of the CR propagation parameters and hence, the 
results should be taken with caution. Here we use the same 
propagation model parameters as for the protons and anti-protons related to 
the hadronic emission (see Appendix \ref{sec:hadronic_emission}) and we assume
the MED scenario mentioned above. The uncertainty in the resulting gamma-ray 
intensity can be quite large, depending on the arrival direction, and the 
results can also change with different $e^+/e^-$ propagation models. 
Nevertheless, we neglect this source of uncertainty noting that IC emission 
is relevant only for a fraction of the energy range considered here and only 
for massive DM candidates (see Fig. \ref{fig:energy_spectra}): in the case of 
the $b$-model, with a mass of 200 GeV, the IC emission is located almost 2 
orders of magnitude below the prompt emission and is dominated by interactions 
with the ultraviolet component of the ISRF. In the case of a decaying DM 
particle, though the mass is higher, the signal gets stronger because it is 
not concentrated around the GC and the average over the whole sky is larger. 
Moreover, for the case of decaying DM, the signal is proportional to the 
inverse of the DM mass, whereas in the annihilating case it is inversely 
proportional to its square. For the $\tau$-model the same difference appears 
between annihilation and decay and it is even stronger since the masses are 
the same for both cases.
Moreover, since in this case the prompt emission is much lower than for the 
$b$-model, IC and prompt emission become of comparable importance, especially 
below 10 GeV.

The DM-induced IC emission is implemented in a different way for the 
different components that constitute the emission: for the smooth halo of the
MW, the complete ISRF given by \citet{Moskalenko:2005ng} is used, solving 
Eq. \ref{eqn:diffusion} and considering the propagation of $e^+/e^-$ produced 
by DM annihilation/decay before they interact with the ISRF. On the other 
hand, for the extragalactic (sub)halos and for the galactic subhalos (both 
resolved and unresolved), we only consider IC scattering with the CMB 
photons, and ignore the propagation of $e^+/e^-$ (spatial diffusion is ignored 
in Eq.~\ref{eqn:diffusion}, and only IC energy losses are considered). In 
principle, the secondary IC emission from massive halos (and some of the most 
massive subhalos) may be more realistically described if a full propagation 
model that includes the effect of baryons and secondary emission contributed 
by starlight and infrared light is applied instead (e.g., see 
\citealt{Colafrancesco:2005ji,Colafrancesco:2006he} for the case of the Coma 
galaxy cluster and the Draco dwarf spheroidal). However, the contribution of 
extragalactic structures and galactic subhalos is dominated by low-mass 
objects where star formation is highly suppressed and thus are expected to 
have a rather small stellar component or be devoid of stars.

Finally, we note that when we use the complete ISRF provided in 
\citet{Moskalenko:2005ng}, the template maps for the IC emission (Cartesian 
maps with 90$\times$180 pixels) have a poorer resolution than the maps 
obtained from the prompt emission (HEALPix Maps with {\ttfamily N\_side=512}) 
due to the substantial numerical effort required in solving 
Eq. \ref{eqn:diffusion}. For our purposes it is enough to re-bin the Cartesian 
maps into a HEALPix pixelization.

\section{Hadronic emission}
\label{sec:hadronic_emission}
The hadronic emission is the mechanism that contributes the least to our 
signal but it has a different spatial morphology with respect to the others 
considered in this work. We only account for it in the case of the smooth MW 
halo. It comes from the interaction of CR protons and anti-protons with 
interstellar gas. To compute such a component one needs to derive the 
$p/\bar{p}$ intensity everywhere in the diffusion halo. To do so, we follow the 
semi-analytical method of \citet{Barrau:2001ev} using the propagation 
parameters of the MED scenario in \citet{Donato:2003xg} which gives a good fit 
to the boron-to-carbon observational data. Once the $p/\bar{p}$ distribution 
has been obtained, it is convolved with the gamma-ray production cross 
section\footnote{Note that, following \citealt{Stecker:1967}, we consider 
here that protons and anti-protons have the same cross-section.} (taken from 
\citealt{Huang:2006bp}) and the interstellar gas distribution (taken from 
\citealt{Pohl:2007dz}). See \citet{Timur:2011vv} for more details on the 
computation. 

Contrary to the IC emission, this component is less dependent on the choice 
of propagation parameters due to the fact that protons propagate much further 
than electrons and tend to smooth out all small scale effects. Moreover, the 
hadronic emission naturally follows the interstellar gas distribution.
A source of uncertainty, which we neglect, may come from the presence of DM 
substructures near the galactic disk, which may alter locally the gamma-ray 
intensity. 

We note that the angular resolution of the hadronic component is mainly 
limited by the resolution of the gas maps: $0.5\deg \times 0.5\deg$ 
Cartesian maps. As for the case of the IC emission, these maps are
transformed into HEALPix maps.

\section{Map making for a generic particle physics model}
\label{sec:Particle_physics_method}
In this section we describe the implementation of an approximate method that 
can be used to obtain a full-sky map of the DM-induced extragalactic 
emission for any particle physics model, given a reference map obtained for a 
specific model. Thanks to this method we only need to run once our map-making 
code, saving computation time. 

For the purposes of computing the DM-induced emission, each particle 
physics model is defined by the mass of the DM candidate, its annihilation 
cross section $(\sigma_{\rm ann}v)$ (or decay lifetime $\tau$), and the 
branching fractions for different annihilation (or decay) channels with the
corresponding photon yields, $B_i$ and $dN_i/dE$. Unless the model has a 
velocity-dependent cross section (a case we do not explore here), 
$(\sigma_{\rm ann} v)$ is constant, as it is $\tau$ in the case of decay, and, 
therefore, it is just a multiplicative factor in 
Eqs. \ref{eqn:annihilation_flux} and \ref{eqn:decay_flux}. The photon yield, 
however, depends on redshift. In the case of the galactic halo emission, the 
redshift variation across the DM sources is negligible, but for the 
extragalactic component, the additional integration over redshift links 
together the astrophysical and the particle physics factors in 
Eqs. \ref{eqn:annihilation_flux} and \ref{eqn:decay_flux}.

We benefit from the fact that the past-light cone in our simulated maps is 
divided in concentric shells with a small redshift width $\Delta z$ (see 
Sec. \ref{sec:Millennium_II}). It is then always possible to find a particular 
redshift value contained within $\Delta z$ (called $z_{\rm ref}$) so that we 
can take the factor $dN_\gamma/dE$ outside the integral in 
Eq. \ref{eqn:annihilation_flux} and write the intensity coming from that shell 
as\footnote{An analogous equation for the case of decaying DM can be written, 
of course.}:
\begin{eqnarray}
\frac{d\Phi}{dE}(E_\gamma,\Psi,\Delta z) & = &
\frac{(\sigma_{\rm ann}v)}{8\pi m_\chi^2} \sum_i B_i 
\frac{dN^i_\gamma(E_\gamma(1+z_{\rm ref}))}{dE} \times\nonumber \\
& & \int_{\Delta z} d\lambda(z) \, \rho^2(\lambda(z),\Psi) \, 
e^{-\tau_{\mbox{\tiny{EBL}}}(z,E_\gamma)}.
\label{eqn:z_ref}
\end{eqnarray}
In principle, each line of sight (pixel) in the sky map will have a different 
value of $z_{\rm ref}$ (for the same shell) since the integrand in the r.h.s. 
of Eq.~\ref{eqn:z_ref} changes according to the DM density field in each 
direction. The set of values $\{z_{\rm ref}^i\}$ corresponding to the pixels in 
a given map and their average $\bar{z}_{\rm ref}$ can be determined by 
comparing, pixel per pixel, a full map of the DM-induced intensity (using 
fully Eq.~\ref{eqn:annihilation_flux}) and a map containing only the result 
of the integral in the r.h.s. of Eq.~\ref{eqn:z_ref} (which we call a $J$-map). 
This needs to be done separately for all the different shells since 
$z_{\rm ref}$ changes shell by shell, and then combined to produce the total
observed emission map.

There are no approximations made up to this point. We argue now that a map
for a generic particle physics model can be reconstructed multiplying the
J-map by the corresponding particle physics factor evaluated at the set of 
$\{z_{\rm ref}^i\}$ obtained for our reference case as described above. Moreover, 
to a very good approximation, the pixel average value $\bar{z}_{\rm ref}$ can be 
used instead of the full set $\{z_{\rm ref}^i\}$. We test these arguments by 
using this technique to reconstruct the gamma-ray map for an annihilating 
DM candidate with a mass of 2 TeV, a cross-section of 
$(\sigma_{\rm ann}v)=3 \times 10^{-26}$cm$^{3}$s$^{-1}$ and annihilating only into
$\tau$ leptons obtained from a sky map for a reference case of a DM candidate 
with a mass of 200~GeV, the same cross section, but with an annihilation 
channel into bottom quarks.
We then compare the reconstructed map with one corresponding to the same 
particle physics scenario but obtained directly from the complete map-making 
code.
The test is restricted to the EG-MSII component for an energy of 10~GeV and 
to the simulation output corresponding to $z=3.1$ and $\Delta z=0.25$, which 
is larger than the shell with the largest redshift ``thickness'' we consider 
in this work. 
We find that the reconstructed map has essentially the same APS as the 
original maps, and the average intensities of the two maps agree at the level 
of 1\%. This reconstruction method is not only precise when the reconstructed 
map is obtained accounting for the pixel dependence of $z_{\rm ref}$, but also 
when the constant average value $\bar{z}_{\rm ref}$ is used for all pixels. 

We are then confident that this procedure can be used to reconstruct maps 
of the extragalactic gamma-ray emission for any particle physics model.

\section{Anisotropy from unresolved Galactic subhalos}
\label{sec:Galactic_unresolved_subhalos}
In the present section we described how we implement the method described
in \cite{Ando:2009fp} to compute the APS of galactic unresolved subhalos.

For the subhalo radial distribution we adopt an Einasto profile with 
parameters chosen to match those of the Aq-A-1 main halo: 
$M_{200}=9.4 \times 10^{-11}$~M$_{\odot}$, $r_{-2}=199$~kpc, $c_{-2}=1.24$, and 
$\alpha=0.678$, with the normalization set by the fraction of the smooth halo 
mass $M_{200}$ in subhalos $f_{\rm sub}$.  We require $f_{\rm sub}=0.136$ for 
subhalo masses in the range $1.7 \times 10^{5}$ to $10^{10}$~M$_{\odot}$, which 
is the fraction of the halo mass found in resolved subhalos in Aq-A-1; 
extrapolating the mass function to $M_{\rm min}$ below the minimum resolved 
subhalo mass leads to larger values of $f_{\rm sub}$.  We take the subhalo mass 
function slope to be $-1.9$, and evaluate the anisotropy for several values 
of $M_{\rm min}$.  

The substructure luminosity function for annihilation is determined by 
assuming the subhalo luminosity is related to the subhalo mass by 
$L(M_{\rm sub})= A K (M_{\rm sub}/{\rm M}_{\odot})^\beta$, with 
$K= b_{\rm sh} (\sigma v) N_{\gamma} / (2 m_{\chi}^2)$ and $A$ a normalization set 
related to the ``astrophysical factor''. We consider two sets of the 
mass-luminosity parameters ($A$ and $\beta$), chosen to reproduce the LOW and 
HIGH cases in the text. The HIGH case extrapolates $L(M_{\rm sub})$ to 
$M_{\rm min}$ using the same relation found to fit the resolved subhalos in 
Aq-A-1; the mass-luminosity relation is calibrated to the measured 
mass-concentration relation and assumes each subhalo is well-described by a 
NFW density profile.  
For the HIGH case we take $A=6.48 \times 10^{9} M_{\rm sub}^{2}$ kpc$^{-3}$ and 
$\beta=0.77$. The LOW case assumes $A=3.21 \times 10^{8} M_{\rm sub}^{2}$ 
kpc$^{-3}$ and $\beta=0.86$ for subhalos with $M_{\rm sub} < 1.7 \times 10^{5}$ 
M$_{\odot}$, and the same parameters as the HIGH case for subhalos with 
$M_{\rm sub} > 1.7 \times 10^{5}$ M$_{\odot}$. The LOW case corresponds to a 
scenario in which subhalo concentrations increase more mildly with decreasing 
subhalo mass, and hence in the LOW case the contribution to the intensity and 
APS from low-mass subhalos is reduced relative to the HIGH case. For decay, 
the subhalo luminosity is always directly proportional to the subhalo mass.

We calculated the APS from unresolved subhalos after masking the region
with $|b| < 30^{\circ}$.
We find that for this subhalo model, the contribution to the total intensity 
APS from unresolved subhalos for both annihilation and decay is small. For 
annihilation this contribution is $\sim 10\%$ of the contribution from the 
resolved subhalos for the HIGH case, and $\sim 5\%$ for the LOW case. For 
both the LOW and HIGH cases the majority of this contribution from unresolved 
subhalos originates from subhalos with masses above $\sim 10^{3}$~M$_{\odot}$.  
For decay we find that the contribution from unresolved subhalos is at most a 
few percent of the resolved subhalo anisotropy. Since these contributions are 
small compared to other sources of uncertainty in the APS, we do not include 
them.

\bibliographystyle{mn2e}
\bibliography{bibliography}

\end{document}